\documentclass[a4paper,twocolumn,11pt,accepted=2026-09-25]{quantumarticle}\pdfoutput=1

\usepackage{ulem}
\usepackage[utf8]{inputenc}
\usepackage[english]{babel}
\usepackage[T1]{fontenc}

\usepackage[colorlinks=true]{hyperref} 

\usepackage{amsmath,amsfonts,amssymb,bm,braket,booktabs,dcolumn,graphicx,multirow,afterpage}

\usepackage[numbers,sort&compress]{natbib}

\newcommand{\mr}{\mathrm}
\newcommand{\Om}{\Omega}
\newcommand{\del}{\delta}
\newcommand{\Del}{\Delta}
\newcommand{\tha}{\theta}
\newcommand{\Lam}{\Lambda}
\newcommand{\gam}{\gamma}
\newcommand{\Gam}{\Gamma}
\newcommand{\sig}{\sigma}

\newcommand{\selective}{\triangle}
\newcommand{\nonselective}{\boxdot}
\newcommand{\ensemble}{\boxtimes}
\newcommand{\bath}{\blacksquare}

\newcommand{\bigb}[1]{\left[#1\right]}
\newcommand{\Bigb}[1]{\Bigl[#1\Bigr]}

\newcommand{\Tr}{\operatorname{Tr}}

\newtheorem{definition}{Definition}

\begin{document}

\title{Tutorial: Understanding quantum Zeno and anti-Zeno regimes}

\author{Sacha Greenfield}
 \email[Email contact: ]{srgreenf@usc.edu}
 \affiliation{%
 Department of Physics and Astronomy, and Center for Quantum Information Science and Technology, University of Southern California, Los Angeles, CA 90089, USA
}
\affiliation{%
Institute for Quantum Studies, Chapman University, Orange, CA 92866, USA}

\author{Archana Kamal}
\affiliation{%
Department of Physics and Astronomy, Northwestern University, Evanston, IL 60208, USA
}%

\author{Justin Dressel}
\affiliation{%
Institute for Quantum Studies, Chapman University, Orange, CA 92866, USA}
\altaffiliation[]{Schmid College of Science and Technology, Chapman University, Orange, CA 92866, USA}

\author{Eli Levenson-Falk}
\email[Email contact: ]{elevenso@usc.edu}
 \affiliation{%
 Department of Physics and Astronomy, and Center for Quantum Information Science and Technology, University of Southern California, Los Angeles, CA 90089, USA
}%
 \altaffiliation[]{Department of Electrical and Computer Engineering, University of Southern California, Los Angeles, CA 90089, USA}%

\date{September 25, 2026}

\begin{abstract}
The quantum Zeno effect is a striking feature of quantum mechanics with foundational implications and practical applications in quantum control, error suppression, and error correction. In recent years, the effect has branched off into a variety of different interpretations, making it easy to miss its underlying unifying features. In particular, the quantum Zeno effect has been studied in the context of both selective and nonselective measurements; for both pulsed and continuous interactions; for suppression and enhancement of decay (Zeno / anti-Zeno effects); and even in the absence of measurement entirely. This tutorial presents a unified picture of these effects by examining how they all arise in the pedagogical example of a driven qubit subjected to measurements or dissipation. Zeno and anti-Zeno effects appear as \textit{regimes} of a unified effect that appears when a measurement-like process competes with non-commuting evolution. The current landscape of Zeno and anti-Zeno effects is examined through this unifying lens, with a focus on experimental applications and implementations. Thus, this tutorial is ideal for new researchers interested in learning about cutting-edge tools in quantum control and measurement, while remaining highly relevant to quantum computing specialists aiming to understand the quantum Zeno effect's applications in adjacent subfields or apply it in a novel way in their own research. The quantum Zeno effect is found to be both ubiquitous and essential for the future of near-term quantum computing. 
\end{abstract}

\maketitle

\section{Introduction}

Since its conception by \citet{vonneumannMathematicalFoundationsQuantum1932} in 1932 and popularization by \citet{misraZenoParadoxQuantum1977} in 1977, the quantum Zeno effect has been a striking feature of quantum measurement theory with foundational implications and practical applications. In essence, the quantum Zeno effect is the inhibition of natural state evolution by frequent measurements, which repeatedly collapse the quantum state back to the nearest eigenstate of the measurement with high probability, thus preventing evolution away from that eigenstate. This is equivalent to suppression of state mixing. Originally a theoretical curiosity framed as a paradox of motion, the effect now sees widespread experimental application in quantum control, open systems engineering, error suppression, and error correction. These diverse applications have correspondingly prompted different communities to develop disparate physical interpretations and domain-specific terminology for the quantum Zeno effect, inadvertently hiding its universality. For example, the anti-Zeno effect --- the \textit{enhancement} of decay by measurement --- has seen extensive study in certain subfields while being largely overlooked by others, despite its prevalence under realistic experimental measurement conditions. In this tutorial, we introduce a unified picture of quantum Zeno and anti-Zeno effects that clarifies key physical features and establishes a cohesive terminology to describe their various manifestations.

The quantum Zeno paradox was first pointed out (albeit not by that name) by \citet{vonneumannMathematicalFoundationsQuantum1932} in his 1932 treatise on quantum measurement theory. After introducing the measurement collapse postulate, von Neumann outlines a scenario in which a quantum system is initialized in a measurement eigenstate and the measurement basis is then rotated adiabatically (i.e., much slower than the measurement rate). The system remains in the measurement eigenstate, a principle now known as ``Zeno dragging'' \cite{aharonovMeaningIndividualFeynman1980,altenmullerDynamicsMeasurementAharonovs1993,hacohen-gourgyIncoherentQubitControl2018,lewalleOptimalZenoDragging2024}.

The phenomenon was largely forgotten until being revisited as a paradox: If measurements of a quantum system inhibit evolution, how can particles move through a bubble chamber that is continuously monitoring them? In 1977, \citet{misraZenoParadoxQuantum1977} popularized this conundrum under the name ``quantum Zeno paradox'', alluding to the paradox of motion formulated by the Greek philosopher Zeno \cite{huggettZenosParadoxes2024,harringtonEngineeredDissipationQuantum2022}. As originally formulated, the paradox is defined in terms of a sequence of identical measurement outcomes. Nevertheless, while early experimental studies \cite{itanoQuantumZenoEffect1990,nagelsQuantumZenoEffect1997} physically performed a measurement, the record of specific measurement results was not used. This sparked debate on whether selection on a measurement record---``selectivity''---is essential to understanding (or even defining) the quantum Zeno effect \cite{peresIncompleteCollapsePartial1990,frerichsQuantumZenoEffect1991,beigeProjectionPostulateAtomic1996,nakazatoUnderstandingQuantumZeno1996,mackDecoherenceGeneralizedMeasurement2014}.  

Recent applications of the quantum Zeno effect fall broadly into two categories according to selectivity. The first---bath-engineering, or autonomous error correction---uses nonselective measurements to suppress ensemble-averaged evolution away from a protected state or subspace. This category encompasses such diverse topics as quantum Zeno subspaces \cite{facchiQuantumZenoSubspaces2002,facchiQuantumZenoDynamics2008,schaferExperimentalRealizationQuantum2014,sorensenQuantumControlMeasurements2018,ruskovQuantumZenoStabilization2006,marcantoniUltrastrongCouplingNonselective2025}\footnote{A Zeno subspace is a linear span of eigenstates of a measurement that share the same eigenvalue.},
decoherence-free subspaces \cite{beigeQuantumComputingUsing2000}, photon-blockading \cite{bretheauQuantumDynamicsElectromagnetic2015,chakramMultimodePhotonBlockade2022}, dissipatively protected cat-state encodings \cite{mirrahimiDynamicallyProtectedCatqubits2014,touzardCoherentOscillationsQuantum2018,lescanneExponentialSuppressionBitflips2020,zapletalStabilizationMultimodeSchrodinger2022}, and other forms of dissipation engineering \cite{harringtonEngineeredDissipationQuantum2022}.  Notably, quantum anti-Zeno studies---both experimental \cite{fischerObservationQuantumZeno2001,alvarezZenoAntiZenoPolarization2010,harringtonCharacterizingStatisticalArrow2019,thorbeckReadoutinducedSuppressionEnhancement2024} and theoretical \cite{kofmanQuantumZenoEffect1996,kofmanAccelerationQuantumDecay2000,kofmanFrequentObservationsAccelerate2001,kofmanZenoAntiZenoEffects2001, facchiQuantumZenoInverse2001,zhangDynamicsQuantumZeno2014,chaudhryZenoAntiZenoEffects2014,chaudhryGeneralFrameworkQuantum2016,zhouQuantumZenoAntiZeno2017,lassalleConditionsAntiZenoeffectObservation2018,zhangCriterionQuantumZeno2018,majeedQuantumZenoAntiZeno2018,heQuantumZenoEffect2019,naRevisitingQuantumZeno2021}---utilize a primarily ensemble-averaged framework. As a bath-engineering technique, quantum Zeno is ubiquitous even as it often goes by other names.

The second category of applications that display quantum Zeno effects---quantum error detection and active error correction---uses selective measurements to suppress state evolution away from a protected subspace in individual realizations, detect escapes, and correct errors via feedback. These selective measurements, in addition to providing information about errors that have already occurred, can passively suppress non-Markovian error rates or unitary evolution between subspaces; this is a manifestation of the selective quantum Zeno effect \cite{erezCorrectingQuantumErrors2004}. Demonstrated in diverse systems such as optical \cite{kwiatInteractionFreeMeasurement1995,kwiatHighefficiencyQuantumInterrogation1999}, ion trap \cite{balzerQuantumZenoEffect2000,balzerRelaxationlessDemonstrationQuantum2002}, and superconducting circuit-based architectures \cite{gambettaQuantumTrajectoryApproach2008,vijayObservationQuantumJumps2011,slichterQuantumZenoEffect2016,guttelGraduallyOpeningSchrodingers2026}, the selective quantum Zeno effect has been applied in real-time error detection and correction \cite{mohseniniaAlwaysQuantumErrorTracking2020,livingstonExperimentalDemonstrationContinuous2022}, entanglement generation \cite{blumenthalDemonstrationUniversalControl2022,gaikwadEntanglementAssistedProbe2024}, and Zeno dragging in the form of a measurement-based gate \cite{hacohen-gourgyIncoherentQubitControl2018}. Selective Zeno effects can also add a passive improvement component to active quantum error correction schemes such as syndrome detection \cite{cramerRepeatedQuantumError2016} when non-Markovian errors are present.

In this tutorial, we offer a unified picture of both selective and nonselective versions of the quantum Zeno effect in both the Zeno regime and the anti-Zeno regime, while highlighting the broad relevance of the effect to state-of-the-art quantum control. We feature a pedagogical model---a driven and damped qubit oscillation---which exemplifies the key features of Zeno physics. We introduce this model in Sec. \ref{sec:minimal-model}. Section \ref{sec:witness} introduces key terminology and witnesses for both Zeno and anti-Zeno regimes of the effect, while Sec. \ref{sec:pulsed-quantum-zeno} introduces the pulsed quantum Zeno effect and analyzes its key features in terms of these witnesses. Section \ref{sec:continuous-quantum-zeno} examines analogous features for continuous measurements. Subsequently, Sec. \ref{sec:zeno-anti-zeno} examines the quantum Zeno effect when the unitary evolution is nonorthogonal to the measurement. We examine informational quantum trajectories in \textit{both} Zeno and anti-Zeno regimes, which we highlight as an original result of this paper\footnote{The anti-Zeno effect has been studied almost exclusively as an ensemble-averaged effect, with \citet{ruseckasQuantumTrajectoryMethod2006} being a notable exception. However, while Ruseckas and Kaulakys studied simulated trajectories of an atom coupled to a fluorescing two-level detector in both Zeno and anti-Zeno regimes, they display and comment on trajectory-level effects only in the Zeno regime; only averages of trajectory ensembles are displayed for anti-Zeno regimes. Thus our focus on trajectories in \textit{both Zeno and anti-Zeno regimes is an important contribution of our paper.}}. In Sec. \ref{sec:experimental-connections}, we reexamine connections to the literature in detail and reflect on the relevance of the quantum Zeno effect for quantum control, error suppression, detection, and correction.

\section{Minimal qubit model}
\label{sec:minimal-model}

Throughout our discussion, we examine a quantum bit (qubit) with oscillatory Hamiltonian evolution being disrupted by repeated quantum non-demolition (QND) measurements, i.e., measurements that distinguish the eigenstates without disturbing them. The quantum Zeno effect arises because the repeated measurements interrupt the natural oscillatory evolution. We distinguish between selective measurements that are conditioned on obtaining specific measurement outcomes and nonselective measurements that are averaged over all possible outcomes that could have occurred (Sec. \ref{sec:witness}).   
This simple abstract system will map to a variety of experimental implementations, which we discuss further in Sec. \ref{sec:experimental-connections}.

\begin{figure*}
\includegraphics[width=\linewidth]{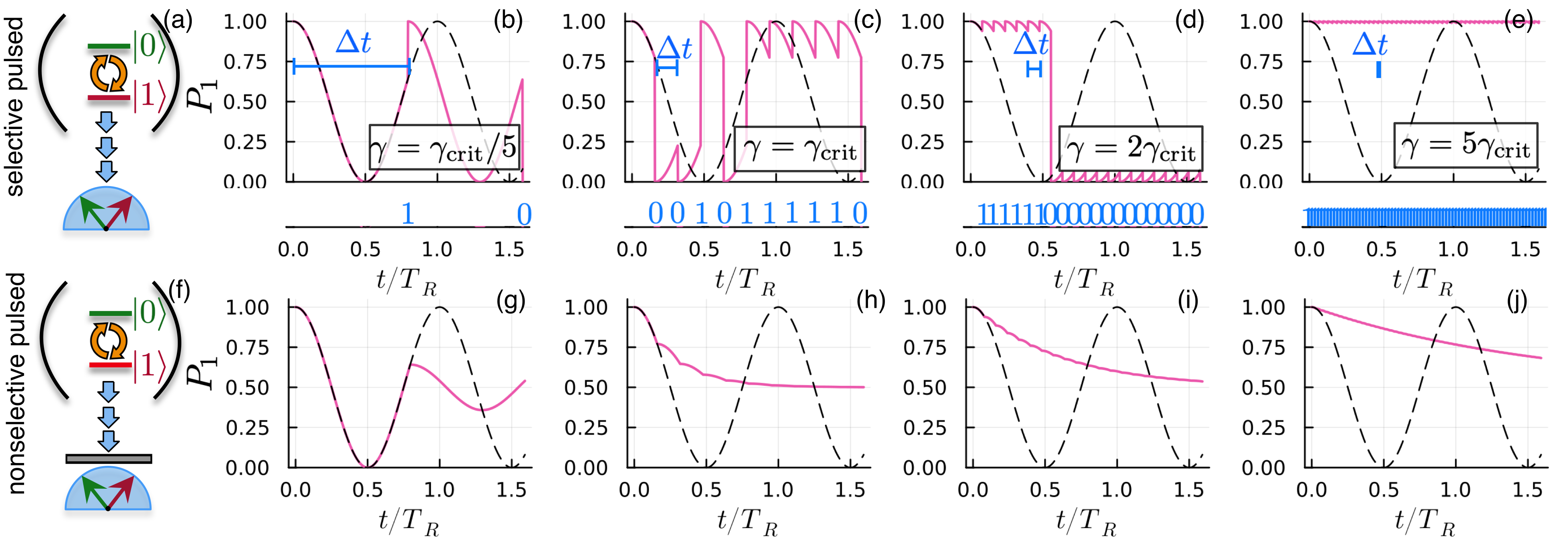}
	\caption{Population evolution $P_1(t)$ of the initial qubit state $\ket{1}$ under selective (top) and nonselective (bottom) pulsed projective measurements of the $\{\ket{0},\ket{1}\}$ basis. Different pulse repetition rates $\gamma = 1/\Del t$ are compared to a critical rate, $\gamma_\mr{crit} = \Omega_R$, where $\Omega_R = 2\pi/T_R$ is the Rabi frequency of Hamiltonian evolution. (a) For selective measurements, the device records the result for each pulse in a sequence, yielding a conditional quantum trajectory determined by the record. (b) Infrequent projections (compared to the critical rate) interrupt the underlying Rabi oscillation (black dashed line) every $\Delta t$, with qubit evolution correlated with the observed measurement record ($1$s and $0$s). This leads to average mixing of the qubit $\ket1$ population (pink trace) that is increased by more frequent measurements (anti-Zeno regime). (c) At $\gam = \gam_\mr{crit}$, the qubit randomly jumps between eigenstates similarly to telegraph noise, and measurement-induced mixing is enhanced compared to (b). (d) As the pulse rate increases past $\gam_\mr{crit}$ (moving into the Zeno regime), the random jumps become less frequent and the evolution remains confined near the eigenstates. (e) For very rapid measurements, the state is pinned to the initial state $\ket1$ and does not evolve, yielding a quantum version of Zeno's paradox. (f) For nonselective measurements, no measurement record is available, so the best estimate of the evolution that one can make is the ensemble average over all possible selective trajectories that could have occurred. (g) Nonselective (ensemble-averaged) evolution resembles a damped oscillation, with loss of coherence occurring at discrete times $\Delta t$. (h) At $\gam_\mathrm{crit}$, the average evolution approximates exponential decay. (i,j) The exponential mixing rate from the initial eigenstate is suppressed with increasing $\gamma$. In the limit $\gam \rightarrow \infty$ of Zeno's paradox, the ensemble-averaged evolution also remains pinned to the initial state $\ket{1}$.}
 \label{fig:selective-nonselective-pulsed}
\end{figure*}

We choose both the oscillation and measurement axes to lie in the $x$-$z$ plane of the qubit Bloch sphere, with the two measurement eigenstates, $\ket1$ and $\ket0$, defining the Pauli operators,
\begin{align}
        \hat \sigma_x &\equiv \ket1 \!\bra0 + \ket0\!\bra1, &
        \hat \sigma_z &\equiv \ket1 \!\bra1 - \ket0\!\bra0,
\end{align}
such that $\hat \sigma_z$ corresponds to the measurement axis. We consider both pulsed projective measurements of $\hat \sigma_z$ with tunable repetition rate $\gam \equiv 1/\Del t$ (Sec. \ref{sec:pulsed-quantum-zeno}, Fig.~\ref{fig:selective-nonselective-pulsed}) and continuous weak measurements characterized by the same information acquisition rate $\gam$ (Sec. \ref{sec:continuous-quantum-zeno}, Fig.~\ref{fig:selective-nonselective-continuous}). For simplicity, we start the qubit in the initial state $\ket{1}$ and will track the dynamics of the time-dependent population $P_1(t)$ of this initial state as a natural witness for the effect.

In absence of measurement, the qubit Hamiltonian,
\begin{subequations}
    \begin{align}
        \hat H/\hbar &= (\Om/2) \hat \sig_\theta \equiv (\Om_R/2)\hat \sig_x + (\Del/2) \hat \sig_z,
     \\
     \hat \sig_\theta &\equiv \hat \sig_z \cos \tha  + \hat \sig_x \sin \tha
    \end{align}
    \label{eq:H-general}
\end{subequations}
generates unitary oscillations,
\begin{equation}
    \hat U(t) = \exp(-i\hat H t/\hbar) = \exp(-i\Omega t\, \hat{\sigma}_\theta/2)
\end{equation}
at frequency $\Om = \sqrt{\Omega_R^2 + \Delta^2}$ around an axis tilted by an angle $\theta = \arctan(\Omega_R / \Delta)$ in the $x$-$z$ plane from the $z$-axis of the Bloch sphere. The quantum Zeno regime will occur when the measurement rate $\gamma$ dominates the unitary oscillation frequency $\Om$, in which case the non-unitary state collapse from the measurement will prevent the Hamiltonian evolution from occurring. 

The (Rabi) frequency component $\Omega_R$ about the $x$-axis has a generator, $\hat{\sigma}_x$, that does \textit{not} commute with $\hat \sigma_z$, so will be affected by the measurement. In contrast, the 
(detuning) frequency component $\Delta$ about the $z$-axis has a generator that \textit{does} commute with $\hat \sigma_z$, so will not be affected by the measurement. To elucidate the basic quantum Zeno effect, we will first consider orthogonal measurement and oscillation axes (i.e., $\Delta = 0$ and $\theta = \pi/2$) in Sections ~\ref{sec:witness}--\ref{sec:continuous-quantum-zeno}. The measurement of $\hat \sigma_z$ suppresses the noncommuting evolution generated by $\hat \sigma_x$. We then consider in Sec.~\ref{sec:zeno-anti-zeno} nonorthogonal measurement and oscillation axes (specifically, $\Delta > \Omega_R$ and $\theta < \pi/4$), which will permit richer structure for the anti-Zeno regime.

\section{Witnessing the quantum Zeno effect}\label{sec:witness}
In the authors' view, conflation of concepts and terminology in quantum Zeno effect literature has led to confusion about how to define and witness the effect, particularly regarding the role of selective and nonselective measurements \cite{nakazatoUnderstandingQuantumZeno1996,mackDecoherenceGeneralizedMeasurement2014}. We review these concepts here in the context of the minimal model and define a witness of the quantum Zeno effect that will be suitable for both cases.

\textbf{Selective measurements - } A \textit{selective measurement} selects on a \textit{particular outcome} from the measurement, inducing measurement backaction from information gain (i.e., quantum state collapse) in addition to the natural Hamiltonian evolution. Repeated measurements yield a sequence of measurement outcomes called a \textit{measurement record}. The full evolution conditioned on all outcomes of a particular measurement sequence is termed the \textit{quantum state trajectory} for a single experimental realization \cite{gardinerInputOutputDamped1985,barchielliMeasurementTheoryStochastic1986,diosiContinuousQuantumMeasurement1988,belavkinQuantumContinualMeasurements1992,carmichaelQuantumTrajectoryTheory1993,wisemanQuantumTheoryFieldquadrature1993,gardinerQuantumNoiseHandbook2000,korotkovSelectiveQuantumEvolution2001,gambettaQuantumTrajectoryApproach2008,carmichaelOpenSystemsApproach1993,wisemanQuantumMeasurementControl2009,korotkovQuantumBayesianApproach2011,korotkovQuantumBayesianApproach2016}. In general, measurements can be \textit{inefficient} and lose information, yielding mixed quantum state trajectories $\hat \rho(t)$ that average over the lost information as a best mean estimate. However, for simplicity, our model considers lossless measurements that lead to pure quantum state trajectories, $\ket{\psi(t)}$.

To illustrate the quantum Zeno effect in their popular 1977 article, \citet{misraZenoParadoxQuantum1977} showed that the probability of measuring the initial state as the eigenstate of \textit{every one} of a sequence of projective measurements converges to $1$ in the limit of infinitely rapid measurements. In terms of our example, this equates to starting the qubit in $\ket{1}$, measuring $\hat \sigma_z$ projectively at a pulse repetition rate $\gam = 1/\Delta t$, and seeing the sequence of eigenvalues $(1,1,\ldots)$ with a probability limiting to 1 as $\gam \to \infty$ (as shown in Fig.~\ref{fig:selective-nonselective-pulsed}(a,b,c,d,e)). Their treatment thus used selective measurements, and identified the probability of a particular measurement sequence as a witness for the suppression of inter-measurement Hamiltonian evolution.

\textbf{Nonselective measurements -} In contrast, a \textit{nonselective measurement}   has no selectable outcome. The outcome is either not recorded (e.g., for ``passive'' measurements by a coupled environment), or is ``actively'' recorded by a selective measurement but then discarded (or lost). The best estimate of the measured system's evolution after a nonselective measurement is the average over all possible outcomes that \textit{could have} occurred. Thus, the mean measurement backaction for a QND nonselective measurement is equivalent to an \textit{incoherent dephasing event} that leaves the measurement basis invariant while decohering superpositions of the measurement eigenstates. Information about the system is lost, so the evolution can no longer be described by the pure state dynamics of a closed system; instead, the ensemble evolution must be described by a density operator $\hat \rho(t)$. 

In absence of a measurement record, the probability of repeatedly finding the same measurement eigenstate is no longer well-defined (or experimentally accessible). Instead, a similar ensemble-compatible witness for the quantum Zeno effect is to check the ensemble-likelihood that a measurement eigenstate (here the initial state $\ket{1}$) would be projectively measured at time $t$,
\begin{equation}\label{eq:survival-probability}
    P_1(t) \equiv \braket{1 | \hat\rho(t) | 1} \equiv (1 +z(t))/2,
\end{equation}
where $z(t) \equiv \mathrm{Tr}(\hat \sig_z ~ \hat \rho(t))$ is the qubit Bloch coordinate. For rapid measurements, the evolution of this ensemble-likelihood $P_1(t)$ is also suppressed and can witness the suppression of evolution characteristic of the quantum Zeno regime. Indeed, in the limit of infinitely rapid measurements, $\gam \to \infty$, $P_1(t)$ also limits to 1.

For slower measurement rates, however, the ensemble-likelihood $P_1(t)$ averages over all jumps to and from distinct eigenstates, unlike witnesses that use selective measurements on individual trajectories. This additional averaging leads to an approximately exponential decay envelope for $z(t)$ that damps to a mixture of the measurement eigenstates for measurements faster than the natural evolution at rate $\Omega$, 
\begin{align}\label{eq:escape-rate}
    z(t) &\propto \exp(-\Gam_\mr{mix}~ t), & \gamma \gg \Omega,
\end{align}
as shown in Fig.~\ref{fig:selective-nonselective-pulsed}(f,g,h,i,j). The associated average rate of escape from the initial eigenstate in this regime, $\Gam_\mr{mix}$, is thus a key figure of merit for the quantum Zeno effect. In the limit of infinitely rapid measurements, $\Gam_\mr{mix} \to 0$ and all evolution ceases. By contrast, for weak measurements $\Gamma_\mathrm{mix}$ generally increases with $\gamma$, coinciding with the intuition that measurement dephases the system. Importantly, this figure of merit also applies to selective measurements after ensemble-averaging.

We thus adopt this ensemble-compatible definition as our primary witness for the quantum Zeno effect:
\begin{definition} \label{def:qze}
\textbf{(Quantum Zeno Effect)} The \textit{average} (nonselective) natural state evolution is modified by the measurement-induced dephasing such that the probability of remaining in an initial eigenstate has an approximately exponential decay envelope with rate $\Gam_\mr{mix}$.
\end{definition}

This definition has the added benefit of distinguishing two \textit{regimes} of the quantum Zeno effect. The quantum Zeno regime of rapid measurements is considered counterintuitive because measurement disruptions stabilize, rather than dephase, the quantum state. By contrast, the intuitive behavior is for mixing to \textit{increase} with measurements. In the case of coupling to a detuned environment, the enhancement of decay with measurement has been dubbed the anti-Zeno effect \cite{fischerObservationQuantumZeno2001,alvarezZenoAntiZenoPolarization2010,harringtonCharacterizingStatisticalArrow2019,thorbeckReadoutinducedSuppressionEnhancement2024,kofmanQuantumZenoEffect1996,kofmanAccelerationQuantumDecay2000,kofmanFrequentObservationsAccelerate2001,kofmanZenoAntiZenoEffects2001, facchiQuantumZenoInverse2001,zhangDynamicsQuantumZeno2014,chaudhryZenoAntiZenoEffects2014,chaudhryGeneralFrameworkQuantum2016,zhouQuantumZenoAntiZeno2017,lassalleConditionsAntiZenoeffectObservation2018,zhangCriterionQuantumZeno2018,majeedQuantumZenoAntiZeno2018,heQuantumZenoEffect2019,naRevisitingQuantumZeno2021}. It is even more ubiquitous in our framework: In general, as $\gamma$ increases from zero there is a regime of measurement-induced mixing enhancement that occurs prior to measurement-induced mixing suppression.  

We quantify these distinct regimes through the \textit{derivative} of the mixing rate with respect to the measurement rate $\gam$, which we term the \textit{Zeno response}:
\begin{equation}
    \mathcal{R}_Z(\gam) \equiv -d\Gam_\mr{mix}/d\gam.
\end{equation}

The two regimes of the quantum Zeno effect are then identified by the sign of the Zeno response:
\begin{definition} \label{def:anti-zeno-regime}
\textbf{(anti-Zeno regime)} $\Gam_\mr{mix}$ is \textbf{enhanced} with increasing measurement rate, i.e. \textbf{negative} Zeno response $\mathcal{R}_Z(\gam) < 0$.
\end{definition}
\begin{definition} \label{def:zeno-regime}
\textbf{(Zeno regime)} $\Gam_\mr{mix}$ is \textbf{suppressed} with increasing measurement rate, i.e. \textbf{positive} Zeno response $\mathcal{R}_Z(\gam) > 0$.
\end{definition}

The definition of the quantum Zeno effect in terms of an overall transition rate that is suppressed with more frequent or stronger measurements is ubiquitous in ensemble studies such as \cite{itanoQuantumZenoEffect1990,streedContinuousPulsedQuantum2006,kakuyanagiObservationQuantumZeno2015,xiaoNMRAnaloguesQuantum2006,palacios-laloyExperimentalViolationBell2010}, and the relevance also extends to trajectories studies where $\Gamma_\mathrm{mix}$ is determined by the rate of quantum jumps between the measurement eigenstates \cite{slichterQuantumZenoEffect2016}. Thus, the explicit definition of the Zeno response introduced here formalizes what has been implicit in the literature\footnote{\citet{zhangCriterionQuantumZeno2018} also note the implicit use of this derivative throughout the literature in defining Zeno and anti-Zeno regimes.}.

Similarly, defining anti-Zeno as a regime of negative Zeno response at a particular measurement rate harkens back to the original conception of these effects as trends \cite{kofmanFrequentObservationsAccelerate2001,facchiQuantumZenoInverse2001,facchiQuantumZenoDynamics2008} and is in accordance with the formal definitions proposed by \citet{chaudhryGeneralFrameworkQuantum2016,zhangCriterionQuantumZeno2018,liUnificationQuantumZenoanti2023}. At the same time, it is consistent with definitions that compare a non-Markovian decay rate undisturbed by measurement to one disturbed by measurement, and see if the latter is larger (``anti-Zeno'') or smaller (``Zeno'') than the former \cite{fischerObservationQuantumZeno2001,harringtonQuantumZenoEffects2017}. The Zeno response is a more general and precise metric than point-wise comparison: more general because it also applies to systems where the concept of an ``undisturbed decay rate'' doesn't exist (i.e., one where the undisturbed evolution is purely unitary, as in our minimal model), and more precise because it is properly defined at a \textit{particular} value of $\gamma$.

Because the anti-Zeno regime generally precedes a Zeno regime, the anti-Zeno regime also functions as a kind of \textit{regime of onset} of the Zeno regime, as in \cite{snizhkoQuantumZenoEffect2020,kumariQubitControlUsing2023,kumarQuantumZenoEffect2020}.

Finally, our definition in terms of a decay \textit{envelope} rate generalizes the effect to additionally encompass regimes that exhibit additional dynamical structure, such as the underdamped oscillations seen in the anti-Zeno regime of our minimal example. This evokes generalized definitions (in terms of decay envelopes or purity) such as \cite{facchiControlDecoherenceAnalysis2005,alvarezZenoAntiZenoPolarization2010,kumarQuantumZenoEffect2020,gaikwadEntanglementAssistedProbe2024}.

\section{Pulsed quantum Zeno effect}
\label{sec:pulsed-quantum-zeno}
We first examine the quantum Zeno effect as it arises for the traditional case of pulsed projective measurements, illustrated in Fig. \ref{fig:selective-nonselective-pulsed}. For simplicity, we start with $\Delta = 0$ in Eq.~\eqref{eq:H-general}, such that the measurement of $\hat\sigma_z$ and the oscillation around $\hat\sigma_x$ do not commute and thus maximally disrupt each other. We term this the orthogonal case, since the two associated Bloch sphere axes are orthogonal. 

Between each projective measurement, the system oscillates between its two states at the Rabi frequency $\Omega_R = 2\pi/T_R$, such that
\begin{subequations}
    \begin{align}
		\ket{\psi(t)} &= \hat U(t) \ket1,\quad \hat{U}(t) = \exp(-i(\hat{H}/\hbar)t), \\
        &=  \cos (\Omega_R t / 2) \ket 1 - i \sin (\Omega_R  t / 2)  \ket 0.
    \end{align}
\end{subequations}
Every $\Delta t$, a measurement interrupts this unitary evolution and projects the state to a measurement eigenstate conditioned on the observed binary readout $r=0,1$. 

In anticipation of the generalization to continuous measurements in Sec. \ref{sec:continuous-quantum-zeno}, we formally describe the backaction of the measurement on the state by a measurement collapse (Kraus) operator.
    \begin{align}
    \label{eq:pulsed-POVM}
        \hat{\Lam}_r &\equiv \ket{r}\!\bra{r}, & r &= 0,\, 1.
    \end{align}
Each possible readout $r$ occurs with probability
\begin{align}
    P(r) &= \braket{\psi |\hat{P}_r |\psi}, & \hat{P}_r &\equiv \hat\Lam_r^\dagger \hat\Lam_r,
    \label{eq:binary-r-sampling}
\end{align}
and induces the (renormalized) state collapse,
\begin{equation}
             \ket{\psi} \mapsto \mathcal{M}_r \ket{\psi} \equiv \hat \Lambda_r  \ket{\psi} / \sqrt{P(r)}
            \label{eq:pulsed-projection}
\end{equation}
for pure state vectors, which generalizes to
\begin{align}
   \hat{\rho} \mapsto \mathcal{M}_r(\hat{\rho}) \equiv \hat{\Lambda}_r\hat{\rho}\hat{\Lambda}_r^\dagger / P(r)
   \label{eq:pulsed-projection-rho}
\end{align}
for mixed state density operators. The operators $\hat{P}_r$ form a \textit{positive operator-valued measure} (POVM), $\sum_r \hat{P}_r = \hat{1}$, over the readout index $r$. Since the readout $r$ is perfectly correlated with the measurement eigenstates in this case, $\hat\Lambda_r$ and $\hat P_r$ are projection operators that coincide.

\textbf{Selective pulsed measurements - } In Fig. \ref{fig:selective-nonselective-pulsed}(a,b,c,d,e) we show the quantum Zeno effect for individual quantum trajectories initialized in state $\ket{1}$ as the measurement repetition rate $\gamma = 1/\Delta t$ increases. 

\renewcommand{\arraystretch}{1.0}
\begin{figure*}[!t]\includegraphics[width=\linewidth]{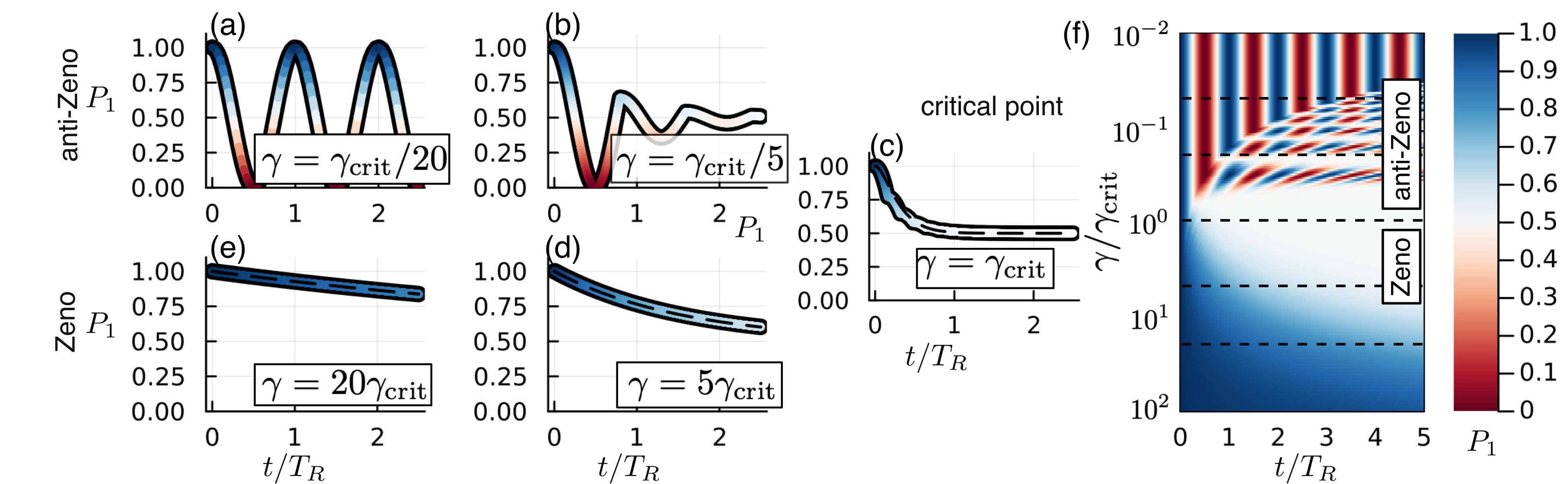}
	\caption{Nonselective measurement population evolution $P_1(t)$ for different measurement rates $\gamma$ that increase clockwise from the upper left plot. The anti-Zeno (a,b), critical (c), and Zeno (d,e) regimes are determined by the size of $\gam$ relative to the critical rate $\gam_{\mr{crit}} = \Omega_R$. (a,b) The periodic dephasing of the Rabi oscillations produces an oscillation envelope that decays approximately exponentially. (c) At the critical point, evolution transitions to approximately exponential decay with rate given by Eq.~\eqref{eq:escape-rate-pulsed} (dashed black line), and the overall decay rate is maximized. (d,e) Faster measurements suppress the exponential decay rate.  (f) The heatmap of $P_1$ (color) versus time (horizontal axis) and $\gam / \gam_\mr{crit}$ (vertical axis in log scale) shows how the transition from decay envelope enhancement (anti-Zeno) to decay envelope suppression (Zeno) also corresponds to the underdamped to overdamped transition, with horizontal dashed lines corresponding to the cases shown in (a-e).}
 \label{fig:pulsed-heatmap-stacked}
\end{figure*}

Starting with initial state $\ket{1}$ and making $N$ repeated measurements at intervals $\Delta t$ yields a quantum state trajectory, 
\begin{equation}
    \ket{\psi(N\Delta t)} = \left[\prod_{i=1}^N \mathcal{M}_{r_i}\hat{U}(\Delta t) \right]\ket{1},
\end{equation} 
labeled by a measurement record consisting of a binary sequence $(r_1,r_2,\ldots,r_N)$ with $r_i\in\{0,1\}$. Each sequence occurs with probability $P(r_1,\ldots,r_N) = ||\prod_{i=1}^N[\hat\Lambda_{r_i}\hat{U}(\Delta t)]\ket{1}||^2$, which for this projective case simplifies to 
\begin{equation}
    P(r_1,\ldots,r_N) = (P_{\rm stay})^{N-m}(P_{\rm flip})^m
\end{equation}
with $m$ counting the number of bit flips in the record (i.e., when $r_i \neq r_{i-1}$), and where
\begin{align}\label{eq:pstay}
    P_{\rm stay} &= \cos^2\left(\Om_R \Del t/2\right), & P_{\rm flip} &= 1 - P_{\rm stay}.
\end{align}
For measurements that are rapid compared to the Rabi oscillation frequency, $\gamma = 1/\Delta t \gg \Omega_R$, $P_{\rm flip} \approx (\Omega_R/2\gamma)^2 = \Gamma_{\rm jump}\Delta t$, and the flip probability is characterized by an asymptotic \textit{jump rate},
\begin{align}
    \Gamma_{\rm jump} &= \Omega_R^2/4\gamma. & (\gamma \gg \Omega_R)
\end{align}
In the rapid measurement limit $\gam \to \infty$, $\Gamma_{\rm jump} \to 0$, so the only trajectory that occurs with non-negligible probability is the one with no flips, $(1,1,\ldots,1)$, in which case the state remains pinned to the initial state $\ket{1}$. This prevention of state evolution by making sufficiently rapid measurements is the selective quantum Zeno's paradox.

We simulate and show example stochastic state trajectories for different repetition rates $\gam$ in Fig. \ref{fig:selective-nonselective-pulsed}(b-e). As the repetition rate increases, the natural oscillatory evolution gets interrupted by measurement (Fig. \ref{fig:selective-nonselective-pulsed}(b)), causing expected measurement-induced dephasing, until it is suppressed and replaced by mostly telegraphic evolution (Fig. \ref{fig:selective-nonselective-pulsed}(c,d)) in the form of random jumps at the rate $\Gamma_{\rm jump}$ between confined regions near the measurement eigenstates. For a sufficiently large measurement rate $\gamma$, both the jumps and intermediate evolution become suppressed and the state remains essentially pinned to the initial eigenstate (Fig. \ref{fig:selective-nonselective-pulsed}(e)).

\textbf{Nonselective pulsed measurements -} In Fig. \ref{fig:selective-nonselective-pulsed}(f,g,h,i,j) we show the corresponding ensemble-averaged (nonselective) perspective of the quantum Zeno effect. This complementary perspective has broader applicability, since it can include ensemble-averaged dephasing sources without an obvious measurement record. 

In the absence of a measurement record, we instead consider the probability in Eq.~\eqref{eq:survival-probability} of measuring the initial state $\ket{1}$ at the end of the sequence of ensemble-dephasing events. For simplicity of calculation, we consider a duration $T = N\Delta t$ in which $N-1$ unrecorded projective measurements have occurred, followed by one final measurement that is checked. The last measurement will yield $1$ if an even number of bit flips have occurred over the total $N$ measurements. Thus, we can calculate the probability $P_1$ as the sum over all possibilities with an even number of flips in an $N$-measurement record. This calculation yields,
\begin{align}\label{eq:pulsed-survival}
  P_1(N\Delta t) &= \sum_{m=0}^{\lfloor N/2 \rfloor} \binom{N}{2m} P_{\rm stay}^{N-2m}P_{\rm flip}^{2m}, \nonumber \\
  &= (1 + \cos^{N}(\Omega_R/\gamma))/2,
\end{align}
using Eqs.~\eqref{eq:pstay} to simplify $P_{\rm stay} - P_{\rm flip} = \cos(\Omega_R/\gamma)$. Including the unitary evolution for times between the measurements, $T = N\Delta t + \delta t$ with $\delta t\in[0,\Delta t)$, yields $P_1(t) = (1 + z(t))/2$ with mean $z$-evolution,
\begin{align}\label{eq:pulsed-z}
  z(N\Delta t + \delta t)  &= \cos(\Omega_R \delta t)\cos^{N}(\Omega_R/\gamma).
\end{align}
Notably, for $\gamma > \Omega_R(2/\pi)$ and in the large $N$ limit such that $\delta t \ll t$ and therefore $N \approx \gamma t$, $\cos^N(\Omega_R/\gamma)$ approximates an exponential decay: 
\begin{subequations}
    \begin{align}
    z(t) &\approx e^{-\Gamma_\mathrm{mix} t} \cos(\Omega_R \delta t),\\
 &   \Gamma_\mathrm{mix} = \gamma \ln\left[\sec(\Omega_R/\gamma) \right].
    \end{align}
\end{subequations}

The corresponding ensemble-averaged state at time $T = N \Delta t$ is obtained by iteratively applying the (dephasing) map,
\begin{align}\label{eq:dephasing-pulsed}
  \mathcal{L}(\hat{\rho}) \equiv \sum_{r\in\{0,1\}}\hat{\Lambda}_r\hat{U}(\Delta t)\,\hat{\rho}\,\hat{U}^\dagger(\Delta t)\hat{\Lambda}^\dagger_r
\end{align} 
that averages over the state collapses for all possible outcomes $r$. For $t = N\Delta t$, computing $\hat{\rho}(N\Delta t) = \mathcal{L}^{N}(\ket1\!\bra1)$ yields the classical mixture,
\begin{align}
 \hat \rho(t) &= P_{1}(t) \ket1 \!\bra1 + (1 - P_{1}(t)) \ket0 \!\bra0,
\end{align}
characterized entirely by the initial state probability $P_1(T)$ in Eq.~\eqref{eq:pulsed-survival}. For times in between the measurements, the state $\hat{\rho}(N\Delta t + \delta t) = \hat{U}(\delta t)\hat{\rho}(N\Delta t)\hat{U}^\dagger(\delta t)$ briefly acquires coherent off-diagonal elements that are quickly dephased by the next measurement.

We illustrate the quantum Zeno effect for the mean population $P_1(t)$ in more detail in Fig. \ref{fig:pulsed-heatmap-stacked}. Notably, the evolution of $z(t)$ in Eq.~\eqref{eq:pulsed-z} that determines the average surviving $\ket 1$ probability resembles an underdamped harmonic oscillation\footnote{Satisfyingly, for the continuous measurement case in Sec. \ref{sec:continuous-quantum-zeno}, the mean $z(t)$ evolution will converge exactly to that of a damped harmonic oscillation with the same critical rate.} for $\Omega_R/\gamma > 1$ (e.g., Fig. \ref{fig:pulsed-heatmap-stacked}(b)), with periodic revivals of Rabi oscillations corresponding to measuring at the peak or trough of a Rabi oscillation (visible in Fig. \ref{fig:pulsed-heatmap-stacked}(f)). Evolution transitions to overdamped exponential decay for $\Omega_R/\gamma < 1$ (e.g., Fig. \ref{fig:selective-nonselective-pulsed}(i,j) and Fig. \ref{fig:pulsed-heatmap-stacked}(d)) at the \textit{critical rate},
\begin{align}\label{eq:critical-repetition-rate}
\gamma_{\rm crit} &= \Omega_R,
\end{align}
shown in Fig. \ref{fig:selective-nonselective-pulsed}(h) and Fig. \ref{fig:pulsed-heatmap-stacked}(c). Here, the resonances disappear and evolution transitions to purely exponential decay with a decreasing rate, making it the key transition point from the anti-Zeno regime to the Zeno regime. 

The heatmap in Fig. \ref{fig:pulsed-heatmap-stacked}(f) shows the time evolution (horizontal axis) of the $\ket 1$ population (color) for different ratios $\gam / \gam_\mr{crit}$ (vertical axis). For $\gam/\gam_\mr{crit} \ll 1$, the coherent Rabi oscillation fringes are essentially unperturbed. For $\gam/\gam_\mr{crit} < 1$, the oscillations get progressively damped on average due to the measurement interruptions, making this an anti-Zeno regime (witnessed by negative Zeno response). The fixed periodicity of the pulses leads to additional amplitude resonances for specific commensurate $\gamma$\footnote{This distinct resonance effect will not appear in the continuous measurement case in Sec. \ref{sec:continuous-quantum-zeno}, so we do not dwell on it here. It becomes important in Sec. \ref{sec:zeno-anti-zeno}, where we identify the regime of periodic revivals as a ``structured'' anti-Zeno regime.}. For $\gam/\gam_\mr{crit} > 1$, the mean evolution no longer oscillates and becomes overdamped exponential decay with a damping rate that decreases with $\gamma$, making this a Zeno regime (witnessed by positive Zeno response). For $\gam/\gam_\mr{crit} \gg 1$ the evolution becomes strongly suppressed and $P_1$ stays near $1.0$ to yield the quantum Zeno's paradox. The transition from the anti-Zeno to the Zeno regime---i.e. the transition from measurement-induced decay-enhancement to decay-suppression---thus also corresponds to the transition from the underdamped to the overdamped regime at $\gam_\mr{crit}$. 

Indeed, for large measurement rates $\gamma \gg \gamma_{\rm crit}$ in the Zeno regime, $\cos(\Omega_R/\gamma) \approx 1 - (\Omega_R^2/2\gamma)\Delta t \approx \exp(-\Omega_R^2 \Delta t/2\gamma)$, so Eq.~\eqref{eq:pulsed-z} yields the anticipated exponential decay of $z(t)$ with an asymptotic \textit{mixing rate},
\begin{align}\label{eq:escape-rate-pulsed}
    \Gamma_{\rm mix} &= \Omega_R^2/2\gamma = 2\Gamma_{\rm jump}, & (\gamma \gg \Omega_R)
\end{align}
that is twice the jump rate since it averages over jumps both out of and into the initial state\footnote{Note that generally this is the sum of \textit{unequal} jump rates into and out of the eigenstate.}. By Eq.~\eqref{eq:survival-probability}, this also implies mixing of the initial $\ket 1$ population, $P_1(t) \approx (1 + \exp(-\Gamma_{\rm mix} t))/2$.
Thus, we witness the features of the quantum Zeno effect in the Zeno regime described in Definitions \ref{def:qze} and \ref{def:zeno-regime}: a suppression of the natural Rabi oscillations and mean exponential decay of the initial state probability with rate $\Gam_\mr{mix}$ that limits to 0 as $\gamma\to \infty$.

This leads to the nonselective quantum Zeno paradox: even in the absence of an intermediate measurement record the average surviving population $P_1(t)$ converges to 1 as $\gamma\to \infty$ for any $t$. Thus we infer that the system \textit{must} remain pinned to the initial state. Evolution becomes impossible with sufficiently rapid measurements, consistent with an ensemble of pinned selective trajectories.

\textbf{Key features of the quantum Zeno effect - } The driven qubit is an ideal model for exhibiting the Zeno effect because it has only two competing rates: the Rabi frequency $\Omega_R = \gam_\mr{crit}$ of Hamiltonian evolution, and the repetition rate $\gam = 1/\Del t$ of measurement. Thus, the entire dynamics is characterized by the ratio $\gam / \gam_\mr{crit}$ that mediates the transition from anti-Zeno to Zeno. In the selective case of Eq.~\eqref{eq:pstay}, this ratio determines the rate $\Gamma_{\rm jump}$ of jumping away from the initial eigenstate, while in the nonselective case of Eq.~\eqref{eq:escape-rate} this ratio determines the mean exponential mixing rate $\Gam_\mr{mix} = 2\Gamma_{\rm jump}$.

The quantum Zeno regime manifests for $\gamma/\gam_\mr{crit}>1$, resulting in several distinct features of the evolution:
\begin{enumerate}
     \item Selective quantum trajectory evolution exhibits \textbf{jump behavior} with a rate $\Gamma_{\rm jump}$ between confined regions around the measurement eigenstates, such that the randomly timed jumps approximate classical telegraph noise; 

     \item Nonselective evolution exhibits  \textbf{exponential population decay} away from the measurement eigenstates characterized by an average mixing rate $\Gam_\mr{mix}$;

     \item With increasing measurement rate $\gam$, the jump and escape \textbf{rates are suppressed}, i.e. $d\Gam_\mr{jump}/d\gam < 0$ and $d\Gam_\mr{mix}/d\gam < 0$ (equivalent to $\mathcal{R}_Z(\gamma) > 0$), and eventually vanish in the limit $\gam \to \infty$ to yield the quantum version of Zeno's paradox. 
 \end{enumerate}
While these qualitative features can all serve as criteria for identifying the Zeno regime of the quantum Zeno effect, they are distinct phenomena that have historically been studied by different communities. Trajectories and quantum error correction communities \cite{vijayObservationQuantumJumps2011,slichterQuantumZenoEffect2016,hacohen-gourgyIncoherentQubitControl2018} emphasize the detection of telegraphic quantum jumps (criterion \#1): in general, the evolution being suppressed by the Zeno effect may be unknown, so the conversion of analog drifts into detectable, correctable digital jump errors is the essential qualitative feature. 
Telegraphic jumps induced by projective measurements also imply sequences of identical measurement outcomes, a more traditional criterion for the quantum Zeno effect measured in, e.g., \cite{kwiatHighefficiencyQuantumInterrogation1999,balzerRelaxationlessDemonstrationQuantum2002}.
 
 In contrast, communities without single-shot measurement, such as those doing nuclear magnetic resonance (NMR)~\cite{alvarezZenoAntiZenoPolarization2010}, lattice gas measurements \cite{patilMeasurementInducedLocalizationUltracold2015}, or dissipation engineering \cite{mirrahimiDynamicallyProtectedCatqubits2014,leghtasConfiningStateLight2015,touzardCoherentOscillationsQuantum2018,chakramMultimodePhotonBlockade2022,markInterplayCoherentDissipative2020,kimchi-schwartzStabilizingEntanglementSymmetrySelective2016,lescanneExponentialSuppressionBitflips2020}, don't have access to measurement records from individual trajectories so instead emphasize the ensemble-averaged exponential decay away from the dephasing-stabilized states (criterion \#2). Moreover, these bath-engineering applications emphasize average evolution suppression (or equivalently, state or subspace stabilization) around specific measurement strengths (criterion \#3): while ensemble-averaged evolution, such as leakage, cannot be easily reversed, it can still be suppressed via the Zeno effect by measuring (or coupling to a bath) more strongly. Bath-engineering techniques and nonselective measurements have been used to similar effect to enhance decay via the quantum Zeno effect in the anti-Zeno regime \cite{fischerObservationQuantumZeno2001,harringtonQuantumZenoEffects2017,thorbeckReadoutinducedSuppressionEnhancement2024}.

\section{Continuous quantum Zeno effect}
\label{sec:continuous-quantum-zeno}
While the preceding treatment using projective and instantaneous pulsed measurements is both intuitive and traditional, it does not quite encompass the case of the system being monitored continuously without pauses. Indeed, many ensemble-dephasing processes are better approximated as an environment or detector that is continuously monitoring the system due to an always-on, weakly coupled interaction with the system. Thus, we now supplement the pulsed analysis with a time-continuous treatment that is easier to generalize to generic dephasing processes and more elegantly captures the idea of the Zeno effect as the suppression of motion due to observation. This eliminates the stroboscopic resonances of the pulsed measurement scenario and more accurately captures the concept of measurement that is concurrent with, rather than interleaved with, unitary evolution. We will find that the same general structure of the quantum Zeno effect arises as in the previous section. 
 \begin{figure*}
\includegraphics[width=\linewidth]{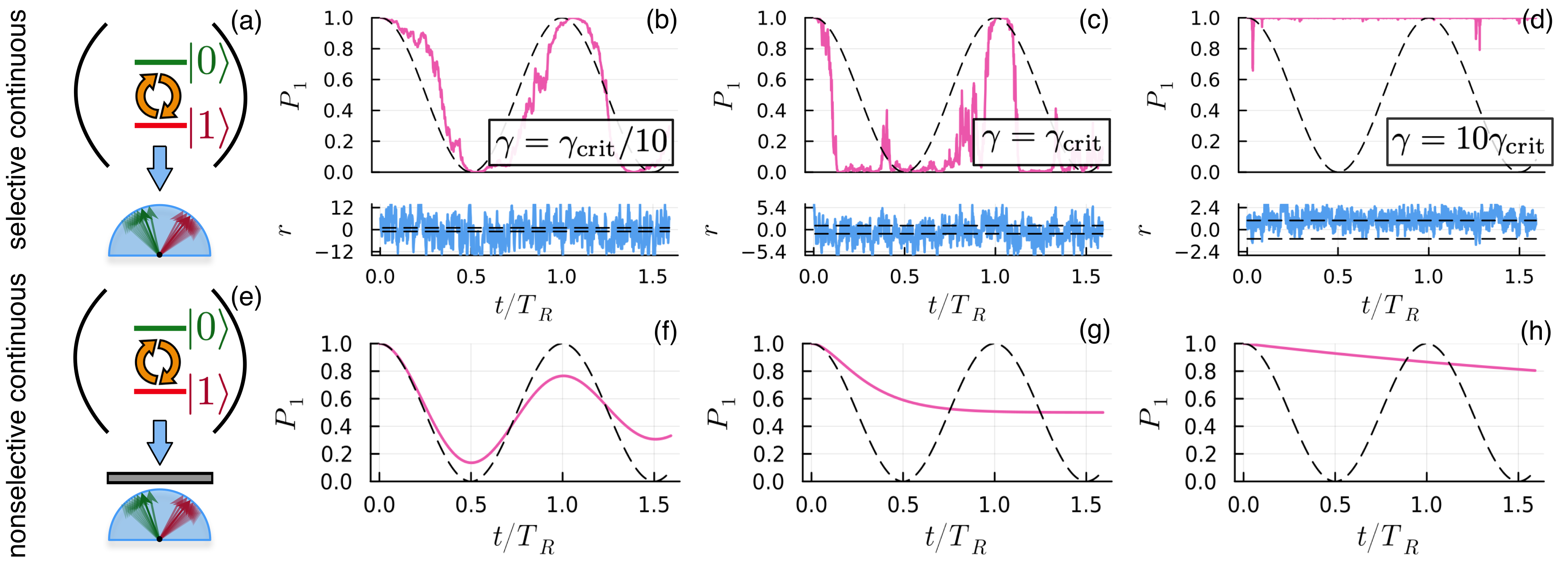}
	\caption{Population evolution $P_1(t)$ of the initial qubit state $\ket 1$ under selective (top) and nonselective (bottom) continuous measurements of the $\{\ket{0},\ket{1}\}$ basis. Different measurement rates $\gam$ are compared to a critical rate $\gam_\mr{crit} = \Om_R$. (a) For selective measurements, the oscillating qubit continuously leaks information as a noisy stream of values $r(t)$ recorded by a measuring apparatus, yielding a conditional quantum trajectory determined by the record. (b) For $\gamma \ll \Omega_R$, the $\ket{1}$ trajectory (pink trace) is stochastically perturbed from the natural Rabi oscillation (black dashed line), and the noisy readout signal (blue trace) is weakly correlated with the state. The displayed noisy records have been exponentially filtered with time constant $1/40\gam_\mr{crit}$ for visual clarity. (c) At the critical point, the qubit stays confined near the eigenstates, with occasional jumps between $\ket1$ and $\ket0$ becoming resolvable by the meter. (d) For strong measurements with $\gam \gg \Omega_R$, the state stays pinned close to $\ket 1$, while the meter continuously registers a signal with a clearly constant mean of $1$. (e) For nonselective measurements, the best estimate of the evolution is the ensemble average over all possibilities of the missing measurement record, which produces continuous dephasing at rate $2\gam$. The average behavior moves through (f) underdamped, (g) critically damped, and (h) overdamped regimes, with increasing measurement strength suppressing the resulting exponential decay.}
\label{fig:selective-nonselective-continuous}
\end{figure*}

As a simple model for a continuous measurement, we generalize the concept of projective measurements to non-projective measurements (Fig. \ref{fig:selective-nonselective-continuous}(a)) that will have a fixed rate $\gamma$ of information gain per unit time. By construction, this \textit{continuous measurement rate} $\gam$ is the continuous generalization of the repetition rate for the pulsed projective measurements used previously. It describes the same rate of measurement information acquisition, now contained in a noisy signal that can be averaged over \textit{tunable} durations without modifying the rate. Formally, we consider a stream of $N$ measurement results separated by a tunable repetition time $\delta t$ distinct from $1/\gamma$ such that averaging the results over a larger duration $T = N \delta t$ yields a signal-to-noise ratio that depends only on the total averaging time $T$ and the information acquisition rate $\gamma$. This construction allows a formal limit to arbitrarily small $\delta t$ and large $N$, while keeping $\gamma$ finite and constant. 

Intuitively, an always-on interaction continuously leaks information from the system, such that the meter registers not individual $1$s and $0$s, but instead a rapid sequence of noisy outcomes $r$ that can take a continuous range of values and are only weakly correlated with the qubit states of $\ket 1$ and $\ket 0$. Formally, we modify the POVM in Eqs.~\eqref{eq:pulsed-POVM} and \eqref{eq:binary-r-sampling} to a more general POVM \cite{jacobsStraightforwardIntroductionContinuous2006,korotkovQuantumBayesianApproach2011}:
\begin{subequations}\label{eq:unsharp-measurement}
	\begin{align}
		\hat M_r &\equiv \sqrt{P(r|1)} \ket1\!\!\bra1 + \sqrt{P(r|0)} \ket0\!\! \bra0, \\
    d\hat{P}_r &\equiv dr\,\hat{M}_r^\dagger \hat{M}_r,
        \end{align}
\end{subequations}
with the measurement collapse operators $\hat{\Lambda}_r$ replaced by the \textit{partial} collapse operators $\hat{M}_r$ in the normalized state update $\mathcal{M}_r$ defined in Eqs.~\eqref{eq:pulsed-projection} and \eqref{eq:pulsed-projection-rho}.

Each readout outcome $r$ can now have any real value, sampled from the probability measure,
\begin{align}
    dP(r) &= \Tr(\hat{\rho}\,d\hat{P}_r) = dr\sum_{\lambda\in\{0,1\}}P(r|\lambda)P_\lambda,
\end{align}
which is characterized by distinct likelihood distributions $P(r|\lambda)$ conditioned on each definite qubit state, $\ket \lambda$. These likelihoods can be measured experimentally as the readout histograms obtained from definite eigenstate preparations. Other state preparations  mix the conditioned likelihoods according to the eigenstate probabilities $P_\lambda = \bra{\lambda}\hat{\rho}\ket{\lambda}$. The relative likelihood $P(r|0)/P(r|1)$ of a particular readout $r$ determines the information gained about the eigenstates, and thus the degree of partial state collapse induced by $\mathcal{M}_r$, which is the quantum generalization of Bayes' rule for probability updates \cite{korotkovQuantumBayesianApproach2011}.

We assume the conditioned likelihoods $P(r|\lambda)$ are Gaussian distributions 
    \begin{align}\label{eq:unsharp-measurement-distributions} 
    P(r|\lambda) &= e^{-2\gamma\delta t (r - \mu_\lambda)^2} \sqrt{2\gamma\delta t/\pi}, 
\end{align}
centered at calibrated means $\mu_1 = 1$ and $\mu_0 = -1$ corresponding to the eigenvalues of $\hat{\sigma}_z$, with a common variance $\sigma^2 = 1/4\gamma\delta t$. The variance must scale inversely with $\delta t$ so that averaging the sum of $N$ successive and independent measurements over a duration $T = N\delta t$ correctly reproduces a similar Gaussian distribution but with a reduced variance $1/4\gamma T$ that sets the signal-to-noise ratio for the averaged signal \cite{jacobsStraightforwardIntroductionContinuous2006}.

Thus, the continuous limit as $\delta t \to 0$ formally yields a moving-mean stochastic process $r(t) = z(t) + \xi(t)/\sqrt{4\gamma}$ centered at $z(t) = P_1(t) - P_0(t)$, where $\langle \xi(t)\xi(t+\tau)\rangle = \delta(t-\tau)$ is delta-correlated zero-mean white noise.  However, any physical detector will have a finite bandwidth that sets the minimum accessible timescale $\delta t$ for the recorded sequence of readout values \cite{korotkovOutputSpectrumDetector2001,korotkovQuantumBayesianApproach2011}.

 \textbf{Selective continuous measurements -} In Fig. \ref{fig:selective-nonselective-continuous}(a,b,c,d) we show the onset of the quantum Zeno effect for individual quantum trajectories initialized in state $\ket{1}$ as the measurement rate $\gamma$ increases. 

Starting with initial state $\ket{1}$ and making $N$ repeated measurements at intervals $\delta t$ yields a quantum state trajectory, $\ket{\psi(N\delta t)} = \prod_{i=1}^N [\mathcal{M}_{r_i}\hat{U}(\delta t)]\ket{1}$, labeled by a stochastic measurement record $(r_1,r_2,\ldots,r_N)$\footnote{This discrete Trotterized form that interleaves measurement and unitary evolution \cite{trotterProductSemiGroupsOperators1959} is valid when the unitary evolution is approximately linear between measurement updates, i.e., $\delta t ~\Omega_R \ll 1$.}. We simulate trajectories using this numerically stable discrete formulation.  
However, it is convenient for analysis to formally take the continuum limit $\delta t \to 0$, which yields a stochastic Schr\"odinger equation \cite{jacobsStraightforwardIntroductionContinuous2006},
\begin{align}\label{eq:sme}
  d\ket{\psi} &= (\hat{H}/i\hbar - \gamma\hat{1}/2)\ket{\psi}dt + \sqrt{\gamma}\,\hat{\sigma}_z\ket{\psi}dW,
\end{align}
in the It\^o picture with Wiener increment 
satisfying $dW^2=dt$, such that $dW/dt = \xi(t) = \sqrt{4\gamma}\,(r(t) - z(t))$ is the zero-mean noise of the stochastic readout $r(t)$ \cite{wisemanQuantumTheoryFieldquadrature1993,wisemanQuantumMeasurementControl2009,korotkovQuantumBayesianApproach2011}.

We simulate and show example stochastic trajectories in Fig. \ref{fig:selective-nonselective-continuous}(b,c,d), using a repetition time of $\delta t = T_R/100$. As in the pulsed 
measurement, continuous measurements cause random fluctuations in the qubit state. However, trajectories now display diffusive random-walk behavior due to the steady rate of information gain in the continuous record. These underdamped trajectories dephase relative to one another, leading to overall mixing that increases with the measurement rate (anti-Zeno regime). Critically damped and overdamped trajectories, paralleling those in Fig. \ref{fig:selective-nonselective-pulsed}, consist of telegraphic jumps between the eigenstates with jump rates suppressed by stronger measurements (Zeno regime). The random jumps are no longer instantaneous and occur over a finite duration set by the inverse measurement rate $1/\gamma = \Delta t$. In the limit $\gamma \to \infty$ for Zeno's paradox, the evolution becomes pinned to the initial state $\ket 1$ while the noise in the readout vanishes to leave the constant value of $r(t)\to 1$.

\textbf{Nonselective continuous measurements -} In Fig. \ref{fig:selective-nonselective-continuous}(e,f,g,h) we show the corresponding ensemble-averaged (nonselective) perspective of the quantum Zeno effect for continuous measurements. 
As before, the ensemble average of selective continuous measurements is the best estimate for a trajectory whose record is absent. Per time step $\delta t$, the ensemble-averaged dephasing map $\mathcal{L}(\hat\rho)$ will thus have a form similar to Eq.~\eqref{eq:dephasing-pulsed}, but using the Gaussian measurement operators $\hat{M}_r$ from Eq.~\eqref{eq:unsharp-measurement} in place of the projective $\hat{\Lambda}_r$, and with an integral over all possible readout values $r$ on the real line. 

\begin{figure*}
\includegraphics[width=\linewidth]{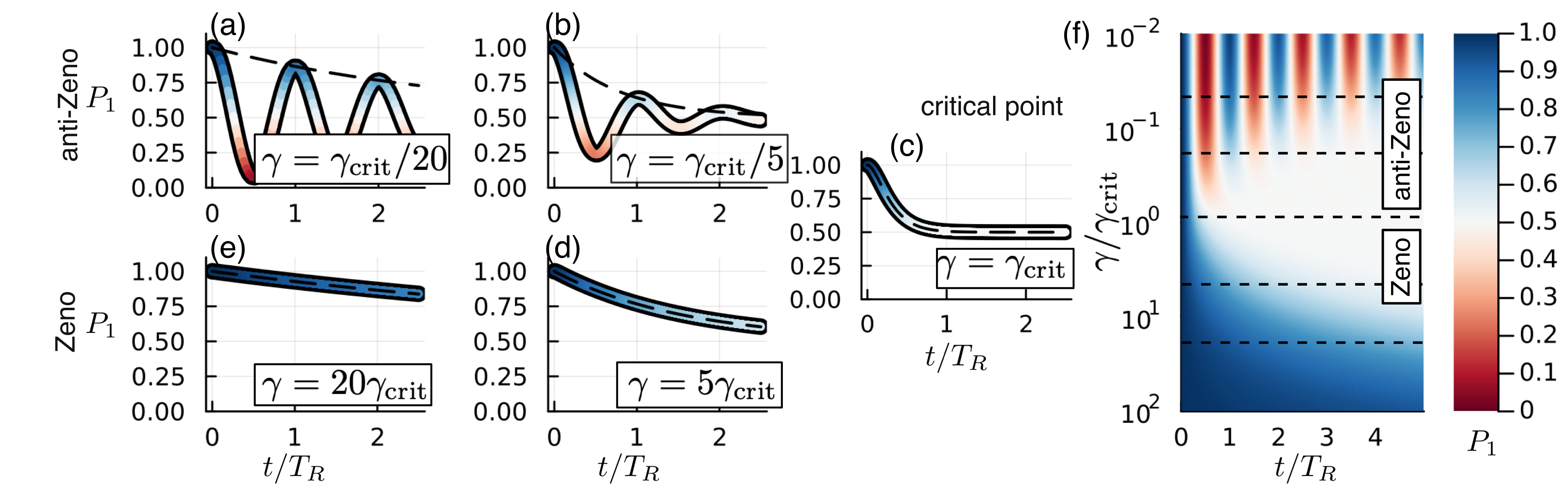}
	\caption{Continuous nonselective measurement population evolution $P_1(t)$ for different measurement rates $\gamma$ that increase clockwise from the upper left plot. The anti-Zeno (a,b), critical (c), and Zeno (d,e) regimes are determined by the size of $\gamma$ relative to the critical rate $\gam_\mr{crit} = \Om_R$. (a,b) The stochastic partial collapses in the underdamped regime produce an oscillation envelope (dashed) that decays exponentially with rate $\gamma$. (c) At the critical point, the evolution transitions to purely exponential decay with rate given by Eq.~\eqref{eq:escape-rate-continuous} (black dashed line), and the overall decay rate is maximized. (d,e) Faster measurement rates suppress the exponential decay. (f)  The heatmap of $P_1$ (color) versus time (horizontal axis) and $\gam / \gam_\mr{crit}$ (vertical axis in log scale) shows how the transition from decay envelope enhancement (anti-Zeno) to decay envelope suppression (Zeno) also corresponds to the smooth transition from underdamped oscillations to overdamped exponential decay, with horizontal dashed lines corresponding to the cases shown in (a-e).}
 \label{fig:continuous-heatmap-stacked}
\end{figure*}

It is simpler, however, to calculate the averaged evolution by first taking the continuum limit $\delta t \to 0$ to find  the It\^o stochastic master equation implied by Eq.~\eqref{eq:sme},
\begin{align}\label{eq:lindblad}
    d\hat{\rho} &= \left([\hat{H},\hat{\rho}]/i\hbar + \gamma(\hat{\sigma}_z\hat{\rho}\hat{\sigma}_z - \hat{\rho})\right)dt \nonumber \\
    &\quad + \sqrt{\gamma}\,(\hat{\sigma}_z\hat{\rho} + \hat{\rho}\hat{\sigma}_z)\,dW,
\end{align}
then averaging over all measurement realizations. Since $\langle dW\rangle = 0$ in the ensemble average, the last term of Eq.~\eqref{eq:lindblad} vanishes, yielding 
Markovian \textit{dephasing} evolution in Lindblad form with a (measurement-induced) ensemble-average dephasing rate $\Gamma_{\rm dephase} = 2\gamma$. This continuum approach more clearly emphasizes the dynamical equivalence between nonselective measurement and other environmental couplings that produce basis-dependent ensemble dephasing, such as those used in bath engineering.

Converting Eq.~\eqref{eq:lindblad} to Bloch coordinates and assuming $\Delta = 0$ as before yields driven and damped harmonic oscillation for the mean evolution of $z(t)$:
\begin{equation}
    \ddot{z} + \Om_R^2 z + 2\gam \dot{z} = 0.
\end{equation}
The characteristic solutions $\exp(-\gamma_\pm t)$ of $z(t)$ thus have the eigenvalues $\gamma_\pm = \gamma \pm \sqrt{\gamma^2 - \Omega_R^2}$. Therefore, in agreement with the pulsed projective measurement case in Sec. \ref{sec:pulsed-quantum-zeno}, the evolution of $z(t)$ under nonselective measurements has three distinct solution regimes separated by the critical measurement rate $\gamma_\mathrm{crit} = \Omega_R$: underdamped oscillations ($\gamma < \gamma_\mr{crit}$), critically damped decay ($\gamma = \gamma_{\rm crit}$), and overdamped exponential decay ($\gamma > \gamma_\mr{crit}$), which correspond to anti-Zeno, critical, and Zeno regimes, respectively. We show these distinct regimes in Fig. \ref{fig:selective-nonselective-continuous}(f,g,h).

The underdamped regime with $\gamma < \Omega_R$ has exponential decay envelope $e^{-\gamma t}$, so $\Gamma_{\rm mix} = \gamma$. The Zeno response, $\mathcal{R}_Z(\gamma) = -1$, is thus negative, witnessing the anti-Zeno regime.
The onset of the quantum Zeno regime corresponds to the transition into the overdamped regime with $\gamma >\Omega_R$. With initial condition $z(0) = 1$, the solution has the form,
\begin{align}
    z(t) &= e^{-\Gamma_\mr{mix} t}\left(1 + (\Gamma_\mr{mix}/\Gamma)\,(1 - e^{-\Gamma t})\right),
\end{align}
where $\Gamma = \gamma_+ - \gamma_- = 2(\gamma - \Gamma_\mr{mix})$, and
\begin{align}\label{eq:escape-rate-continuous}
   \Gamma_\mr{mix} = \gamma_- &= \gamma\left(1 - \sqrt{1 - (\Omega_R/\gamma)^2}\right) \nonumber \\
   &\xrightarrow{\gamma \gg \gamma_\mr{crit}} \Omega_R^2/2\gamma.
\end{align}
is the corresponding mixing rate with precisely the same asymptotic limit as found in Eq.~\eqref{eq:escape-rate-pulsed} for the pulsed projective measurement case\footnote{The Zeno effect for pulsed and continuous measurements was compared theoretically in \citet{schulmanContinuousPulsedObservations1998} and experimentally in \citet{streedContinuousPulsedQuantum2006}. In their systems, which include asymmetric measurement or decay, the scaling of the pulsed and continuous limits differ. In our simple model, the symmetry between states and the choice of $\Delta t = 1/\gamma$ causes the Zeno regime onset points $\gamma_\mathrm{crit}$ \textit{and} the asymptotic limits to coincide.}. At the critical rate, $\lim_{\gamma\to\gamma_\mr{crit}}z(t) = e^{-\gamma_{\rm crit} t}(1 + \gamma_\mathrm{crit} t)$. The Zeno response is monotonically positive for the entire overdamped range $\gamma > \Omega_R$, \begin{align}
    \mathcal{R}_Z(\gamma) = -1 + 1/\sqrt{1 - (\Omega_R/\gamma)^2} > 0,
\end{align}
thus witnessing the Zeno regime.

In Fig. \ref{fig:continuous-heatmap-stacked} we show more detailed simulation results analogous to those in Fig. \ref{fig:pulsed-heatmap-stacked}. The structure of the dynamics as $\gamma$ is varied is broadly similar to the pulsed measurement case, but is smoother due to the lack of resonances between the periodicity of the projective measurements and the oscillating evolution. Indeed, the continuous measurement case is equivalent on average to randomizing the time interval between pulsed measurements according to a Poisson process with rate $2\gamma$, similar to photodetection models \cite{gardinerQuantumNoiseHandbook2000,jacobsStraightforwardIntroductionContinuous2006,pellegriniDiffusionApproximationStochastic2009}. Notably, for $\gamma \gg \gamma_\mr{crit}$, the mixing rate is $\Gamma_\mr{mix}/\Omega_R = (1/2)(\gamma_\mr{crit}/\gamma)$, yielding the slight asymmetry of the exponential decay envelopes about the critical rate observed in Fig. \ref{fig:continuous-heatmap-stacked}, with underdamped decay envelopes at $\gamma = \zeta \gamma_\mr{crit}$ and $\zeta < 1$ decaying twice as fast as the corresponding overdamped evolution with $\gamma = (1/\zeta)\gamma_\mr{crit}$.

\begin{figure*}
\centering
\includegraphics[width=1.0\linewidth]{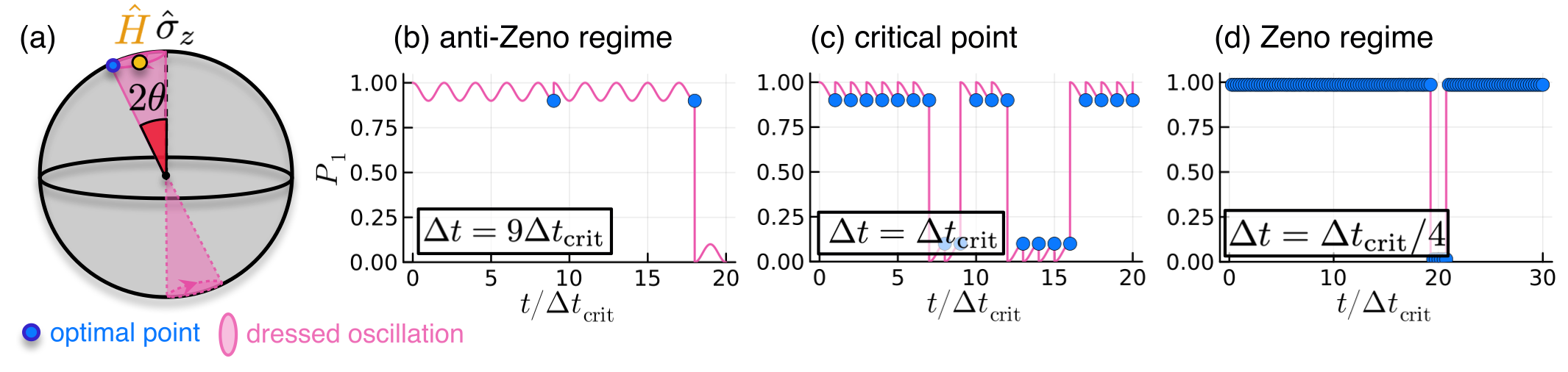}
	\caption{Pulsed trajectories in the stabilized case where the Hamiltonian oscillation axis is at angle $\theta < \pi/4$ to the measurement axis, demonstrating how stabilization gives rise to telegraphic structure even in the anti-Zeno regime.  For these particular simulations, $\Delta = 3 \Omega_R$. (a) The stabilized regime is defined by an oscillation axis that is tilted by an angle $\theta < \pi/4$ from the measurement axis, with smaller $\theta$ indicating greater stabilization. In this regime, Hamiltonian evolution of the measurement eigenstates yields two disconnected orbits that stay confined to a narrow range of polar angles $2\theta$ near the eigenstates. The instantaneous jump probability for any given measurement is maximized by measuring at the optimal point (blue circle), which is half an oscillation $\Delta t_\mathrm{crit} = (2\pi/\Omega)/2$. The pulse repetition rate for $\Delta t = 1/\gamma$ is expressed in terms of $\Delta t_\mathrm{crit} = (2\pi/\Omega) / 2$ in panels (b-d).  (b) Measurements enable rare telegraphic jumps between these dynamically stabilized regions around the eigenstates. While measuring at the optimal point (blue circles) maximizes the instantaneous jump probability, more frequent measurements translate directly to more opportunities for a jump, leading to an anti-Zeno regime. (c) The overall jump rate is maximized when the system is measured every half-oscillation point, i.e., at $\Delta t = \Delta t_\mathrm{crit}$, indicating the critical point. (d) Beyond the critical point, measuring even more rapidly again suppresses the overall jump rate, indicating a Zeno regime. Notably, telegraphic behavior is present in both anti-Zeno and Zeno regimes of the stabilized case, in contrast to the standard case $\Delta = 0$ which supported telegraphic trajectories only in the Zeno regime.} 
 \label{fig:anti-Zeno-pulsed-intuition}
\end{figure*}

\section{Stabilized quantum Zeno effect}
\label{sec:zeno-anti-zeno}

Thus far, we have explored how measurement suppresses competing evolution (i.e., evolution that does not commute with the measurement). Intuitively, the regime with underdamped oscillations on average had negative Zeno response, thus corresponding to the anti-Zeno regime, while the regime with overdamped exponential decay on average had positive Zeno response, thus corresponding to the Zeno regime. Notably, only the Zeno regime exhibited telegraphic jump-like behavior between otherwise confined regions of state space, while the anti-Zeno regime had less perturbed quasi-oscillatory evolution. 

Historically, however, the anti-Zeno regime has been witnessed and studied in systems that \textit{do} show strong confinement near the measurement eigenstates, typically as confinement of excitations due to strong system-bath detuning \cite{kofmanZenoAntiZenoEffects2001,zhangCriterionQuantumZeno2018,harringtonQuantumZenoEffects2017,thorbeckReadoutinducedSuppressionEnhancement2024}. In order to include this possibility in our simple model, we now consider evolution with a (detuning) frequency $\Del > \Omega_R$ that \textit{commutes} with the measurement in the Hamiltonian in Eq.~\eqref{eq:H-general}. In this regime, the overall unitary evolution is along an axis $\theta = \arctan(\Omega_R/\Delta)<\pi/4$ from the measurement axis $\hat \sigma_z$. As a result, for a system initialized in a measurement eigenstate, the unitary evolution never takes that system outside of the corresponding Bloch hemisphere. Whereas unitary evolution along an orthogonal axis $\theta = \pi/2$ naturally \textit{destabilizes} the measurement eigenstates, evolution along $\theta < \pi/4$ causes \textit{stabilization} in the neighborhoods of the $\hat{\sigma}_z$ eigenstates that increases as $\theta \rightarrow 0$.\footnote{In the extreme stabilized limit $\theta \rightarrow 0$ ($\Delta \gg \Omega_R$), the oscillations out of the eigenstates become essentially negligible.} For slower measurement rates $\gamma$, the stochastic measurement backaction can then disrupt this stabilization and increase the transition rate between these dynamically-protected regions, yielding a nontrivial trajectory-level manifestation of the quantum Zeno effect in the anti-Zeno regime which has not been explored in previous literature. 
\begin{figure*}
\includegraphics[width=\linewidth]{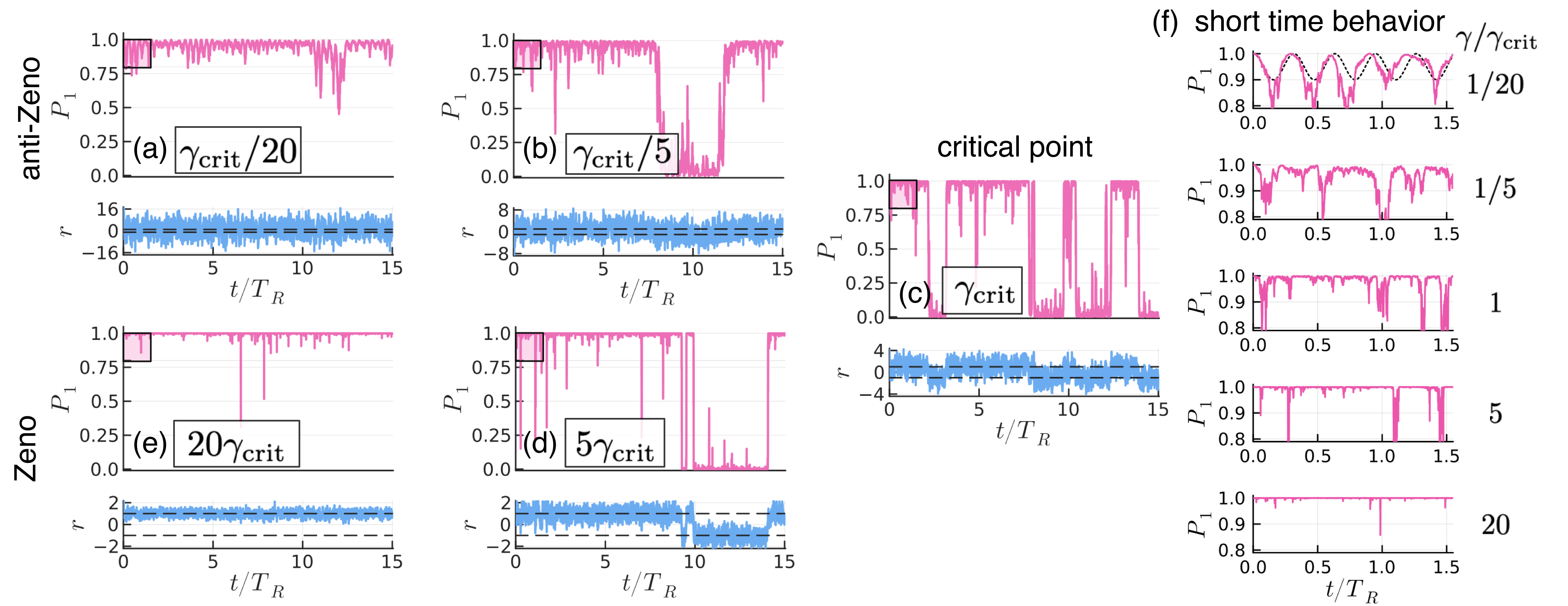}
	\caption{Continuously measured population trajectories (pink) of the nonorthogonally driven qubit with $\Delta = 3 \Omega_R$, for different measurement rates $\gam$ defined relative to $\gam_\mr{crit} = \Om/2 = \sqrt{\Del^2 + \Om_R^2}/2$. $\gam$ increases clockwise from the upper left plot, moving through anti-Zeno (a,b), critical (c), and Zeno (d,e) regimes. Short-time behavior (boxed highlights) is shown in panel (f) with $\gam$ increasing downwards. An exponential filter with time constant $1/8\gamma_\mathrm{crit}$ is used to display each trajectory's record (blue). (a) For $\gam \ll \gam_\mr{crit}$, measurements weakly perturb the natural small-amplitude oscillations at rate $\Om$, with the state remaining dynamically stabilized near the measurement eigenstates. (b) As $\gam$ increases, measurement fluctuations increasingly enable jumps between the dynamically stabilized regions, leading to telegraphic evolution decorated by coherent oscillation fringes. (c) At $\gam_{\rm crit}$, the rate of jumps is maximum. (d) For $\gam > \gam_\mr{crit}$, the coherent oscillations are suppressed and the jump rate decreases to a value similar to the corresponding anti-Zeno trajectory in (b). (e) Further increasing $\gam$ leads to stronger pinning with rare jumps, similar to the corresponding anti-Zeno trajectory in (a). (f) While long-time jump behavior in (a-e) is qualitatively symmetric about the critical point, short-time behavior and the confinement mechanisms are not. As $\gam$ increases (right panel, top to bottom), short-time noisy oscillations (compared to unperturbed oscillations, dashed black line) are steadily suppressed.} 
 \label{fig:zeno-anti-zeno-trajectories}
\end{figure*}

\textbf{Pulsed measurements -} Fig.~\ref{fig:anti-Zeno-pulsed-intuition} depicts the case just described,
where the oscillation axis is tilted by only a small angle $\theta = \arctan(\Omega_R/\Delta)$ from the measurement axis $z$. The eigenstates of $\hat{\sigma}_z$ then follow Hamiltonian orbits confined to a region of polar angles either between $[0,2\theta]$ or $[\pi-2\theta,\pi]$. Measuring $\hat{\sigma}_z$ both reinforces these dynamical basins of attraction near the measurement eigenstates and induces stochastic jumps between them \cite{chantasriActionPrincipleContinuous2013}, and yields the quantum Zeno paradox of frozen dynamics when $\gamma \to \infty$. 

A nonzero measurement rate $\gamma$ enables the state to traverse the otherwise forbidden region of polar angles $[2\theta,\pi-2\theta]$ that separate the Hamiltonian orbits of the measurement eigenstates. Increasing $\gamma$ thus initially increases the rate of jumps, leading to more rapid mixing of the measurement eigenstates on average (i.e., negative Zeno response signaling the anti-Zeno regime). We highlight that this can be understood as a measurement-induced resonant tunneling effect. Continuing to increase $\gamma$ will then suppress the Hamiltonian dynamics and thus decrease the rate of jumps and the mixing rate on average (i.e., positive Zeno response signaling the Zeno regime). 

More precisely, the Hamiltonian evolution of a state initialized in $\ket 1$ has the form
\begin{align}
    \ket{\psi(t)} &= \cos(\Omega t/2)\ket{1} - i\sin(\Omega t/2)(\hat{\sigma}_\theta\ket1),
\end{align}
with $\Omega = \sqrt{\Omega_R^2 + \Delta^2}$ the orbit frequency, yielding the jump probability to $\ket 0$ at $\Delta t = 1/\gamma$,
\begin{align}
    P_{\rm jump}(\Delta t) &= |\braket{0|\psi(t)}|^2 = \sin^2(\Omega/2\gamma)\sin^2\theta,
\end{align}
which is modified by a Lorentzian (detuning) factor,
\begin{align}
  \sin^2\theta &= \Omega_R^2/(\Omega_R^2 + \Delta^2)
\end{align}
relative to the orthogonal case with $\Delta = 0$. The $\sin^2(\Omega/2\gamma)$ factor produces resonances between the measurement repetition rate and the oscillation period, with  $P_\mathrm{jump}(\Delta t) \in [0, \sin^2 \theta]$ depending on whether the qubit is measured at the top of its oscillation ($P_\mathrm{jump}=0$), the bottom ($P_\mathrm{jump} = \sin^2\theta$), or somewhere in between.

These resonances appear directly in the exact time-evolution of the $z$ coordinate, derived analogously to Eq.~\eqref{eq:pulsed-z}:
\begin{subequations}
\label{eq:exact-pulsed-stabilized-z}
    \begin{align}
            z(N \Delta t+\delta t) &= \left[1-2 P_\mathrm{jump}(\Delta t)\right]^N A(\delta t), \\ 
    A(\delta t) &\equiv 1-2 \sin ^2(\Omega \delta t / 2) \sin ^2 \theta, \label{eq:exact-mixing-rate-pulsed}
    \end{align}
\end{subequations}
where $N$ represents the number of measurements (and thus $z(N \Delta t)$ is the state immediately after the $N^\mathrm{th}$ measurement) and $A(\delta t)$ is the oscillation envelope in terms of $\delta t$, the time elapsed since the $N^\mathrm{th}$ measurement. In the stabilized regime $\theta = \arctan(\Omega_R/\Delta) < \pi/4$ (equivalently, $\Delta > \Omega_R$) and taking the large $N$ limit so that $\delta t \ll t$ and thus $N \approx \gamma t$, Eq.~\eqref{eq:exact-pulsed-stabilized-z} can be expressed directly as an exponential decay modulated by the oscillation:
\begin{subequations}
    \begin{align}
        z(t) &= e^{-\Gamma_\mathrm{mix}\, t} A(t - N \Delta t), \\ 
        \Gamma_\mathrm{mix} &= \gamma \ln \left[1 / (1 - 2P_\mathrm{jump}(\Delta t))\right]. \label{eq:pulsed-gamma-mix}
    \end{align}
\end{subequations}
Here, we highlight that the condition $1 - 2 P_\mathrm{jump}(\Delta t) > 0$ $\Longleftrightarrow$ $\sin^2(\Omega/2\gamma)\sin^2\theta < 1/2$ for having a well-defined natural log is equivalent to the geometric condition that the evolution stay in the upper Bloch hemisphere. Thus, $\theta < \pi/4$ is a sufficient condition for this inequality to be satisfied for all $\gamma$, defining an exponential envelope regardless of measurement strength. This is a key difference from the $\Delta = 0$ pulsed case, in which an exponential envelope was only well-defined just before the Zeno regime, and motivates our definition of the stabilized regime as $0 < \theta < \pi/4$.\footnote{While we don't explore the intermediate regime $\pi/4 < \theta < \pi/2$ in this paper, we note that the geometric argument above suggests an approach to understanding this regime. Solving the condition $\sin^2(\Omega/2\gamma)\sin^2\theta < 1/2$ for $2\gamma/\Omega > 1/\arcsin(\csc \theta/\sqrt{2}\,)$ naturally defines a stabilized exponential decay threshold $\gamma_\mathrm{stab}$ that interpolates between $\gamma_\mathrm{stab} = \Omega/\pi$ for $\theta = \pi/4$ (near stabilized regime) to $\gamma_\mathrm{stab} = \Omega (2/\pi)$ for $\theta = \pi/2$ (standard case). Understanding the distinction between the stability threshold $\gamma_\mathrm{stab}$ and the anti-Zeno / Zeno critical point $\gamma_\mathrm{crit}$ is an interesting direction for future work.}

For measurement rates $\gamma$ slow compared to the orbit frequency $\Omega$ and in the strongly stabilized regime $\Omega_R \ll \Delta$, the $\sin^2(\Omega/2\gamma)$ term, which causes $P_\mathrm{jump}(\Delta t)$ to vary periodically in $\Omega/2\gamma$ between $0$ and $\sin^2 \theta$, oscillates rapidly relative to changes in $\gamma$. Coarse-graining over these oscillations by taking $\sin^2(\Omega/2\gamma) \sim 1/2$ yields a mixing rate
\begin{equation}
    \Gamma_\mathrm{mix} = \gamma (\Omega_R/\Omega)^2.
\end{equation}
Since the average state mixing \textit{increases} with $\gamma$, this witnesses an anti-Zeno regime. This is in accordance with 
the mean jump rate computed as $\Gamma_{\rm jump} = \gamma \overline{P}_{\rm jump} \approx (\gamma/2)\sin^2\theta$, leading to the same mean mixing rate that also conforms to Fermi's Golden Rule:
\begin{align}\label{eq:golden-rule}
    \Gamma_{\rm mix} = 2\Gamma_{\rm jump} = \gamma\,|\bra{0}\hat{\sigma}_\theta\ket{1}|^2 = \gamma\,(\Omega_R/\Omega)^2.
\end{align}
As mentioned, the rate of rare jumps between the separate regions near the measurement eigenstates increases with $\gamma$, which is a negative Zeno response witnessing the anti-Zeno regime. However, increasing the detuning $\Delta$ independently decreases the mixing rate due to increased confinement of the oscillation orbit, which strengthens the dynamical stabilization of the measurement eigenstates. 

Furthermore, because of the periodic nature of $P_\mathrm{jump}(\Delta t)$, $\Gamma_\mathrm{mix}(\gamma)$ is highly non-monotonic in the region that the coarse-grained Eq.~\eqref{eq:golden-rule} would otherwise lead us to identify as a purely anti-Zeno regime. Thus, we can identify this as a kind of \textit{structured anti-Zeno regime}, i.e. one that has $\Gamma_\mathrm{mix}$ increasing with $\gamma$ as an overall trend but has additional structure for smaller variations in $\gamma$\footnote{Note that a similar structure appears in the pulsed $\Delta = 0$ quantum Zeno effect, but is less prominent due to lack of stabilization.}.

For measurement rates $\gamma \gg \Omega$ fast compared to the orbit frequency, $\sin^2(\Omega/2\gamma) \approx (\Omega^2/4\gamma)\Delta t$, so the jump rate $\Gamma_{\rm jump} = (\Omega^2/4\gamma)\sin^2\theta = (\Omega^2/4\gamma)(\Omega_R^2/\Omega^2) = \Omega_R^2/4\gamma$ converges to the same result as the $\Delta = 0$ case, thus yielding the same $\Delta$-independent asymptotic mixing rate $\Gamma_{\rm mix} = 2\Gamma_{\rm jump} \to \Omega_R^2/2\gamma$. This is consistent with taking the large $\gamma$ limit of Eq.~\eqref{eq:pulsed-gamma-mix}. The Hamiltonian evolution, and thus the rate of jumps, is suppressed with increasing $\gamma$, which is a positive Zeno response, witnessing the Zeno regime. Thus, the quantum Zeno paradox with $\Gamma_{\rm mix} \to 0$ still occurs in the limit $\gamma\to\infty$ even in the stabilized regime $\Delta > \Omega_R$.

\begin{figure*}
\includegraphics[width=\linewidth]{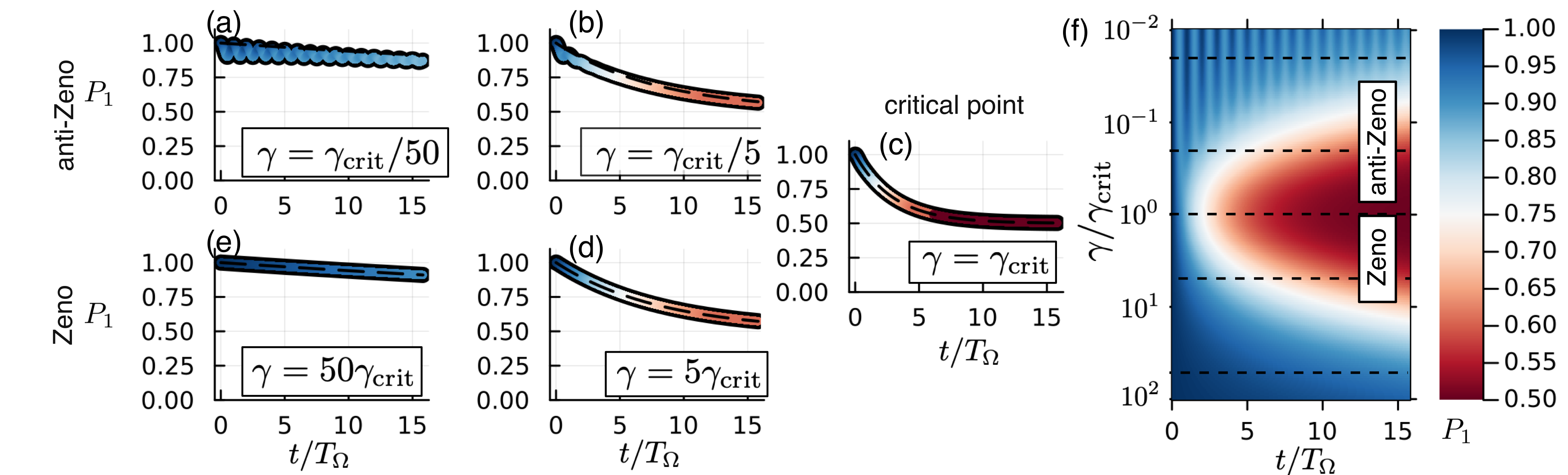}
	\caption{Continuous nonselective measurement population evolution $P_1(t)$ (color) for the nonorthogonally driven qubit with $\Del = 3 \Om_R$, for different measurement rates $\gam$ relative to $\gam_\mr{crit} = \Om/2$, with $\Om = \sqrt{\Om_R^2 + \Delta^2}$. $\gam$ increases clockwise from the upper left, moving through anti-Zeno (a,b), critical (c), and Zeno (d,e) regimes. (a,b) In the anti-Zeno regime, confined oscillations are steadily damped as $\gam$ approaches $\gam_\mr{crit}$, and the evolution envelope decays exponentially at rate $\Gam_\mr{mix} \approx \Gamma_0$ (Eq.~\eqref{eq:anti-Zeno-gam}, black dashed line). (c) At the critical point, the exponential decay rate is maximum. (d,e) For measurements faster than the natural evolution, evolution is purely exponential and the decay rate is suppressed with increasing $\gam$. (f) The heatmap of $P_1$ (color) versus time (horizontal axis) and $\gam / \gam_\mr{crit}$ (vertical axis in log scale) shows a clear transition from decay envelope enhancement (anti-Zeno) to decay envelope suppression (Zeno), with symmetric decay rates mirrored across the critical point. The horizontal dashed lines correspond to (a--e).}
 \label{fig:continuous-anti-Zeno-heatmap}
\end{figure*}

\textbf{Continuous measurements -} For continuous measurements, there is no free Hamiltonian evolution between measurements and no resonances with unmeasured evolution. Instead, when $\Delta \gg \Omega_R$ and $\gamma < \Omega$ the linewidth of the oscillation  becomes broadened on average by the measurement-induced dephasing rate $2\gamma$.
This creates a measurement-broadened version of Eq.~\eqref{eq:golden-rule}, 
\begin{subequations}
    \begin{align}
    \Gamma_\mathrm{mix} &= \frac{\Gamma_0}{1-2(\Gamma_0/\Omega_R)^2}\label{eq:stabilized-gam},\\
	\Gam_0 &\equiv 2\gam\, \frac{\Om_R^2}{ \Om^2 + (2\gam)^2}. \label{eq:stabilized-gam-0}
    \end{align}
    \label{eq:anti-Zeno-gam}
\end{subequations}
The zeroth order decay rate $\Gamma_0$ (derived in Appendix~\ref{appen:zeno-anti-zeno}) is weakly perturbed by the dimensionless factor $1 / (1-2(\Gamma_0/\Omega_R)^2)$. As long as this decay rate $\Gamma_0$ is much slower than the Rabi rate $\Omega_R$---as is typical in the stabilized regime of small $\theta$---the decay $\Gamma_\mathrm{mix} \approx \Gamma_0$ again takes the form of a Lorentzian broadened by the measurement-induced dephasing $2\gamma$. Notably, this result yields a small $\gamma$ limit $\Gamma_\mathrm{mix} \approx 2 \gamma (\Omega_R/\Omega)^2$ that is twice the mean pulsed decay rate Eq.~\eqref{eq:golden-rule} in the anti-Zeno regime due to the lack of resonances (see Appendix \ref{appen:pulsed-continuous}).

Eq.~\eqref{eq:anti-Zeno-gam} is derived more formally in Appendix~\ref{appen:zeno-anti-zeno} from the ensemble-averaged master Eq.~\eqref{eq:lindblad}. Notably, the mixing rate has the same form as a Purcell-limited energy-relaxation ($T_1$) rate when coupling with rate $\Om_R$ to an auxiliary system detuned by $\Del$. Indeed, both the traditional Purcell effect and this nonselective decay rate can be understood as applications of Fermi's Golden Rule.

The Zeno response in this regime is
\begin{equation}
    \mathcal{R}_Z = -d\Gamma_\mathrm{mix}/d\gamma = (d\Gamma_\mathrm{mix}/d\Gamma_0)(d\Gamma_0/d\gamma).
\end{equation}
As shown in Appendix \ref{appen:zeno-anti-zeno}, $d\Gamma_\mathrm{mix}/d\Gamma_0 > 0$, so that the sign and turning points of the Zeno response depend entirely on $d\Gamma_0/d\gamma$. Thus,
\begin{align}
	\mathcal{R}_Z &\propto - d\Gam_0 /d\gamma \\
    &= - 2 \left(\frac{\Omega_R}{\Omega^2 + (2\gamma)^2}\right)^2 (2\gamma - \Omega)(2\gamma+\Omega),
     \nonumber
    \label{eq:anti-Zeno-response}
\end{align}
which has a critical transition point at 
\begin{align}
    \gamma_{\rm crit} = \Omega/2.
\end{align}
Thus, $\gamma < \gamma_{\rm crit}$ corresponds to a negative Zeno response, witnessing the anti-Zeno regime, and $\gamma > \gamma_{\rm crit}$ corresponds to a positive Zeno response, witnessing the Zeno regime. Since the continuous stabilized case lacks the resonant structure of the pulsed stabilized case, there is a single anti-Zeno to Zeno transition at the turning point.

The selectively measured quantum trajectories shown in Fig. \ref{fig:zeno-anti-zeno-trajectories} confirm the expected behavior. For $\gamma < \gamma_{\rm crit}$, the trajectories exhibit quantum jumps between the dynamically stabilized regions of local oscillations around the eigenstates. Since these transitions are otherwise forbidden, this is a measurement-induced tunneling effect and is closely related to that recently predicted for tunneling through a triple quantum dot \cite{singhCapturingElectronVirtual2025}. The short-time behavior clearly shows the localized oscillations being perturbed by the measurement. For $\gamma > \gamma_{\rm crit}$, the trajectories instead exhibit quantum jumps between measurement-pinned eigenstates that have the Hamiltonian oscillations fully suppressed. Thus, while there is a qualitative symmetry of the jump behavior around the critical measurement rate, the nature of the short-time behavior and the mechanism for state confinement in the anti-Zeno and Zeno regimes are fundamentally different.

The ensemble-averaged evolution shown in Fig. \ref{fig:continuous-anti-Zeno-heatmap} confirms the exponential decay envelopes with the broadened mixing rate in Eq.~\eqref{eq:anti-Zeno-gam}. The symmetry of the exponential decay rates (and the corresponding symmetry of jump rates visible in Fig. \ref{fig:zeno-anti-zeno-trajectories}) around the critical rate $\gamma_{\rm crit} = \Omega/2$ also becomes more clear. 
Indeed, Eq.~\eqref{eq:stabilized-gam-0} can be written as an even function of a dimensionless parameter,
\begin{subequations}\label{eq:dimless-anti-Zeno-Gamma}
    \begin{align}
        \Gam_0 &= (\Om_R^2 /2\Om) \mr{sech}(\chi), &
        \chi &\equiv \log(\gam/\gam_\mr{crit}),
    \end{align}
\end{subequations}
so that it is manifestly symmetric about $\gam=\gam_\mr{crit}$ on the logarithmic axis used in Fig. \ref{fig:continuous-anti-Zeno-heatmap}(f).  

For fast measurement rates Eq.~\eqref{eq:anti-Zeno-gam} limits to the same asymptotic rate in Eq.~\eqref{eq:escape-rate-pulsed} for $\Delta = 0$ in both pulsed and continuous cases,
\begin{align}
\label{eq:strong-measurement-detuned-gam}
    \Gam_\mr{mix} &\approx \Om_R^2/2\gam,
    & \gam \gg \Delta, \Om_R
\end{align}
underscoring the universality of the quantum Zeno paradox where all evolution becomes suppressed by rapid measurement.
In contrast, anti-Zeno effects have typically been studied in dynamically stabilized regimes $\Del \gg \Om_R$ with slower measurement rates $\gam \ll \Del$. In these regimes, $\Delta$ dominates, so that $\Omega \approx \Delta$ and we obtain
\begin{align}
     \Gam_\mr{mix} &\approx 2\gam\,(\Om_R /\Delta )^2,
     & \gam \lesssim \Om_R \ll \Del
\end{align}
with a form that highlights the similarity to Purcell decay \cite{auffevesControllingDynamicsCoupled2010,setePurcellEffectMicrowave2014}.

\begin{figure}[!ht]
	\includegraphics[width=\linewidth]{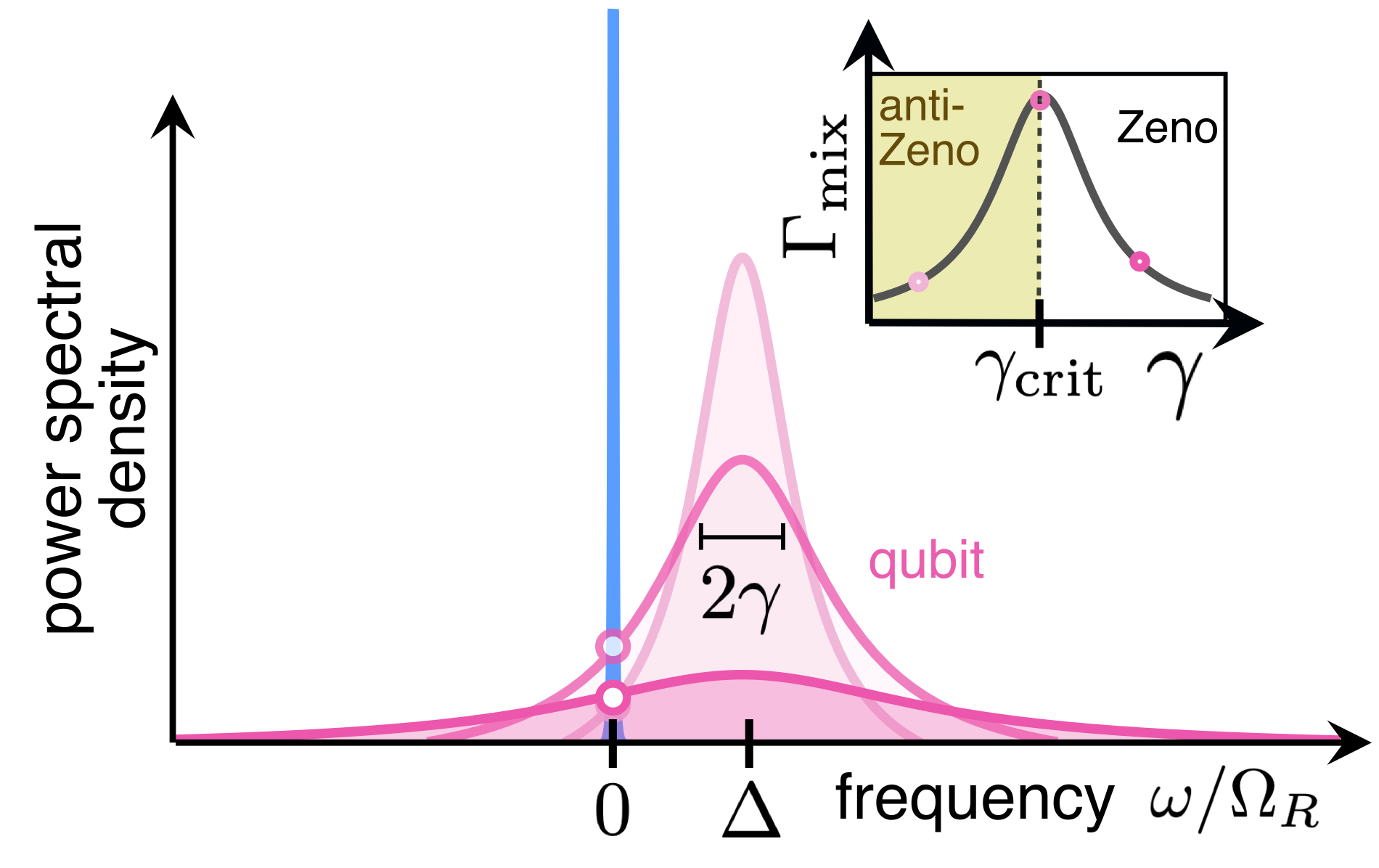}
	\caption{Spectral picture for the transition between Zeno and anti-Zeno regimes for the highly stabilized ($\Delta \gg \Omega_R$) case. The exponential mixing rate $\Gamma_{\rm mix}$ (inset) is proportional to the overlap between a zero-frequency delta-function (blue) and a Lorentzian (pink) centered at $\Delta/\Omega_R$ with width $2\gamma/\Omega_R$. Increasing the measurement rate $\gamma$ initially increases the overlap (anti-Zeno regime), reaching a maximum at $\gamma_{\rm crit}$, then decreases the overlap (Zeno regime). We illustrate three $\gamma$ (pink circles) to highlight these regimes.}
 \label{fig:anti-Zeno-spectral-picture}
\end{figure}

For the nonselective evolution, the Lorentzian form of Eq.~\eqref{eq:stabilized-gam-0} suggests an intuitive spectral picture, which we illustrate in Fig. \ref{fig:anti-Zeno-spectral-picture} for the strongly stabilized region $\Delta \gg \Omega_R$ typically studied in anti-Zeno literature. The mixing rate $\Gamma_{\rm mix}$ is proportional to the overlap between a zero-frequency delta function (corresponding roughly to a pure tone drive spectral line) and a Lorentzian (corresponding to the broadened qubit spectral line) centered at $\Delta/\Omega_R$ with width $2\gamma/\Omega_R$. Increasing $\Delta$ decreases this overlap, yielding dynamical stabilization. Increasing $\gamma$ initially \textit{increases} the overlap as the Lorentzian tail increases (anti-Zeno regime), reaches a maximum overlap at $\gamma_{\rm crit}$, then decreases as the Lorentzian flattens into a uniform distribution (Zeno regime). This picture is the equivalent in our simple model to the spectral picture of the anti-Zeno effect used in, e.g., \cite{zhangCriterionQuantumZeno2018,kofmanQuantumZenoEffect1996,chaudhryGeneralFrameworkQuantum2016,thorbeckReadoutinducedSuppressionEnhancement2024}, in which a dephasing bath (i.e., a broadened drive spectral line) is detuned from the system by a frequency $\Del$ and $\Gam_\mr{mix}$ corresponds to the effective energy-relaxation ($T_1$) rate for the system.

\section{Experimental connections}
\label{sec:experimental-connections}

To emphasize the broad applicability of our minimal model that exhibits the main features of the quantum Zeno effect, including the Zeno and anti-Zeno regimes, we now review how it applies to key experimental demonstrations of the quantum Zeno effect in the literature, along with essential caveats.

Our model has three components, all of which may be modified while maintaining the essence of the Zeno effect:
\begin{enumerate}
    \item \textbf{Qubit $\mathbf{\ket1}$ and $\mathbf{\ket0}$}, where $\ket1$ is the \textit{initial} or \textit{desired} state  stabilized by the measurement, and $\ket0$ is the leakage state;

    \item \textbf{Unitary evolution} between $\ket1$ and $\ket0$;

    \item \textbf{QND measurement} distinguishing $\ket1$ and $\ket0$.
\end{enumerate}
We discuss modifications to each of these in turn, while also highlighting the important limitations of applying our framework in each case.

\textbf{1. Qubit states $\mapsto$ Zeno subspaces - } In general, quantum measurements distinguish \textit{subspaces}, rather than individual states, when all states in the same subspace have the same measurement eigenvalue. Thus, it is possible to suppress or enhance transitions \textit{between subspaces} via measurements. When two subspaces are decoupled via strong dissipation or measurement, they are referred to as \textit{Zeno subspaces} \cite{facchiQuantumZenoSubspaces2002}, and the evolution \textit{within} a subspace (left invariant by the measurement) is \textit{Zeno dynamics} \cite{facchiQuantumZenoDynamics2008}\footnote{We emphasize that while the corresponding literature has focused on the Zeno regime, the same principles should apply to subspaces stabilized by unitary effects (i.e. dynamical decoupling) that can then be recoupled via measurements by operating in the anti-Zeno regime. This could be applied to develop measurement-induced gates between naturally decoupled states or subspaces, similar to \citet{singhCapturingElectronVirtual2025}.}. Bath-engineering, entanglement generation, and syndrome detection frequently stabilize Zeno subspaces as the \textit{code space}, and Zeno dynamics are the \textit{logical dynamics}. As pointed out by \citet{burgarthExponentialRiseDynamical2014}, the resulting dynamics can often be exponentially more complex, allowing for universal quantum computation on the resulting Zeno subspace even when such is lacking on the larger Hilbert space.

\textbf{Limitations -} Because our model is the simplest possible quantum system, including only two states, the measurement and the unitary it interferes with both generate transitions between these two states. In the more general case of two subspaces (e.g., a code subspace and an error subspace) whose transitions are suppressed or enhanced by measurements,  these transitions are unlikely to be pure rotations between subspaces, but instead may map specific states (or subspaces) within the code space to specific states (or subspaces) within the error space with different rates. Likewise, while an ideal syndrome measurement will perfectly distinguish a code and error space, the ability to engineer such a measurement between any two subspaces for a particular physical implementation is not guaranteed, and realistic measurements often distinguish different states at different rates.

Our model presents Zeno and anti-Zeno regimes as mutually exclusive; however, the existence of multiple error rates and multiple measurement rates in practice creates the possibility of being in a Zeno regime with respect to one error process while being in an anti-Zeno regime with respect to another process (e.g., $\gamma_{1,\mathrm{crit}}< \gamma < \gamma_{2,\mathrm{crit}}$).

\enlargethispage*{\baselineskip}   

\textbf{2. Unitary evolution $\mapsto$ non-Markovian evolution - } As highlighted in this paper, unitary evolution can be suppressed or enhanced by measurements. However, non-unitary evolution can also be suppressed or enhanced so long as it is non-Markovian and non-commuting with the measurement\footnote{Unitary evolution is connected to non-Markovian decay because we can obtain a non-Markovian decay process by modeling unitary coupling to a bath subsystem that is traced out. Notably, no physical coupling is truly Markovian, since that would require infinitely fast environmental response.}. Commuting evolution is insensitive to the measurement by definition: the time-ordering of the maps is irrelevant. Similarly, Markovian decay is memoryless and instantaneous, so measurements (themselves Markovian) cannot interrupt it. There is no Zeno response for pure Markovian decay (see Appendix \ref{appen:markovian-decay}).\footnote{Nevertheless, Markovian decay can itself modify the \textit{onset} of the Zeno regime by playing a role similar to that of the measurement \cite{guttelGraduallyOpeningSchrodingers2026}.}

Anti-Zeno studies have emphasized generalized non-Markovian evolution by suppressing or enhancing loss to a detuned bath. Furthermore, syndrome detection will passively suppress non-Markovian errors (unknown unitary drifts, correlated noise, or finite-memory bath effects) via the quantum Zeno effect.

Our model is readily extended to include non-Markovian decay by sampling $\Om_R(t)$ as a random variable with a non-white spectral line (classical bath), or by including loss from $\ket{0}$ to a third state, thus forcing the probability of jump-back events from $\ket{0}$ to $\ket{1}$ to decay exponentially (quantum bath).

\textbf{Limitations -} The main limitation to applying our framework to non-unitary evolution is the inability of the Zeno effect to suppress Markovian loss into environmental modes, a common source of errors in quantum computation. In general, all experimental noise may be understood as a mixture of non-Markovian (colored) and Markovian (white) noise to varying degrees. Repeated measurements must be faster than the noise process in order to suppress it, which manifests in our model as the existence of a $\gamma_\mathrm{crit}$ defining the Zeno regime. Since white noise has components at all frequencies, it is impossible to suppress the higher frequency components. The same principle applies to dynamical decoupling.

However, we emphasize that non-Markovian error sources are more common than the ubiquity of Lindblad models of decoherence would suggest: all physical processes that give rise to errors must have an associated timescale, it is simply a question of whether that timescale is experimentally accessible. Assuming a Markovian decay model outright precludes the possibility of Zeno suppression and dynamical decoupling where it may actually be possible; this is reflected in the significantly improved performance of error-suppressed systems \citep{alvarezDynamicalDecouplingNoise2011,yuanPreservingMultilevelQuantum2022,tripathiQuditDynamicalDecoupling2025,vezvaeeDemonstrationHighfidelityEntangled2026}. This fact highlights the importance of choosing a model that accurately captures the mixture of Markovian and non-Markovian noise sources present in one's system, and error suppression techniques such as the quantum Zeno effect and dynamical decoupling can serve as non-Markovian noise spectrometers because of this very feature.

\textbf{3. QND measurements $\mapsto$ non-commuting competing rate - } While we have explored selective and nonselective QND measurements, QND-ness is not necessary to witness the quantum Zeno effect. In general, \textit{any} competing evolution that is non-commuting with the (non-Markovian) evolution is sufficient \cite{burgarthQuantumZenoDynamics2020}. The unitary analog to measurement-based quantum Zeno is dynamical decoupling \cite{facchiControlDecoherenceAnalysis2005,facchiUnificationDynamicalDecoupling2004}.

Whereas a QND (quantum non-demolition) measurement distinguishes both measurement eigenstates without perturbing them\footnote{Note that a superposition of the two eigenstates is, by the nature of quantum measurement, perturbed; the key point is that the QND measurement does not perturb the measurement eigenstates themselves.}, a half-QND (a.k.a. asymmetric) measurement leaves only one eigenstate invariant. This is common in fluorescence measurements and absorptive measurements, and resembles the introduction of $T_1$ effects on certain states.

\textbf{Limitations -} Because $\ket0$ and $\ket1$ are treated on equal footing in our model, symmetric jump rates from $\ket0 \rightarrow \ket1$ and $\ket1 \rightarrow \ket0$ are observed. However, introducing a fluorescence measurement (e.g., measurement of $\hat \sigma_-$ or $\hat a$) breaks this symmetry by removing the possibility of jump-back events. In this case, the mixing rate $\Gamma_\mathrm{mix}$ must be generalized to a decay rate $\Gamma_\mathrm{loss}$. Slightly more complicated is the case (e.g., \cite{thorbeckReadoutinducedSuppressionEnhancement2024}) where swaps between two states $\ket d \leftrightarrow \ket b$ are suppressed or enhanced by fluorescence measurement of the bright state $\ket b$, which emits a photon when it decays to the ground state $\ket g$. In this case, jump-back events are rare but possible. The existence of asymmetric jump rates for non-QND measurements will suppress the appearance of the symmetric back-and-forth tunneling effects and instead give rise to suppression and enhancement of decay to a ground state, which is more closely aligned with previous studies of anti-Zeno effects in the literature.

\subsection{Foundational experiments}
In Table \ref{tab:model-experiment-connection}, we highlight a cross-section of experiments studying the quantum Zeno effect, mapping them to the three components of the minimal model.

\textbf{Asymmetric measurements in early experiments -} Early tests of the quantum Zeno effect in the Zeno regime used asymmetric measurements in trapped ion \cite{itanoQuantumZenoEffect1990,balzerQuantumZenoEffect2000,balzerRelaxationlessDemonstrationQuantum2002} (fluorescence) and optical \cite{kwiatInteractionFreeMeasurement1995,kwiatHighefficiencyQuantumInterrogation1999} (photon absorption) setups, while the stabilized quantum Zeno effect in the anti-Zeno regime was first demonstrated by \citet{fischerObservationQuantumZeno2001} in atomic states (fluorescence). In fluorescence measurements, detection of a photon means the system \textit{was} in the fluorescing level but has since relaxed, while a single-photon detector absorbs the optical photon, ending the interferometric experiment. The first selective demonstrations, by \citet{kwiatInteractionFreeMeasurement1995,kwiatHighefficiencyQuantumInterrogation1999}, leverage this ``null'' or ``interaction-free'' aspect: if we measure a photon absorption event, it cannot have been absorbed up until that time, thus heralding pinning for the entire prior trajectory. By contrast, \citet{balzerQuantumZenoEffect2000,balzerRelaxationlessDemonstrationQuantum2002} measure the probabilities of identical sequences of outcomes in their fluorescence detection setup, thus detecting pinning probabilities despite the presence of jump-and-jump-back events.\footnote{QND measurements will also generally have jump-and-jump-back events \cite{khezriQubitMeasurementError2015}.}

Because of the challenges of engineering a selective QND measurement, QND demonstrations of quantum Zeno (primarily in superconducting platforms) came more than a decade later \cite{palacios-laloyExperimentalViolationBell2010,vijayObservationQuantumJumps2011,kakuyanagiObservationQuantumZeno2015,slichterQuantumZenoEffect2016}. 

\textbf{Stabilized quantum Zeno experiments -} Similar subtleties come into play in the stabilized anti-Zeno literature. Early proposals \cite{kofmanAccelerationQuantumDecay2000,kofmanZenoAntiZenoEffects2001, kofmanFrequentObservationsAccelerate2001,facchiQuantumZenoInverse2001} diverged from prior studies by considering suppression of nonexponential decay from \textit{unstable systems}, which parallel null measurements: jump-back events are rare for a system weakly coupled to a large bath. Therefore, if the state is measured in the initial state at time $t$, it has probably been in the initial state continuously, making $P_1(t)$ approximate the probability of measuring a trajectory of all 1s. This allows \cite{kofmanAccelerationQuantumDecay2000,kofmanZenoAntiZenoEffects2001, kofmanFrequentObservationsAccelerate2001,facchiQuantumZenoInverse2001} to reason about decay rates and spectral overlaps (which are ensemble properties) using trajectory probabilities. However, while rare, jump-back events \textit{must} be possible for short times to create initial nonexponential decay \cite{fischerObservationQuantumZeno2001}. Otherwise, evolution would be Markovian, and there could be no Zeno response.

Along these lines, key anti-Zeno experiments were nonselective, with measurements performed in parallel on a physical ensemble (\citet{fischerObservationQuantumZeno2001}) or implemented as non-informational or passive measurements by an engineered bath (\citet{harringtonQuantumZenoEffects2017,thorbeckReadoutinducedSuppressionEnhancement2024}). These experiments suppressed loss to a detuned, non-Markovian bath: a noisy drive with finite spectral linewidth \cite{harringtonQuantumZenoEffects2017} or decay processes allowing jump-back events for short times \cite{fischerObservationQuantumZeno2001,thorbeckReadoutinducedSuppressionEnhancement2024}. The system-bath detuning introduces nonorthogonality, decoupling the two until measurements are introduced. Because there is no concept of detuning for a white-noise (flat spectrum) bath, this is a non-Markovian feature.

\subsection{Recent applications}

\textbf{Incoherent qubit control and stabilization - } Zeno dragging and entanglement stabilization combine quantum error detection with leakage suppression caused by the quantum Zeno effect as a form of measurement-based control. By rotating the measurement axis much more slowly than the measurement rate, \citet{hacohen-gourgyIncoherentQubitControl2018} dragged the qubit state while simultaneously detecting escapes to the opposite state. This allowed postselection on the desired state, heralding high preparation fidelity. In \citet{blumenthalDemonstrationUniversalControl2022}, a single-qubit operation was turned into a multi-qubit entangling gate by operating in the Zeno regime to forbid certain transitions. Like \cite{hacohen-gourgyIncoherentQubitControl2018}, \citet{blumenthalDemonstrationUniversalControl2022} post-select on the state staying within the desired Zeno subspace to improve gate fidelities.

\textbf{Bath engineering - } Even without post-selection or feedback, similar techniques can be used to forbid certain state transitions and stabilize target subspaces. While not included in Table \ref{tab:model-experiment-connection}, we highlight these as applications of the nonselective quantum Zeno effect implemented via passive, bath-induced measurements. The effect is commonly used to stabilize cat states using dissipation engineering \cite{gautierDesigningHighFidelityZeno2023,mirrahimiDynamicallyProtectedCatqubits2014,leghtasConfiningStateLight2015,lescanneExponentialSuppressionBitflips2020}, to localize many-body systems \cite{barontiniControllingDynamicsOpen2013,patilMeasurementInducedLocalizationUltracold2015}, and to engineer pure $N$-body interactions  via photon-blockading \cite{chakramMultimodePhotonBlockade2022}. Understanding these within the quantum Zeno framework may inspire hybrid protocols that implement passive stabilization via active syndrome detection, enabling further improvements in preparation fidelity and state heralding.

\textbf{Syndrome detection and quantum error correction -} In syndrome detection---a foundational building block of quantum error correction---a measurement that \textit{monitors} for errors (so that they may be corrected actively via feedback) also passively \textit{suppresses} non-Markovian errors via the quantum Zeno effect \cite{erezCorrectingQuantumErrors2004}. This is an underappreciated, but significant, feature of quantum error correction.

Because error correction studies focus on improvements to state and process fidelity due to active correction, there are not many existing comparisons of the improvements due to the active correction component versus passive improvements from making selective measurements. An important exception is \citet{cramerRepeatedQuantumError2016}, who perform real-time feedback based on the outcomes of stabilizer measurements. While the paper does not explicitly highlight the Zeno effect, the data indicate significant suppression of errors by the syndrome measurement even \textit{before} applying feedback. 

Because of this, we believe that the differentiation of passive (detection-induced) and active (correction-induced) improvements is essential to benchmarking performance gains. The quantum Zeno effect in the Zeno regime of strong measurements may be an essential feature of quantum error correction. Far more than simply detecting errors, syndrome measurements can have a twofold effect in protecting the code subspace:
\begin{enumerate}
    \item Analog drifts are \textbf{converted to digital errors} (quantum jumps) that can then be actively corrected via simple gate operations;

    \item Non-Markovian \textbf{digital error rates are suppressed} by more rapid syndrome measurements due to positive Zeno response, so that fewer corrections are needed in the first place.
\end{enumerate}
Thus, for systems dominated by analog (unitary or non-Markovian) errors, this analog-to-digital conversion of errors enables digital quantum error correction in the first place.\footnote{By contrast, Markovian errors such as photon loss appear instantaneous within the bandwidth of the measuring device, so do not require this analog-to-digital conversion.}

\setlength{\tabcolsep}{5pt}
\renewcommand{\arraystretch}{0.5}
\begin{table*}[] 
\caption[]{Connection of the minimal model to a variety of experimental implementations. A representative sample of the literature is presented to illustrate the broad applicability of the model, but the list is not comprehensive. The legend distinguishes several aspects of the measurement: \textbf{(1)} selective vs. nonselective;
 \textbf{(2)} active vs. passive, which captures whether an information-acquiring measurement takes place (active) or a measurement-mimicking bath is used (passive); and \textbf{(3)} sequential vs. parallel, which captures whether a single quantum system was sequentially measured or a physical ensemble was simultaneously measured.  \\ \\
{$\!\begin{aligned}[t]
    &\text{\underline{Measurement type}:}                  \\     
    &\selective~      \text{selective measurements}                                      && \nonselective~   \text{active nonselective measurements (sequential)}  \\
    &\bath~         \text{passive nonselective measurements}            && \ensemble~      \text{active nonselective measurements (parallel)} 
\end{aligned}$}}
\label{tab:model-experiment-connection}
\small
\begin{tabular}{@{} p{0.20\linewidth}      
                    p{0.03\linewidth}      
                    p{0.005\linewidth}       
                    p{0.25\linewidth}       
                    p{0.20\linewidth}       
                    p{0.20\linewidth}@{}}   
\toprule
                    Citation  
                    & Year &
                     & 
                    Qubit $\ket1$ and $\ket0$ & 
                    Competing drive / loss & 
                    Measurement \\
\midrule \midrule
\multicolumn{4}{p{0.40\linewidth}}{Leakage suppression }\\
    \midrule  \citet{itanoQuantumZenoEffect1990} & 1990 & $\ensemble$          &              ${ }^9 \mathrm{Be}^{+}$ ground-state hyperfine levels     & optical pump & fluorescence measurement on a third bright state \\ \\
                      \citet{palacios-laloyExperimentalViolationBell2010} & 2010 & $\nonselective$ &      transmon $\ket g$ and $\ket e$    & microwave pump & dispersively coupled cavity \\ \\\citet{kakuyanagiObservationQuantumZeno2015}  & 2015  & $\nonselective$   &          flux qubit $\ket \circlearrowright$ and $\ket \circlearrowleft$             & microwave pump & Josephson bifurcation amplifier \\ 
\midrule
\multicolumn{4}{p{0.40\linewidth}}{Pinning}\\
    \midrule                 \multirow{2}{0.7\linewidth}{\citet{kwiatInteractionFreeMeasurement1995}\cite{kwiatHighefficiencyQuantumInterrogation1999}} & \multirow{2}{0.7\linewidth}{1995, 1999} & $\selective$ &            single photon spatial modes in Mach-Zehnder interferometer, $\ket L$ and $\ket R$ & beamsplitter & single-photon detector \\ \\
                    &  &        &   laser polarization modes $\ket H$ and $\ket V$                                 &   double-pass through quarter waveplate &  polarization interferometer and single-photon detector \\ \\
                      \multirow{2}{0.7\linewidth}{\citet{balzerQuantumZenoEffect2000,balzerRelaxationlessDemonstrationQuantum2002}} & 2000, 2002 & $\selective$        &               ${ }^{172} \mr{Yb}^{+}$ $S_{1/2}$ and $D_{5/2}$ states    & optical pump & \multirow{2}{0.8\linewidth}{fluorescence measurement on $P_{1/2}$ bright state}  \\ 
                    &   &      & ${ }^{171} \mr{Yb}^{+}$ ground-state hyperfine levels  &    & \\   
\midrule
\multicolumn{4}{p{0.40\linewidth}}{Quantum jumps}\\
    \midrule    Vijay et al.   \cite{vijayObservationQuantumJumps2011} and \citet{slichterQuantumZenoEffect2016} & 2011, 2016 &  $\selective$ &       transmon $\ket g$ and $\ket e$ & microwave pump & \multirow{2}{0.8\linewidth}{dispersively-coupled cavity} \\ \\
                
\midrule
\multicolumn{4}{p{0.40\linewidth}}{Leakage enhancement and suppression}\\
    \midrule Fischer et al.
    \cite{fischerObservationQuantumZeno2001} & 2001 & $\ensemble$       &              optically trapped states and continuum states of Na atom ensemble  & tunneling due to instability & fluorescence imaging of atom position via charge-coupled-device camera \\ \\
Harrington et al. \cite{harringtonQuantumZenoEffects2017} & 2017 & $\blacksquare$    &                   transmon $\ket g$ and $\ket e$                                                  & microwave pump with Lorentzian spectrum & Berry phase randomization ``quasi-measurement''\\ \\
\citet{thorbeckReadoutinducedSuppressionEnhancement2024} & 2024 & $\blacksquare$  &    single-excitation states $\ket{01}$ and $\ket{10}$ of two transmons   & vacuum Rabi oscillations into lossy $\ket{01}$ state & dispersive measurement on first qubit  \\  
\midrule

\end{tabular}
\end{table*}

\setlength{\tabcolsep}{5pt}
\renewcommand{\arraystretch}{0.5}
\begin{table*}[t]
\caption[]{(Continued from previous page) Connection of the minimal model to a variety of experimental implementations. A representative sample of the literature is presented to illustrate the broad applicability of the model, but the list is not comprehensive. The legend distinguishes several aspects of the measurement: \textbf{(1)} selective vs. nonselective;
 \textbf{(2)} active vs. passive, which captures whether an information-acquiring measurement takes place (active) or a measurement-mimicking bath is used (passive); and \textbf{(3)} sequential vs. parallel, which captures whether a single quantum system was sequentially measured or a physical ensemble was simultaneously measured.  \\ \\
{$\!\begin{aligned}[t]
    &\text{\underline{Measurement type}:}                  \\     
    &\selective~      \text{selective measurements}                                      && \nonselective~   \text{active nonselective measurements (sequential)}  \\
    &\bath~         \text{passive nonselective measurements}            && \ensemble~      \text{active nonselective measurements (parallel)} 
\end{aligned}$}}
\label{tab:model-experiment-connection-2}
\small
\begin{tabular}{@{} p{0.20\linewidth}      
                    p{0.03\linewidth}      
                    p{0.005\linewidth}       
                    p{0.25\linewidth}       
                    p{0.20\linewidth}       
                    p{0.20\linewidth}@{}}   
\toprule
                    Citation  
                    & Year &
                     & 
                    Qubit $\ket1$ and $\ket0$ & 
                    Competing drive / loss & 
                    Measurement \\
\midrule \midrule

\multicolumn{4}{p{0.40\linewidth}}{Syndrome detection}\\
    \midrule
  \citet{cramerRepeatedQuantumError2016} & 2016 & $\selective$&  identical-phase states of three $^{13}C$ nuclear spins & natural phase errors & stabilizer measurements on an NV center ancilla \\ 

\midrule
\multicolumn{4}{p{0.40\linewidth}}{Zeno dragging}\\
    \midrule      \citet{hacohen-gourgyIncoherentQubitControl2018} & 2018 & $\selective$ &   time-dependent transmon states $(\ket g \pm e^{i \del(t)} \ket e)/\sqrt2$ & adiabatic change of measurement axis & amplitude measurement of longitudinally coupled cavity with sideband phase $\del(t)$ \\ \\ \midrule
\multicolumn{4}{p{0.40\linewidth}}{Entanglement stabilization}\\
    \midrule 
    \citet{blumenthalDemonstrationUniversalControl2022} & 2022    & $\selective$                  & transmon qutrit-qubit subspaces $\hat I - \ket{fe}\bra{fe}$ and $\ket{fe}\bra{fe}$ & microwave pump on qutrit $\ket f \leftrightarrow \ket e$  & joint qutrit-qubit dispersive coupling to cavity  \\ \\
  \citet{gaikwadEntanglementAssistedProbe2024} & 2024 & $\blacksquare$                          & transmon $\ket g$ and $\ket e$ & resonant transverse coupling to neighboring transmon & noisy microwave drive on dispersively coupled resonator \\
\bottomrule
\end{tabular}
\end{table*}
\pagestyle{empty}
 \pagestyle{plain}
\renewcommand{\arraystretch}{1.0}
\section{Conclusion}
\label{sec:conclusions}
In this tutorial, we have explored how the quantum Zeno effect arises in myriad experimental scenarios. 

Using the Rabi driven qubit as an exemplar, we showed how the quantum Zeno effect appears for realistic, finite-strength measurements that compete with natural evolution, with the quantum Zeno paradox arising in the limit of infinitely strong or infinitely fast measurements. This revealed the importance of the Zeno response in quantifying regimes of measurement-induced decay envelope  enhancement and suppression, corresponding respectively to anti-Zeno and Zeno regimes. 

We explored the consequences of this picture for standard ($\Delta = 0$) and stabilized ($\Delta > \Omega_R$) quantum Zeno effects, and emphasized how the anti-Zeno to Zeno transition is mediated by the competition of the measurement and dressed oscillation rates. We also identified ensemble- and trajectory-level signatures of the quantum Zeno effect in both anti-Zeno and Zeno regimes, the latter being a new result of this paper.

We highlighted the quantum Zeno effect's historical development and extensive applications in quantum control and error suppression, detection, and correction, emphasizing that the effect can be deliberately engineered or may emerge as a natural consequence of error detection and bath-engineering. Given its profound and wide-ranging impact, we envision that a unified framework for understanding the quantum Zeno effect will provide important insights for leveraging it in near-term and universal quantum computation.

\section*{Acknowledgments}
SG and ELF thank Danielle Gov and Barkay Guttel for useful discussions regarding the onset of the Zeno regime due to half-QND measurements. SG acknowledges funding from the Graduate Fellowships for STEM Development (GFSD).
SG and ELF were supported by the Office of Naval Research under grant N00014-21-1-2688 and by Research Corp under Cottrell Scholarship 27550. AK was supported by the National Science Foundation under grants DMR-2047357 and DMR-2508447. JD was supported by the U.S. Army Research Office under grant W911NF-22-1-0258. 
\linebreak

\section*{Author contributions}
SG wrote code, performed simulations, and wrote the first draft. SG and JD derived results. JD, AK, and ELF guided the research project. All authors contributed to writing the manuscript.
An LLM was used in proofreading, formatting, and checking results in the revised manuscript.

\begin{widetext}
\newpage 

\appendix

\section{Derivation of decay rates for nonorthogonal competing Hamiltonian}
\label{appen:zeno-anti-zeno}
We consider the case where the measurement is along $z$, and include both $x$ and $z$ terms in the Hamiltonian. While the orthogonal Hamiltonian $\hat H = (\Om_R/2) \hat \sig_x$ represents a rotation about the $x$-axis with the $\ket{\pm}$ states as eigenstates, the tilted Hamiltonian $\hat H = (\Om_R/2) \hat \sig_x + (\Del/2) \hat \sig_z \equiv (\Om/2) \hat \sig_\theta$ [Eq.~\eqref{eq:H-general}] represents a rotation around an axis $z'= z \cos \tha + x \sin \tha$ tilted by $\tha$ away from the original $z$ axis such that
\begin{subequations}
    \begin{align}
        \cos \tha &= \Del / \Om \\
        \sin \tha &= \Om_R / \Om \\
        \tan \tha &= \Om_R / \Del \\
        \Om &\equiv \sqrt{\Del^2 + \Om_R^2}.
    \end{align}
\end{subequations}

It will be convenient to work in rotated coordinates
\begin{subequations}
    \begin{align}
        \hat \sig_x ' &= \hat \sig_x \cos \tha -\hat \sig_z \sin \tha \\
        \hat \sig_y ' &= \hat \sig_y \\
        \hat \sig_z' &= \hat \sig_z \cos \tha  + \hat \sig_x \sin \tha, 
    \end{align}
\end{subequations}
where $\hat \sig_z' \equiv \hat \sig_\tha$ in Eq.~\eqref{eq:H-general}. Notably, the $y$ coordinate remains unchanged by the transformation.

We can obtain differential equations for the tilted Bloch coordinates by expanding Eq.~\eqref{eq:lindblad} in terms of
\begin{equation}
    \hat \rho \equiv (1/2) \big[\hat I + x' \hat \sig_x' + y' \hat \sig_y' + z'\hat\sig_z' \big].
\end{equation}
This yields
\begin{subequations}
    \begin{align}
        \dot x' &= - \Om y' - 2 \gam(x' \cos^2 \tha + z' \sin \tha \cos \tha) \\
        \dot y' &= \Om x' - 2\gam y'\\
        \dot z' &= -2 \gam (z' \sin^2 \tha + x' \sin \tha \cos \tha).
    \end{align}
\end{subequations}
We can identify the coherence in the dressed oscillation plane as 
\begin{equation}
    \rho_{01}' \equiv (x' + i y')/2,
\end{equation}
so that
\begin{subequations}
    \begin{align}
    \dot \rho_{01}' &= i(\Om + 2i \gam)\rho_{01}'  - \gam \sin \tha \left[- x'\sin \tha + z' \cos \tha \right]\\
    &= i(\Om + 2i \gam)\rho_{01}' - (\gam \Om_R/\Om) z,    
    \end{align}
\end{subequations}
where we have identified $z = - x'\sin \tha + z' \cos \tha$ as the bare $z$ Bloch coordinate. Using the integrating factor method, we can obtain an integro-differential solution
\begin{equation}
    \rho_{01}'(t) = \rho_{01}'(0) - (\gam \Om_R/\Om) \int_{0}^{t} e^{i(\Om + 2i \gam)\tau} z(t - \tau) d\tau.
    \label{eq:integro-differential}
\end{equation}

Recall that the bare $z$ evolution is coupled only to $y$ via the $\Om_R\hat \sig_x$ component of the drive, since it is an eigenstate of both the measurement and the $\Del \hat \sig_z$ drive component. Thus, the imaginary part of Eq.~\eqref{eq:integro-differential} directly implies a solution for $z$:
\begin{subequations}
    \begin{align}
        \dot z = \Om_R ~y = 2 \Om_R ~\mr{Im} (\rho_{01}') &= -A \int_{0}^{t} e^{-2\gam \tau} \sin (\Om \tau)~ z(t - \tau) d\tau, \\ 
        A &\equiv 2 \gamma \Omega_R^2 / \Omega.
        \label{eq:z-solution}
    \end{align}
\end{subequations}
Eq.~\eqref{eq:z-solution} is an exact expression for the time-evolution of the bare $z$ coordinate. To examine successive contributions of this convolution to an overall decay, we can expand $z(t - \tau)$ in orders of $\tau$ as
\begin{equation}
    z(t - \tau) = z(t) - \frac{\partial z}{\partial t}(t) \tau + \frac{1}{2} \frac{\partial^2 z}{\partial t^2}(t) \tau^2 + \ldots.
\end{equation}
This allows us to expand Eq.~\eqref{eq:z-solution} as follows:
\begin{equation}
        \dot z = -A z(t) \int_0^t f(\tau) d\tau + A \dot z\int_0^t ~ \tau f(\tau) d\tau - \frac{A}{2} \ddot z \int_0^t \tau^2 f(\tau) d\tau + \ldots
\end{equation}
where $f(\tau) = e^{-2\gamma \tau} \sin \Omega \tau$ is the memory kernel of the convolution. Because we expect evolution to approximate exponential decay, we expect higher order derivatives of $z$ to be small, and truncate the series at the first derivative:
\begin{equation}
 \dot z = -z(t) \left[\frac{A\int_0^t f(\tau) d\tau}{1 - A \int_0^t \tau f(\tau) d\tau}\right].
 \label{eq:effective-z-dot}
\end{equation}
We can show that
\begin{subequations}
    \begin{align}
    A \int_0^t f(\tau) d\tau &= \Omega_R \xi_0 + e^{-2\gamma t}  \left[~\ldots~\right] \\  
    A \int_0^t \tau f(\tau)  d\tau & \approx 2 \xi_0^2 + e^{-2\gamma t} \left[~\ldots~\right]
    \end{align}
    \label{eq:integral-equations}
\end{subequations}
where $\xi_0$ is a dimensionless rate defined as
\begin{equation}
        \xi_0 \equiv \left(\frac{\tilde{\gamma}}{1 + \tilde{\gamma}^2}\right) \sin \theta, \qquad
        \tilde{\gamma} \equiv 2\gamma/\Omega,  
        \qquad\sin \theta \equiv \Omega_R/\Omega,
\end{equation}
and the exponentially decaying oscillatory terms are denoted as $e^{-2\gamma t}\left[~\ldots~\right]$. Focusing only on the time-independent parts of Eq.~\eqref{eq:integral-equations}, Eq.~\eqref{eq:effective-z-dot} implies an effective decay 
\begin{equation}
    z(t) = z(0)\exp(-\Gamma_\mathrm{eff} ~t),  \qquad
    \frac{\Gamma_\mathrm{eff}}{\Omega_R} \equiv \xi= \frac{\xi_0}{1 - 2\xi_0^2},
\end{equation}
which appears in dimensionful form in Eq.~\eqref{eq:anti-Zeno-gam}. Note that by neglecting the time-dependent parts, we ignore oscillatory behavior that would show up as an effective time-dependent decay rate. This is consistent with our interest in the behavior of the overall decay envelope specifically. It also means that this approximation will be increasingly accurate as $\gamma$ increases, since oscillations in larger $\gamma$ regimes will damp out more quickly. 

The dimensionless decay rate $\xi$ has a few key features. First of all, we can see that the turning points of $\xi$ relative to the dimensionless measurement rate $\tilde{\gamma}$ depend only on the turning points of $\xi_0$ and appear at $\tilde{\gamma} = \pm 1$:
\begin{equation}
        \frac{d\xi}{d\tilde{\gamma}} = \frac{d\xi}{d\xi_0} \frac{d \xi_0}{d\tilde{\gamma}}         = \left[\frac{1+2\xi_0^2}{(1-2\xi_0^2)^2}\right] \left[-\frac{\sin \theta}{(1 + \tilde{\gamma}^2)^2} (\tilde \gamma+ 1)  (\tilde \gamma- 1)\right].
\end{equation}
Both $\xi$ and $\xi_0$ thus have turning points at $\tilde{\gamma} = 2\gamma/\Omega=1$ $\implies$ $\gamma_\mathrm{crit} = \Omega/2$. Furthermore, when $\xi_0$ is small, $\xi \approx \xi_0$. In other words, in regimes where evolution is already significantly suppressed (e.g. for $\tilde \gamma \ll 1$, $\tilde \gamma \gg 1$, and / or $\Omega_R \ll \Omega$), $\xi_0$ is a good approximation for the dimensionless decay rate. This is significant because $\xi_0(\tilde \gamma)$ is a Lorentzian, which is clearer when written in its dimensional form:
\begin{equation}
    \Gamma_0 = \xi_0 \Omega_R = \frac{2\gamma \Omega_R^2}{\Omega^2 + (2\gamma)^2},
\end{equation}
which is the form that appears in Eq.~\eqref{eq:anti-Zeno-gam}.


\section{The role of measurement in suppressing T1 effects} \label{appen:markovian-decay} Here, we briefly note the possibility of extending our model to include an average $T_1$ effect, such that Eq.~\eqref{eq:lindblad} becomes
\begin{equation}
    \dot{\hat \rho} = -\frac{i}{\hbar} [\hat H,\hat \rho ] + \gam \bigb{\hat \sig_z \hat \rho \hat \sig_z - \hat \rho} \nonumber \\
    + \gamma_1 \Bigb{\hat \sig_+ \hat \rho \hat \sig_- - \frac12 \{\ket0\bra0, \hat \rho \} }.
    \label{eq:lindblad-T1}
\end{equation}
This models qubit loss as a Markovian process with no presumption as to the cause of that loss. In this case the decay due to $\gamma_1$ is completely independent of $\gamma$ and there is no Zeno response. This is in contrast to models where the decay is induced by an auxiliary system \cite{thorbeckReadoutinducedSuppressionEnhancement2024} or bath \cite{harringtonQuantumZenoEffects2017} that then is traced over, leading to a Markovian $\gamma_1$ with an explicit functional dependence on $\gamma$. We note that the lack of Zeno response is a feature of the simple model only, and that more microscopically-accurate models including bath modes do show Zeno effects, which has also been confirmed experimentally \cite{gaikwadEntanglementAssistedProbe2024}. The key point is that the competing evolution must be either unitary or non-Markovian to be suppressed or enhanced by the (Markovian) measurement-induced dephasing.

\begin{figure}
	\centering \includegraphics[width=0.50\linewidth]{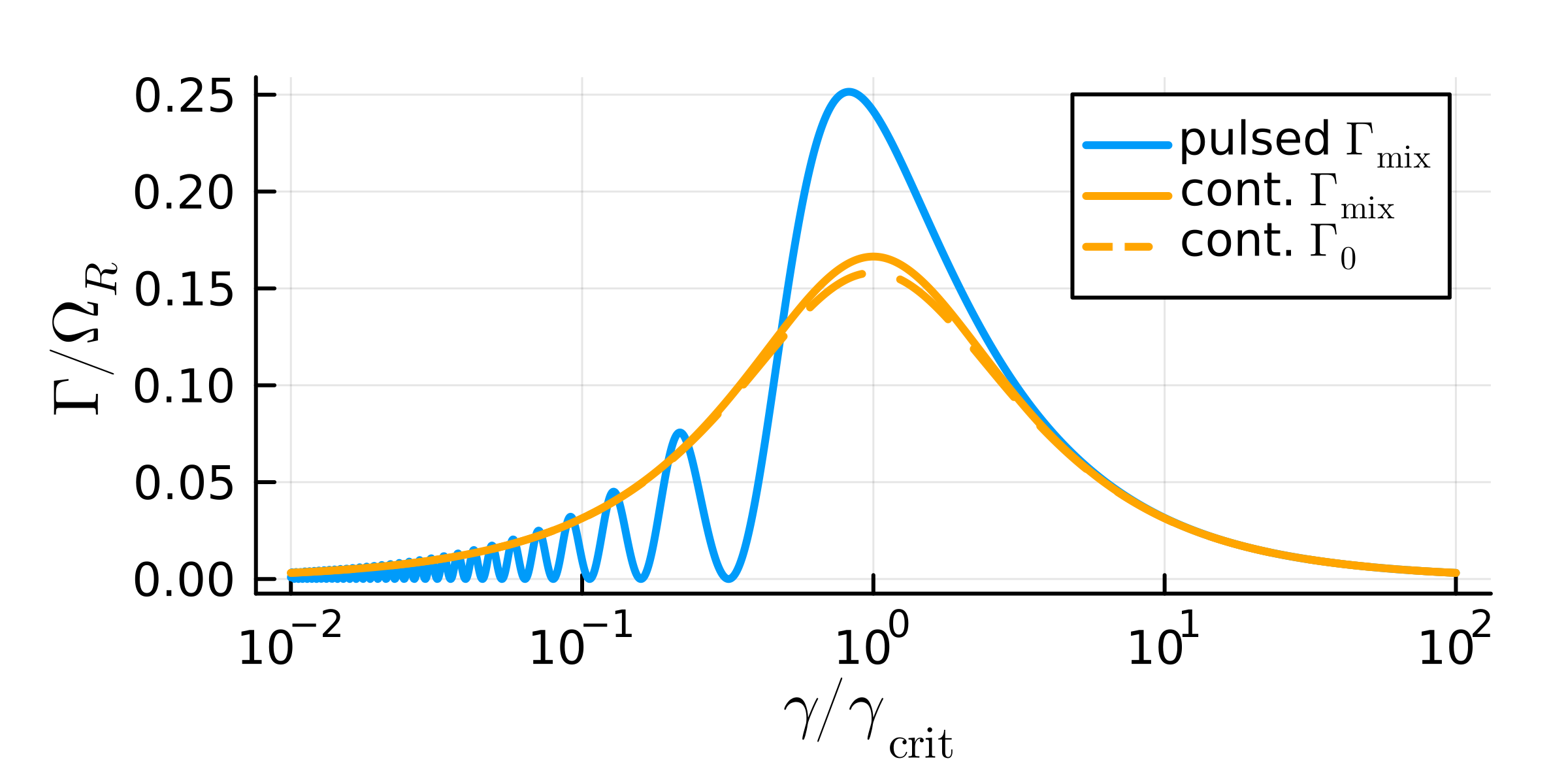}
	\caption{Comparison of the mixing rates in the stabilized regime (with $\Delta = 3\Omega_R$) for pulsed [Eq.~\eqref{eq:pulsed-gamma-mix}] and continuous [Eqs.~\eqref{eq:anti-Zeno-gam}] cases. The mixing rates are plotted in units of the Rabi rate $\Omega_R$, while the measurement rate $\gamma$ is plotted in log-scale relative to $\gamma_\mathrm{crit}$. In the anti-Zeno regime $\gamma < \gamma_\mathrm{crit}$, the pulsed rate is highly structured relative to $\gamma$, with local extrema corresponding to resonances with the dressed evolution rate $\Omega$. Notably, the pulsed curve generally falls below the continuous curve up until $\gamma \lesssim \gamma_\mathrm{crit}$. In the strong measurement regime, as predicted by Eq.~\eqref{eq:strong-measurement-detuned-gam}, pulsed and continuous rates converge to the Zeno limit $\Gamma_\mathrm{mix} =\Omega_R^2/2\gamma$. The presence of distinct correspondences between the continuous and pulsed cases in the two limits $\gamma \ll \gamma_\mathrm{crit}, \gamma \gg \gamma_\mathrm{crit}$ explains why coarse-graining over $\gamma$ to achieve the Fermi's golden rule expression Eq.~\eqref{eq:golden-rule} yields a factor of 2 difference in the $\gamma \ll \gamma_\mathrm{crit}$ limit, even though the two rates converge in the limit $\gamma \gg \gamma_\mathrm{crit}$.}
 \label{fig:pulsed-continuous-rates}
\end{figure}

\section{Correspondence of continuous and pulsed rates}
\label{appen:pulsed-continuous}
In the pulsed measurement case, resonances between the measurement spacing $\Del t = 1/\gamma$ and the unitary rate $\Omega$ lead to a structured anti-Zeno regime in which the mixing rate $\Gamma_\mathrm{mix}$ rapidly oscillates with $\gamma$ (Fig. \ref{fig:pulsed-continuous-rates}).  These oscillations do not appear in the average of continuous diffusive trajectories due to the lack of resonances, leading to distinctive pulsed behavior in the anti-Zeno regime that disappears in the extreme Zeno limit. We can eliminate these resonances by randomizing the projective measurement spacing according to a Poisson process, yielding stochastic evolution that averages to the same master equation as the continuous case.

To do this, we define a Poisson process $dT$ such that, in an infinitesimal time interval $dt$,
\begin{equation}
    dT = \begin{cases}
        1 & \text{with probability } 2\gamma ~dt \\ 
        0 & \text{otherwise}.
    \end{cases}
\end{equation}
We interpret $dT$ as a random variable that tells us whether or not a projective measurement is made in the interval $dt$. By consequence, the previously fixed measurement spacing $\Delta t$ now follows an exponential waiting time distribution, while the rate $\gamma$ remains fixed (as in the continuous case) to determine the effective measurement rate.

In the time interval $dt$, the state either (a) freely evolves as
\begin{equation}
    d\rho = - \frac{i}{\hbar} [\hat H, \rho] ~dt,
    \label{eq:free}
\end{equation}
or (b) is projectively measured such that
\begin{equation}
    \rho(t + dt) = \ket{e}\!\bra{e} \rho_{ee}(t) + \ket{g}\!\bra{g}\rho_{gg}(t).
\end{equation}
In differential form $d \rho = \rho(t + dt) - \rho$, this is
\begin{equation}
    d\rho = -\ket{e}\!\bra{g} \rho_{eg} - \ket{g}\!\bra{e}\rho_{ge} .
    \label{eq:proj}
\end{equation}
Eqs.~\eqref{eq:free} and ~\eqref{eq:proj} can be combined into a single equation using the Poisson process which determines which one occurs:
\begin{subequations}
    \begin{align}
        d\rho &=  \left(- \frac{i}{\hbar} [\hat H, \rho] ~ dt\right) (1- dT) - \left(\ket{e}\!\bra{g} \rho_{eg} + \ket{g}\!\bra{e}\rho_{ge} \right) ~ dT \\ 
&\approx - \frac{i}{\hbar} [\hat H, \rho] ~ dt  - \left(\ket{e}\!\bra{g} \rho_{eg} + \ket{g}\!\bra{e}\rho_{ge} \right)  dT,
    \end{align}
    \label{eq:stochastic-poisson}
\end{subequations}
where we have used $(1 - dT)dt = dt + \mathcal{O}(dt^2)$.

Eq.~\eqref{eq:stochastic-poisson} is a stochastic master equation because $dT$ takes random values. Noting that the ensemble average of $dT$ is $2 \gamma ~dt$, this equation averages to the master equation
\begin{subequations}
    \begin{align}
d \rho &= -\frac{i}{\hbar}[\hat{H}, \hat{\rho}] - 2 \gamma\left(\ket{e}\!\bra{g} \rho_{eg} + \ket{g}\!\bra{e} \rho_{ge}\right)d t \\ 
&= -\frac{i}{\hbar}[\hat{H}, \hat{\rho}] + \gamma\left(\hat{\sigma}_z \rho \hat{\sigma}_z-\rho\right)d t,
\end{align}
\end{subequations}
which is exactly the average of Eq.~\eqref{eq:lindblad}. Thus, the continuous results all have an alternate interpretation as the average of randomized projective measurements. This also clarifies why the continuous results look like a smoothing of the pulsed results.

\end{widetext}

\bibliographystyle{unsrtnat}
\bibliography{zeno_tutorial}

\begin{thebibliography}{106}
\providecommand{\natexlab}[1]{#1}
\providecommand{\url}[1]{\texttt{#1}}
\expandafter\ifx\csname urlstyle\endcsname\relax
  \providecommand{\doi}[1]{doi: #1}\else
  \providecommand{\doi}{doi: \begingroup \urlstyle{rm}\Url}\fi

\bibitem[{von Neumann}(1932)]{vonneumannMathematicalFoundationsQuantum1932}
John {von Neumann}.
\newblock \emph{Mathematical {{Foundations}} of {{Quantum Mechanics}}}.
\newblock Princeton University Press, Princeton, NJ, 1932.
\newblock ISBN 978-0-691-17856-1.
\newblock \doi{10.2307/j.ctt1wq8zhp}.

\bibitem[Misra and Sudarshan(1977)]{misraZenoParadoxQuantum1977}
B.~Misra and E.~C.~G. Sudarshan.
\newblock The {{Zeno}}'s paradox in quantum theory.
\newblock \emph{Journal of Mathematical Physics}, 18\penalty0 (4):\penalty0
  756--763, April 1977.
\newblock ISSN 0022-2488, 1089-7658.
\newblock \doi{10.1063/1.523304}.

\bibitem[Aharonov and Vardi(1980)]{aharonovMeaningIndividualFeynman1980}
Y.~Aharonov and M.~Vardi.
\newblock Meaning of an individual "{{Feynman}} path".
\newblock \emph{Physical Review D}, 21\penalty0 (8):\penalty0 2235--2240, April
  1980.
\newblock ISSN 0556-2821.
\newblock \doi{10.1103/PhysRevD.21.2235}.

\bibitem[Altenm{\"u}ller and
  Schenzle(1993)]{altenmullerDynamicsMeasurementAharonovs1993}
Thomas~P. Altenm{\"u}ller and Axel Schenzle.
\newblock Dynamics by measurement: {{Aharonov}}'s inverse quantum {{Zeno}}
  effect.
\newblock \emph{Physical Review A}, 48\penalty0 (1):\penalty0 70--79, July
  1993.
\newblock ISSN 1050-2947, 1094-1622.
\newblock \doi{10.1103/PhysRevA.48.70}.

\bibitem[{Hacohen-Gourgy} et~al.(2018){Hacohen-Gourgy}, {Garc{\'i}a-Pintos},
  Martin, Dressel, and Siddiqi]{hacohen-gourgyIncoherentQubitControl2018}
S.~{Hacohen-Gourgy}, L.~P. {Garc{\'i}a-Pintos}, L.~S. Martin, J.~Dressel, and
  I.~Siddiqi.
\newblock Incoherent {{Qubit Control Using}} the {{Quantum Zeno Effect}}.
\newblock \emph{Physical Review Letters}, 120\penalty0 (2):\penalty0 020505,
  January 2018.
\newblock ISSN 0031-9007, 1079-7114.
\newblock \doi{10.1103/PhysRevLett.120.020505}.

\bibitem[Lewalle et~al.(2024)Lewalle, Zhang, and
  Whaley]{lewalleOptimalZenoDragging2024}
Philippe Lewalle, Yipei Zhang, and K.~Birgitta Whaley.
\newblock Optimal {{Zeno Dragging}} for {{Quantum Control}}: {{A Shortcut}} to
  {{Zeno}} with {{Action-Based Scheduling Optimization}}.
\newblock \emph{PRX Quantum}, 5\penalty0 (2):\penalty0 020366, June 2024.
\newblock ISSN 2691-3399.
\newblock \doi{10.1103/PRXQuantum.5.020366}.

\bibitem[Huggett(2024)]{huggettZenosParadoxes2024}
Nick Huggett.
\newblock Zeno's {{Paradoxes}}.
\newblock In Edward~N. Zalta and Uri Nodelman, editors, \emph{The {{Stanford
  Encyclopedia}} of {{Philosophy}}}. Metaphysics Research Lab, Stanford
  University, fall 2024 edition, 2024.

\bibitem[Harrington et~al.(2022)Harrington, Mueller, and
  Murch]{harringtonEngineeredDissipationQuantum2022}
Patrick~M. Harrington, Erich Mueller, and Kater Murch.
\newblock Engineered {{Dissipation}} for {{Quantum Information Science}}.
\newblock \emph{Nature Reviews Physics}, 4\penalty0 (10):\penalty0 660--671,
  August 2022.
\newblock ISSN 2522-5820.
\newblock \doi{10.1038/s42254-022-00494-8}.

\bibitem[Itano et~al.(1990)Itano, Heinzen, Bollinger, and
  Wineland]{itanoQuantumZenoEffect1990}
Wayne~M. Itano, D.~J. Heinzen, J.~J. Bollinger, and D.~J. Wineland.
\newblock Quantum {{Zeno}} effect.
\newblock \emph{Physical Review A}, 41\penalty0 (5):\penalty0 2295--2300, March
  1990.
\newblock ISSN 1050-2947, 1094-1622.
\newblock \doi{10.1103/PhysRevA.41.2295}.

\bibitem[Nagels et~al.(1997)Nagels, Hermans, and
  Chapovsky]{nagelsQuantumZenoEffect1997}
B.~Nagels, L.~J.~F. Hermans, and P.~L. Chapovsky.
\newblock Quantum {{Zeno Effect Induced}} by {{Collisions}}.
\newblock \emph{Physical Review Letters}, 79\penalty0 (17):\penalty0
  3097--3100, October 1997.
\newblock ISSN 0031-9007, 1079-7114.
\newblock \doi{10.1103/PhysRevLett.79.3097}.

\bibitem[Peres and Ron(1990)]{peresIncompleteCollapsePartial1990}
Asher Peres and Amiram Ron.
\newblock Incomplete ``collapse'' and partial quantum {{Zeno}} effect.
\newblock \emph{Physical Review A}, 42\penalty0 (9):\penalty0 5720--5722,
  November 1990.
\newblock ISSN 1050-2947, 1094-1622.
\newblock \doi{10.1103/PhysRevA.42.5720}.

\bibitem[Frerichs and Schenzle(1991)]{frerichsQuantumZenoEffect1991}
Vera Frerichs and Axel Schenzle.
\newblock Quantum {{Zeno}} effect without collapse of the wave packet.
\newblock \emph{Physical Review A}, 44\penalty0 (3):\penalty0 1962--1968,
  August 1991.
\newblock \doi{10.1103/PhysRevA.44.1962}.

\bibitem[Beige and Hegerfeldt(1996)]{beigeProjectionPostulateAtomic1996}
Almut Beige and Gerhard~C. Hegerfeldt.
\newblock Projection postulate and atomic quantum {{Zeno}} effect.
\newblock \emph{Physical Review A}, 53\penalty0 (1):\penalty0 53--65, January
  1996.
\newblock \doi{10.1103/PhysRevA.53.53}.

\bibitem[Nakazato et~al.(1996)Nakazato, Namiki, Pascazio, and
  Rauch]{nakazatoUnderstandingQuantumZeno1996}
Hiromichi Nakazato, Mikio Namiki, Saverio Pascazio, and Helmut Rauch.
\newblock Understanding the quantum {{Zeno}} effect.
\newblock \emph{Physics Letters A}, 217\penalty0 (4):\penalty0 203--208, July
  1996.
\newblock ISSN 0375-9601.
\newblock \doi{10.1016/0375-9601(96)00350-7}.

\bibitem[Mack et~al.(2014)Mack, Wallentowitz, and
  Toschek]{mackDecoherenceGeneralizedMeasurement2014}
Gerhard Mack, Sascha Wallentowitz, and Peter~E. Toschek.
\newblock Decoherence in generalized measurement and the quantum {{Zeno}}
  paradox.
\newblock \emph{Physics Reports}, 540\penalty0 (1):\penalty0 1--23, July 2014.
\newblock ISSN 0370-1573.
\newblock \doi{10.1016/j.physrep.2014.02.004}.

\bibitem[Facchi and Pascazio(2002)]{facchiQuantumZenoSubspaces2002}
P.~Facchi and S.~Pascazio.
\newblock Quantum {{Zeno}} subspaces.
\newblock \emph{Physical Review Letters}, 89\penalty0 (8):\penalty0 080401,
  August 2002.
\newblock ISSN 0031-9007, 1079-7114.
\newblock \doi{10.1103/PhysRevLett.89.080401}.

\bibitem[Facchi and Pascazio(2008)]{facchiQuantumZenoDynamics2008}
P.~Facchi and S.~Pascazio.
\newblock Quantum {{Zeno}} dynamics: Mathematical and physical aspects.
\newblock \emph{Journal of Physics A: Mathematical and Theoretical},
  41\penalty0 (49):\penalty0 493001, October 2008.
\newblock ISSN 1751-8121.
\newblock \doi{10.1088/1751-8113/41/49/493001}.

\bibitem[Sch{\"a}fer et~al.(2014)Sch{\"a}fer, Herrera, Cherukattil, Lovecchio,
  Cataliotti, Caruso, and Smerzi]{schaferExperimentalRealizationQuantum2014}
F.~Sch{\"a}fer, I.~Herrera, S.~Cherukattil, C.~Lovecchio, F.S. Cataliotti,
  F.~Caruso, and A.~Smerzi.
\newblock Experimental realization of quantum zeno dynamics.
\newblock \emph{Nature Communications}, 5\penalty0 (1):\penalty0 3194, May
  2014.
\newblock ISSN 2041-1723.
\newblock \doi{10.1038/ncomms4194}.

\bibitem[S{\o}rensen et~al.(2018)S{\o}rensen, Dalgaard, Kiilerich, M{\o}lmer,
  and Sherson]{sorensenQuantumControlMeasurements2018}
Jens~Jakob S{\o}rensen, Mogens Dalgaard, Alexander~Holm Kiilerich, Klaus
  M{\o}lmer, and Jacob Sherson.
\newblock Quantum {{Control}} with {{Measurements}} and {{Quantum Zeno
  Dynamics}}.
\newblock \emph{Physical Review A}, 98\penalty0 (6):\penalty0 062317, December
  2018.
\newblock ISSN 2469-9926, 2469-9934.
\newblock \doi{10.1103/PhysRevA.98.062317}.

\bibitem[Ruskov et~al.(2006)Ruskov, Korotkov, and
  Mizel]{ruskovQuantumZenoStabilization2006}
Rusko Ruskov, Alexander~N. Korotkov, and Ari Mizel.
\newblock Quantum {{Zeno}} stabilization in weak continuous measurement of two
  qubits.
\newblock \emph{Physical Review B}, 73\penalty0 (8):\penalty0 085317, February
  2006.
\newblock \doi{10.1103/PhysRevB.73.085317}.

\bibitem[Marcantoni and
  Merkli(2025)]{marcantoniUltrastrongCouplingNonselective2025}
Stefano Marcantoni and Marco Merkli.
\newblock Ultrastrong coupling, nonselective measurement and quantum {{Zeno}}
  dynamics.
\newblock \emph{Quantum}, 9:\penalty0 1656, March 2025.
\newblock \doi{10.22331/q-2025-03-06-1656}.

\bibitem[Beige et~al.(2000)Beige, Braun, Tregenna, and
  Knight]{beigeQuantumComputingUsing2000}
A.~Beige, D.~Braun, B.~Tregenna, and P.~L. Knight.
\newblock Quantum {{Computing Using Dissipation}} to {{Remain}} in a
  {{Decoherence-Free Subspace}}.
\newblock \emph{Physical Review Letters}, 85\penalty0 (8):\penalty0 1762--1765,
  August 2000.
\newblock ISSN 0031-9007, 1079-7114.
\newblock \doi{10.1103/PhysRevLett.85.1762}.

\bibitem[Bretheau et~al.(2015)Bretheau, {Campagne-Ibarcq}, Flurin, Mallet, and
  Huard]{bretheauQuantumDynamicsElectromagnetic2015}
L.~Bretheau, P.~{Campagne-Ibarcq}, E.~Flurin, F.~Mallet, and B.~Huard.
\newblock Quantum dynamics of an electromagnetic mode that cannot contain {{N}}
  photons.
\newblock \emph{Science}, 348\penalty0 (6236):\penalty0 776--779, May 2015.
\newblock \doi{10.1126/science.1259345}.

\bibitem[Chakram et~al.(2022)Chakram, He, Dixit, Oriani, Naik, Leung, Kwon, Ma,
  Jiang, and Schuster]{chakramMultimodePhotonBlockade2022}
Srivatsan Chakram, Kevin He, Akash~V. Dixit, Andrew~E. Oriani, Ravi~K. Naik,
  Nelson Leung, Hyeokshin Kwon, Wen-Long Ma, Liang Jiang, and David~I.
  Schuster.
\newblock Multimode photon blockade.
\newblock \emph{Nature Physics}, 18\penalty0 (8):\penalty0 879--884, August
  2022.
\newblock ISSN 1745-2473, 1745-2481.
\newblock \doi{10.1038/s41567-022-01630-y}.

\bibitem[Mirrahimi et~al.(2014)Mirrahimi, Leghtas, Albert, Touzard, Schoelkopf,
  Jiang, and Devoret]{mirrahimiDynamicallyProtectedCatqubits2014}
Mazyar Mirrahimi, Zaki Leghtas, Victor~V. Albert, Steven Touzard, Robert~J.
  Schoelkopf, Liang Jiang, and Michel~H. Devoret.
\newblock Dynamically protected cat-qubits: A new paradigm for universal
  quantum computation.
\newblock \emph{New Journal of Physics}, 16\penalty0 (4):\penalty0 045014,
  April 2014.
\newblock ISSN 1367-2630.
\newblock \doi{10.1088/1367-2630/16/4/045014}.

\bibitem[Touzard et~al.(2018)Touzard, Grimm, Leghtas, Mundhada, Reinhold,
  Axline, Reagor, Chou, Blumoff, Sliwa, Shankar, Frunzio, Schoelkopf,
  Mirrahimi, and Devoret]{touzardCoherentOscillationsQuantum2018}
S.~Touzard, A.~Grimm, Z.~Leghtas, S.~O. Mundhada, P.~Reinhold, C.~Axline,
  M.~Reagor, K.~Chou, J.~Blumoff, K.~M. Sliwa, S.~Shankar, L.~Frunzio, R.~J.
  Schoelkopf, M.~Mirrahimi, and M.~H. Devoret.
\newblock Coherent {{Oscillations}} inside a {{Quantum Manifold Stabilized}} by
  {{Dissipation}}.
\newblock \emph{Physical Review X}, 8\penalty0 (2):\penalty0 021005, April
  2018.
\newblock ISSN 2160-3308.
\newblock \doi{10.1103/PhysRevX.8.021005}.

\bibitem[Lescanne et~al.(2020)Lescanne, Villiers, Peronnin, Sarlette, Delbecq,
  Huard, Kontos, Mirrahimi, and
  Leghtas]{lescanneExponentialSuppressionBitflips2020}
Rapha{\"e}l Lescanne, Marius Villiers, Th{\'e}au Peronnin, Alain Sarlette,
  Matthieu Delbecq, Benjamin Huard, Takis Kontos, Mazyar Mirrahimi, and Zaki
  Leghtas.
\newblock Exponential suppression of bit-flips in a qubit encoded in an
  oscillator.
\newblock \emph{Nature Physics}, 16\penalty0 (5):\penalty0 509--513, May 2020.
\newblock ISSN 1745-2473, 1745-2481.
\newblock \doi{10.1038/s41567-020-0824-x}.

\bibitem[Zapletal et~al.(2022)Zapletal, Nunnenkamp, and
  Brunelli]{zapletalStabilizationMultimodeSchrodinger2022}
Petr Zapletal, Andreas Nunnenkamp, and Matteo Brunelli.
\newblock Stabilization of {{Multimode Schr}}\textbackslash "odinger {{Cat
  States Via Normal-Mode Dissipation Engineering}}.
\newblock \emph{PRX Quantum}, 3\penalty0 (1):\penalty0 010301, January 2022.
\newblock \doi{10.1103/PRXQuantum.3.010301}.

\bibitem[Fischer et~al.(2001)Fischer, {Guti{\'e}rrez-Medina}, and
  Raizen]{fischerObservationQuantumZeno2001}
M.~C. Fischer, B.~{Guti{\'e}rrez-Medina}, and M.~G. Raizen.
\newblock Observation of the {{Quantum Zeno}} and {{Anti-Zeno Effects}} in an
  {{Unstable System}}.
\newblock \emph{Physical Review Letters}, 87\penalty0 (4):\penalty0 040402,
  July 2001.
\newblock ISSN 0031-9007, 1079-7114.
\newblock \doi{10.1103/PhysRevLett.87.040402}.

\bibitem[{\'A}lvarez et~al.(2010){\'A}lvarez, Rao, Frydman, and
  Kurizki]{alvarezZenoAntiZenoPolarization2010}
Gonzalo~A. {\'A}lvarez, D.~D.~Bhaktavatsala Rao, Lucio Frydman, and Gershon
  Kurizki.
\newblock Zeno and {{Anti-Zeno Polarization Control}} of {{Spin Ensembles}} by
  {{Induced Dephasing}}.
\newblock \emph{Physical Review Letters}, 105\penalty0 (16):\penalty0 160401,
  October 2010.
\newblock \doi{10.1103/PhysRevLett.105.160401}.

\bibitem[Harrington et~al.(2019)Harrington, Tan, Naghiloo, and
  Murch]{harringtonCharacterizingStatisticalArrow2019}
P.~M. Harrington, D.~Tan, M.~Naghiloo, and K.~W. Murch.
\newblock Characterizing a {{Statistical Arrow}} of {{Time}} in {{Quantum
  Measurement Dynamics}}.
\newblock \emph{Physical review letters}, 123\penalty0 (2):\penalty0
  020502--020502, 2019.
\newblock ISSN 0031-9007.
\newblock \doi{10.1103/PhysRevLett.123.020502}.

\bibitem[Thorbeck et~al.(2024)Thorbeck, Xiao, Kamal, and
  Govia]{thorbeckReadoutinducedSuppressionEnhancement2024}
Ted Thorbeck, Zhihao Xiao, Archana Kamal, and Luke C.~G. Govia.
\newblock Readout-{{Induced Suppression}} and {{Enhancement}} of
  {{Superconducting Qubit Lifetimes}}.
\newblock \emph{Physical Review Letters}, 132\penalty0 (9):\penalty0 090602,
  February 2024.
\newblock \doi{10.1103/PhysRevLett.132.090602}.

\bibitem[Kofman and Kurizki(1996)]{kofmanQuantumZenoEffect1996}
A.~G. Kofman and G.~Kurizki.
\newblock Quantum {{Zeno}} effect on atomic excitation decay in resonators.
\newblock \emph{Physical Review A}, 54\penalty0 (5):\penalty0 R3750--R3753,
  November 1996.
\newblock \doi{10.1103/PhysRevA.54.R3750}.

\bibitem[Kofman and Kurizki(2000)]{kofmanAccelerationQuantumDecay2000}
A.~G. Kofman and G.~Kurizki.
\newblock Acceleration of quantum decay processes by frequent observations.
\newblock \emph{Nature}, 405\penalty0 (6786):\penalty0 546--550, June 2000.
\newblock ISSN 0028-0836, 1476-4687.
\newblock \doi{10.1038/35014537}.

\bibitem[Kofman and Kurizki(2001)]{kofmanFrequentObservationsAccelerate2001}
A.~G. Kofman and G.~Kurizki.
\newblock Frequent observations accelerate decay: {{The}} anti-{{Zeno}} effect.
\newblock \emph{Zeitschrift f\"ur Naturforschung A}, 56\penalty0
  (1-2):\penalty0 83--90, February 2001.
\newblock ISSN 1865-7109, 0932-0784.
\newblock \doi{10.1515/zna-2001-0113}.

\bibitem[Kofman et~al.(2001)Kofman, Kurizki, and
  Opatrn{\'y}]{kofmanZenoAntiZenoEffects2001}
A.~G. Kofman, G.~Kurizki, and T.~Opatrn{\'y}.
\newblock Zeno and anti-{{Zeno}} effects for photon polarization dephasing.
\newblock \emph{Physical review. A, Atomic, molecular, and optical physics},
  63\penalty0 (4), 2001.
\newblock ISSN 1050-2947.
\newblock \doi{10.1103/PhysRevA.63.042108}.

\bibitem[Facchi et~al.(2001)Facchi, Nakazato, and
  Pascazio]{facchiQuantumZenoInverse2001}
P.~Facchi, H.~Nakazato, and S.~Pascazio.
\newblock From the {{Quantum Zeno}} to the {{Inverse Quantum Zeno Effect}}.
\newblock \emph{Physical Review Letters}, 86\penalty0 (13):\penalty0
  2699--2703, March 2001.
\newblock ISSN 0031-9007, 1079-7114.
\newblock \doi{10.1103/PhysRevLett.86.2699}.

\bibitem[Zhang et~al.(2014)Zhang, Ai, Li, Xu, and
  Sun]{zhangDynamicsQuantumZeno2014}
Peng Zhang, Qing Ai, Yong Li, DaZhi Xu, and ChangPu Sun.
\newblock Dynamics of quantum zeno and anti-zeno effects in an open system.
\newblock \emph{Science China. Physics, mechanics \& astronomy}, 57\penalty0
  (2):\penalty0 194--207, 2014.
\newblock ISSN 1674-7348.
\newblock \doi{10.1007/s11433-013-5377-x}.

\bibitem[Chaudhry and Gong(2014)]{chaudhryZenoAntiZenoEffects2014}
Adam~Zaman Chaudhry and Jiangbin Gong.
\newblock Zeno and anti-{{Zeno}} effects on dephasing.
\newblock \emph{Physical Review A}, 90\penalty0 (1):\penalty0 012101, July
  2014.
\newblock \doi{10.1103/PhysRevA.90.012101}.

\bibitem[Chaudhry(2016)]{chaudhryGeneralFrameworkQuantum2016}
Adam~Zaman Chaudhry.
\newblock A general framework for the {{Quantum Zeno}} and anti-{{Zeno}}
  effects.
\newblock \emph{Scientific Reports}, 6\penalty0 (1):\penalty0 29497, July 2016.
\newblock ISSN 2045-2322.
\newblock \doi{10.1038/srep29497}.

\bibitem[Zhou et~al.(2017)Zhou, L{\"u}, Zheng, and
  Goan]{zhouQuantumZenoAntiZeno2017}
Zixian Zhou, Zhiguo L{\"u}, Hang Zheng, and Hsi-Sheng Goan.
\newblock Quantum {{Zeno}} and anti-{{Zeno}} effects in open quantum systems.
\newblock \emph{Physical Review A}, 96\penalty0 (3):\penalty0 032101, September
  2017.
\newblock \doi{10.1103/PhysRevA.96.032101}.

\bibitem[Lassalle et~al.(2018)Lassalle, Champenois, Stout, Debierre, and
  Durt]{lassalleConditionsAntiZenoeffectObservation2018}
E.~Lassalle, C.~Champenois, B.~Stout, V.~Debierre, and T.~Durt.
\newblock Conditions for anti-{{Zeno-effect}} observation in free-space atomic
  radiative decay.
\newblock \emph{Physical Review A}, 97\penalty0 (6), 2018.
\newblock ISSN 2469-9926.
\newblock \doi{10.1103/PhysRevA.97.062122}.

\bibitem[Zhang et~al.(2018)Zhang, Jing, Wang, and
  Zhu]{zhangCriterionQuantumZeno2018}
J.-M. Zhang, J.~Jing, L.-G. Wang, and S.-Y. Zhu.
\newblock Criterion for quantum {{Zeno}} and anti-{{Zeno}} effects.
\newblock \emph{Physical Review A}, 98\penalty0 (1), 2018.
\newblock ISSN 2469-9926.
\newblock \doi{10.1103/PhysRevA.98.012135}.

\bibitem[Majeed and Chaudhry(2018)]{majeedQuantumZenoAntiZeno2018}
Mehwish Majeed and Adam~Zaman Chaudhry.
\newblock The quantum {{Zeno}} and anti-{{Zeno}} effects with non-selective
  projective measurements.
\newblock \emph{Scientific Reports}, 8\penalty0 (1):\penalty0 14887, October
  2018.
\newblock ISSN 2045-2322.
\newblock \doi{10.1038/s41598-018-33181-9}.

\bibitem[He et~al.(2019)He, Duan, Wang, and Chen]{heQuantumZenoEffect2019}
Shu He, Li-Wei Duan, Chen Wang, and Qing-Hu Chen.
\newblock Quantum {{Zeno}} effect in a circuit-{{QED}} system.
\newblock \emph{Physical Review A}, 99\penalty0 (5):\penalty0 052101, May 2019.
\newblock ISSN 2469-9926, 2469-9934.
\newblock \doi{10.1103/PhysRevA.99.052101}.

\bibitem[Na(2021)]{naRevisitingQuantumZeno2021}
Kyungsun Na.
\newblock Revisiting quantum {{Zeno}} effect and anti-{{Zeno}} effect:
  {{Universality}} vs non-universality.
\newblock \emph{Journal of Mathematical Physics}, 62\penalty0 (12):\penalty0
  122101, December 2021.
\newblock ISSN 0022-2488.
\newblock \doi{10.1063/5.0050473}.

\bibitem[Erez et~al.(2004)Erez, Aharonov, Reznik, and
  Vaidman]{erezCorrectingQuantumErrors2004}
Noam Erez, Yakir Aharonov, Benni Reznik, and Lev Vaidman.
\newblock Correcting quantum errors with the {{Zeno}} effect.
\newblock \emph{Physical Review A}, 69\penalty0 (6):\penalty0 062315, June
  2004.
\newblock ISSN 1050-2947, 1094-1622.
\newblock \doi{10.1103/PhysRevA.69.062315}.

\bibitem[Kwiat et~al.(1995)Kwiat, Weinfurter, Herzog, Zeilinger, and
  Kasevich]{kwiatInteractionFreeMeasurement1995}
Paul Kwiat, Harald Weinfurter, Thomas Herzog, Anton Zeilinger, and Mark~A.
  Kasevich.
\newblock Interaction-{{Free Measurement}}.
\newblock \emph{Physical Review Letters}, 74\penalty0 (24):\penalty0
  4763--4766, June 1995.
\newblock ISSN 0031-9007, 1079-7114.
\newblock \doi{10.1103/PhysRevLett.74.4763}.

\bibitem[Kwiat et~al.(1999)Kwiat, White, Mitchell, Nairz, Weihs, Weinfurter,
  and Zeilinger]{kwiatHighefficiencyQuantumInterrogation1999}
P.~G. Kwiat, A.~G. White, J.~R. Mitchell, O.~Nairz, G.~Weihs, H.~Weinfurter,
  and A.~Zeilinger.
\newblock High-efficiency quantum interrogation measurements via the quantum
  {{Zeno}} effect.
\newblock \emph{Physical Review Letters}, 83\penalty0 (23):\penalty0
  4725--4728, December 1999.
\newblock ISSN 0031-9007, 1079-7114.
\newblock \doi{10.1103/PhysRevLett.83.4725}.

\bibitem[Balzer et~al.(2000)Balzer, Huesmann, Neuhauser, and
  Toschek]{balzerQuantumZenoEffect2000}
{\relax Chr}.~Balzer, R.~Huesmann, W.~Neuhauser, and P.~E. Toschek.
\newblock The quantum {{Zeno}} effect -- evolution of an atom impeded by
  measurement.
\newblock \emph{Optics Communications}, 180\penalty0 (1):\penalty0 115--120,
  June 2000.
\newblock ISSN 0030-4018.
\newblock \doi{10.1016/S0030-4018(00)00716-1}.

\bibitem[Balzer et~al.(2002)Balzer, Hannemann, Rei{\ss}, Wunderlich, Neuhauser,
  and Toschek]{balzerRelaxationlessDemonstrationQuantum2002}
{\relax Chr}.~Balzer, {\relax Th}.~Hannemann, D.~Rei{\ss}, {\relax
  Chr}.~Wunderlich, W.~Neuhauser, and P.~E. Toschek.
\newblock A relaxationless demonstration of the {{Quantum Zeno}} paradox on an
  individual atom.
\newblock \emph{Optics Communications}, 211\penalty0 (1):\penalty0 235--241,
  October 2002.
\newblock ISSN 0030-4018.
\newblock \doi{10.1016/S0030-4018(02)01859-X}.

\bibitem[Gambetta et~al.(2008)Gambetta, Blais, Boissonneault, Houck, Schuster,
  and Girvin]{gambettaQuantumTrajectoryApproach2008}
Jay Gambetta, Alexandre Blais, M.~Boissonneault, A.~A. Houck, D.~I. Schuster,
  and S.~M. Girvin.
\newblock Quantum trajectory approach to circuit {{QED}}: {{Quantum}} jumps and
  the {{Zeno}} effect.
\newblock \emph{Physical Review A}, 77\penalty0 (1):\penalty0 012112, January
  2008.
\newblock ISSN 1050-2947, 1094-1622.
\newblock \doi{10.1103/PhysRevA.77.012112}.

\bibitem[Vijay et~al.(2011)Vijay, Slichter, and
  Siddiqi]{vijayObservationQuantumJumps2011}
R.~Vijay, D.~H. Slichter, and I.~Siddiqi.
\newblock Observation of {{Quantum Jumps}} in a {{Superconducting Artificial
  Atom}}.
\newblock \emph{Physical Review Letters}, 106\penalty0 (11):\penalty0 110502,
  March 2011.
\newblock ISSN 0031-9007, 1079-7114.
\newblock \doi{10.1103/PhysRevLett.106.110502}.

\bibitem[Slichter et~al.(2016)Slichter, M{\"u}ller, Vijay, Weber, Blais, and
  Siddiqi]{slichterQuantumZenoEffect2016}
D.~H. Slichter, C.~M{\"u}ller, R.~Vijay, S.~J. Weber, A.~Blais, and I.~Siddiqi.
\newblock Quantum {{Zeno}} effect in the strong measurement regime of circuit
  quantum electrodynamics.
\newblock \emph{New Journal of Physics}, 18\penalty0 (5):\penalty0 053031, May
  2016.
\newblock ISSN 1367-2630.
\newblock \doi{10.1088/1367-2630/18/5/053031}.

\bibitem[Guttel et~al.(2026)Guttel, Gov, Netzer, Goldblatt, Hazanov, Joshi,
  Romito, Gefen, Kumar, Snizhko, Lafont, and
  Rosenblum]{guttelGraduallyOpeningSchrodingers2026}
Barkay Guttel, Danielle Gov, Noam Netzer, Uri Goldblatt, Sergey Hazanov,
  Lalit~M. Joshi, Alessandro Romito, Yuval Gefen, Parveen Kumar, Kyrylo
  Snizhko, Fabien Lafont, and Serge Rosenblum.
\newblock Gradually opening {{Schr\"odinger}}'s box reveals a cascade of sharp
  dynamical transitions.
\newblock arXiv:2602.02672, February 2026.
\newblock \doi{10.48550/arXiv.2602.02672}.

\bibitem[Mohseninia et~al.(2020)Mohseninia, Yang, Siddiqi, Jordan, and
  Dressel]{mohseniniaAlwaysQuantumErrorTracking2020}
Razieh Mohseninia, Jing Yang, Irfan Siddiqi, Andrew~N. Jordan, and Justin
  Dressel.
\newblock Always-{{On Quantum Error Tracking}} with {{Continuous Parity
  Measurements}}.
\newblock \emph{Quantum}, 4:\penalty0 358, November 2020.
\newblock ISSN 2521-327X.
\newblock \doi{10.22331/q-2020-11-04-358}.

\bibitem[Livingston et~al.(2022)Livingston, Blok, Flurin, Dressel, Jordan, and
  Siddiqi]{livingstonExperimentalDemonstrationContinuous2022}
William~P. Livingston, Machiel~S. Blok, Emmanuel Flurin, Justin Dressel,
  Andrew~N. Jordan, and Irfan Siddiqi.
\newblock Experimental demonstration of continuous quantum error correction.
\newblock \emph{Nature Communications}, 13\penalty0 (1):\penalty0 2307, April
  2022.
\newblock ISSN 2041-1723.
\newblock \doi{10.1038/s41467-022-29906-0}.

\bibitem[Blumenthal et~al.(2022)Blumenthal, Mor, Diringer, Martin, Lewalle,
  Burgarth, Whaley, and
  {Hacohen-Gourgy}]{blumenthalDemonstrationUniversalControl2022}
E.~Blumenthal, C.~Mor, A.~A. Diringer, L.~S. Martin, P.~Lewalle, D.~Burgarth,
  K.~B. Whaley, and S.~{Hacohen-Gourgy}.
\newblock Demonstration of universal control between non-interacting qubits
  using the {{Quantum Zeno}} effect.
\newblock \emph{npj Quantum Information}, 8\penalty0 (1):\penalty0 88, July
  2022.
\newblock ISSN 2056-6387.
\newblock \doi{10.1038/s41534-022-00594-4}.

\bibitem[Gaikwad et~al.(2024)Gaikwad, Kowsari, Brame, Song, Zhang, Esposito,
  Ranadive, Cappelli, Roch, {Levenson-Falk}, and
  Murch]{gaikwadEntanglementAssistedProbe2024}
Chandrashekhar Gaikwad, Daria Kowsari, Carson Brame, Xingrui Song, Haimeng
  Zhang, Martina Esposito, Arpit Ranadive, Giulio Cappelli, Nicolas Roch,
  Eli~M. {Levenson-Falk}, and Kater~W. Murch.
\newblock Entanglement assisted probe of the non-{{Markovian}} to {{Markovian}}
  transition in open quantum system dynamics.
\newblock \emph{Physical Review Letters}, 132\penalty0 (20):\penalty0 200401,
  May 2024.
\newblock \doi{10.1103/PhysRevLett.132.200401}.

\bibitem[Cramer et~al.(2016)Cramer, Kalb, Rol, Hensen, Blok, Markham, Twitchen,
  Hanson, and Taminiau]{cramerRepeatedQuantumError2016}
J.~Cramer, N.~Kalb, M.~A. Rol, B.~Hensen, M.~S. Blok, M.~Markham, D.~J.
  Twitchen, R.~Hanson, and T.~H. Taminiau.
\newblock Repeated quantum error correction on a continuously encoded qubit by
  real-time feedback.
\newblock \emph{Nature Communications}, 7\penalty0 (1):\penalty0 11526, May
  2016.
\newblock ISSN 2041-1723.
\newblock \doi{10.1038/ncomms11526}.

\bibitem[Ruseckas and Kaulakys(2006)]{ruseckasQuantumTrajectoryMethod2006}
J.~Ruseckas and B.~Kaulakys.
\newblock Quantum trajectory method for the quantum {{Zeno}} and anti-{{Zeno}}
  effects.
\newblock \emph{Physical Review A}, 73\penalty0 (5):\penalty0 052101, May 2006.
\newblock \doi{10.1103/PhysRevA.73.052101}.

\bibitem[Gardiner and Collett(1985)]{gardinerInputOutputDamped1985}
C.~W. Gardiner and M.~J. Collett.
\newblock Input and output in damped quantum systems: {{Quantum}} stochastic
  differential equations and the master equation.
\newblock \emph{Physical Review A}, 31\penalty0 (6):\penalty0 3761--3774, June
  1985.
\newblock \doi{10.1103/PhysRevA.31.3761}.

\bibitem[Barchielli(1986)]{barchielliMeasurementTheoryStochastic1986}
Alberto Barchielli.
\newblock Measurement theory and stochastic differential equations in quantum
  mechanics.
\newblock \emph{Physical Review A}, 34\penalty0 (3):\penalty0 1642--1649,
  September 1986.
\newblock \doi{10.1103/PhysRevA.34.1642}.

\bibitem[Di{\'o}si(1988)]{diosiContinuousQuantumMeasurement1988}
L.~Di{\'o}si.
\newblock Continuous quantum measurement and it\^o formalism.
\newblock \emph{Physics Letters A}, 129\penalty0 (8):\penalty0 419--423, June
  1988.
\newblock ISSN 0375-9601.
\newblock \doi{10.1016/0375-9601(88)90309-X}.

\bibitem[Belavkin(1992)]{belavkinQuantumContinualMeasurements1992}
V.~P. Belavkin.
\newblock Quantum continual measurements and a posteriori collapse on {{CCR}}.
\newblock \emph{Communications in Mathematical Physics}, 146\penalty0
  (3):\penalty0 611--635, June 1992.
\newblock ISSN 1432-0916.
\newblock \doi{10.1007/BF02097018}.

\bibitem[Carmichael(1993{\natexlab{a}})]{carmichaelQuantumTrajectoryTheory1993}
H.~J. Carmichael.
\newblock Quantum trajectory theory for cascaded open systems.
\newblock \emph{Physical Review Letters}, 70\penalty0 (15):\penalty0
  2273--2276, April 1993{\natexlab{a}}.
\newblock ISSN 0031-9007.
\newblock \doi{10.1103/PhysRevLett.70.2273}.

\bibitem[Wiseman and Milburn(1993)]{wisemanQuantumTheoryFieldquadrature1993}
H.~M. Wiseman and G.~J. Milburn.
\newblock Quantum theory of field-quadrature measurements.
\newblock \emph{Physical Review A}, 47\penalty0 (1):\penalty0 642--662, January
  1993.
\newblock ISSN 1050-2947, 1094-1622.
\newblock \doi{10.1103/PhysRevA.47.642}.

\bibitem[Gardiner(2000)]{gardinerQuantumNoiseHandbook2000}
C.~W. Gardiner.
\newblock \emph{Quantum Noise: A Handbook of {{Markovian}} and
  Non-{{Markovian}} Quantum Stochastic Methods with Applications to Quantum
  Optics}.
\newblock Springer Series in Synergetics. Springer, Berlin ;, 2nd enlarged ed.
  edition, 2000.
\newblock ISBN 978-3-540-66571-7.
\newblock \doi{10.1007/978-3-662-04103-1}.

\bibitem[Korotkov(2001{\natexlab{a}})]{korotkovSelectiveQuantumEvolution2001}
Alexander~N. Korotkov.
\newblock Selective quantum evolution of a qubit state due to continuous
  measurement.
\newblock \emph{Physical Review B}, 63\penalty0 (11):\penalty0 115403, February
  2001{\natexlab{a}}.
\newblock \doi{10.1103/PhysRevB.63.115403}.

\bibitem[Carmichael(1993{\natexlab{b}})]{carmichaelOpenSystemsApproach1993}
Howard Carmichael.
\newblock \emph{An Open Systems Approach to Quantum Optics: Lectures Presented
  at the {{Universit\'e}} Libre de {{Bruxelles}}, {{October}} 28 to
  {{November}} 4, 1991}.
\newblock Number m18 in Lecture Notes in Physics. Springer-Verlag, Berlin,
  1993{\natexlab{b}}.
\newblock ISBN 978-3-540-56634-2.
\newblock \doi{10.1007/978-3-540-47620-7}.

\bibitem[Wiseman and Milburn(2009)]{wisemanQuantumMeasurementControl2009}
Howard~M. Wiseman and Gerard~J. Milburn.
\newblock \emph{Quantum {{Measurement}} and {{Control}}}.
\newblock Cambridge University Press, November 2009.
\newblock ISBN 978-1-139-48291-2.
\newblock \doi{10.1017/CBO9780511813948}.

\bibitem[Korotkov(2011)]{korotkovQuantumBayesianApproach2011}
Alexander~N. Korotkov.
\newblock Quantum {{Bayesian}} approach to circuit {{QED}} measurement.
\newblock \emph{arXiv:1111.4016 [cond-mat, physics:quant-ph]}, November 2011.
\newblock \doi{10.48550/arXiv.1111.4016}.

\bibitem[Korotkov(2016)]{korotkovQuantumBayesianApproach2016}
Alexander~N. Korotkov.
\newblock Quantum {{Bayesian}} approach to circuit {{QED}} measurement with
  moderate bandwidth.
\newblock \emph{Physical Review A}, 94\penalty0 (4):\penalty0 042326, October
  2016.
\newblock \doi{10.1103/PhysRevA.94.042326}.

\bibitem[Streed et~al.(2006)Streed, Mun, Boyd, Campbell, Medley, Ketterle, and
  Pritchard]{streedContinuousPulsedQuantum2006}
Erik~W. Streed, Jongchul Mun, Micah Boyd, Gretchen~K. Campbell, Patrick Medley,
  Wolfgang Ketterle, and David~E. Pritchard.
\newblock Continuous and {{Pulsed Quantum Zeno Effect}}.
\newblock \emph{Physical Review Letters}, 97\penalty0 (26):\penalty0 260402,
  December 2006.
\newblock ISSN 0031-9007, 1079-7114.
\newblock \doi{10.1103/PhysRevLett.97.260402}.

\bibitem[Kakuyanagi et~al.(2015)Kakuyanagi, Baba, Matsuzaki, Nakano, Saito, and
  Semba]{kakuyanagiObservationQuantumZeno2015}
K.~Kakuyanagi, T.~Baba, Y.~Matsuzaki, H.~Nakano, S.~Saito, and K.~Semba.
\newblock Observation of quantum {{Zeno}} effect in a superconducting flux
  qubit.
\newblock \emph{New Journal of Physics}, 17\penalty0 (6):\penalty0 063035, June
  2015.
\newblock ISSN 1367-2630.
\newblock \doi{10.1088/1367-2630/17/6/063035}.

\bibitem[Xiao and Jones(2006)]{xiaoNMRAnaloguesQuantum2006}
Li~Xiao and Jonathan~A. Jones.
\newblock {{NMR}} analogues of the quantum {{Zeno}} effect.
\newblock \emph{Physics Letters A}, 359\penalty0 (5):\penalty0 424--427,
  December 2006.
\newblock ISSN 0375-9601.
\newblock \doi{10.1016/j.physleta.2006.06.086}.

\bibitem[{Palacios-Laloy} et~al.(2010){Palacios-Laloy}, Mallet, Nguyen, Bertet,
  Vion, Esteve, and Korotkov]{palacios-laloyExperimentalViolationBell2010}
Agustin {Palacios-Laloy}, Fran{\c c}ois Mallet, Fran{\c c}ois Nguyen, Patrice
  Bertet, Denis Vion, Daniel Esteve, and Alexander~N. Korotkov.
\newblock Experimental violation of a {{Bell}}'s inequality in time with weak
  measurement.
\newblock \emph{Nature Physics}, 6\penalty0 (6):\penalty0 442--447, June 2010.
\newblock ISSN 1745-2481.
\newblock \doi{10.1038/nphys1641}.

\bibitem[Li et~al.(2023)Li, Wang, Luo, Vemuri, and
  Joglekar]{liUnificationQuantumZenoanti2023}
Jiaming Li, Tishuo Wang, Le~Luo, Sreya Vemuri, and Yogesh~N. Joglekar.
\newblock Unification of quantum {{Zeno--anti Zeno}} effects and parity-time
  symmetry breaking transitions.
\newblock \emph{Physical Review Research}, 5\penalty0 (2):\penalty0 023204,
  June 2023.
\newblock \doi{10.1103/PhysRevResearch.5.023204}.

\bibitem[Harrington et~al.(2017)Harrington, Monroe, and
  Murch]{harringtonQuantumZenoEffects2017}
P.~M. Harrington, J.~T. Monroe, and K.~W. Murch.
\newblock Quantum {{Zeno Effects}} from {{Measurement Controlled Qubit-Bath
  Interactions}}.
\newblock \emph{Physical Review Letters}, 118\penalty0 (24):\penalty0 240401,
  June 2017.
\newblock \doi{10.1103/PhysRevLett.118.240401}.

\bibitem[Snizhko et~al.(2020)Snizhko, Kumar, and
  Romito]{snizhkoQuantumZenoEffect2020}
Kyrylo Snizhko, Parveen Kumar, and Alessandro Romito.
\newblock Quantum {{Zeno}} effect appears in stages.
\newblock \emph{Physical review research}, 2\penalty0 (3), 2020.
\newblock ISSN 2643-1564.
\newblock \doi{10.1103/PhysRevResearch.2.033512}.

\bibitem[Kumari et~al.(2023)Kumari, Rajpoot, Joshi, and
  Jain]{kumariQubitControlUsing2023}
Komal Kumari, Garima Rajpoot, Sandeep Joshi, and Sudhir~Ranjan Jain.
\newblock Qubit control using quantum {{Zeno}} effect: {{Action}} principle
  approach.
\newblock \emph{Annals of physics}, 450:\penalty0 169222--, 2023.
\newblock ISSN 0003-4916.
\newblock \doi{10.1016/j.aop.2023.169222}.

\bibitem[Kumar et~al.(2020)Kumar, Romito, and
  Snizhko]{kumarQuantumZenoEffect2020}
Parveen Kumar, Alessandro Romito, and Kyrylo Snizhko.
\newblock Quantum {{Zeno}} effect with partial measurement and noisy dynamics.
\newblock \emph{Physical Review Research}, 2\penalty0 (4):\penalty0 043420,
  December 2020.
\newblock ISSN 2643-1564.
\newblock \doi{10.1103/PhysRevResearch.2.043420}.

\bibitem[Facchi et~al.(2005)Facchi, Tasaki, Pascazio, Nakazato, Tokuse, and
  Lidar]{facchiControlDecoherenceAnalysis2005}
P.~Facchi, S.~Tasaki, S.~Pascazio, H.~Nakazato, A.~Tokuse, and D.~A. Lidar.
\newblock Control of decoherence: {{Analysis}} and comparison of three
  different strategies.
\newblock \emph{Physical Review A}, 71\penalty0 (2):\penalty0 022302, February
  2005.
\newblock \doi{10.1103/PhysRevA.71.022302}.

\bibitem[Patil et~al.(2015)Patil, Chakram, and
  Vengalattore]{patilMeasurementInducedLocalizationUltracold2015}
Y.~S. Patil, S.~Chakram, and M.~Vengalattore.
\newblock Measurement-{{Induced Localization}} of an {{Ultracold Lattice Gas}}.
\newblock \emph{Physical Review Letters}, 115\penalty0 (14):\penalty0 140402,
  October 2015.
\newblock ISSN 0031-9007, 1079-7114.
\newblock \doi{10.1103/PhysRevLett.115.140402}.

\bibitem[Leghtas et~al.(2015)Leghtas, Touzard, Pop, Kou, Vlastakis, Petrenko,
  Sliwa, Narla, Shankar, Hatridge, Reagor, Frunzio, Schoelkopf, Mirrahimi, and
  Devoret]{leghtasConfiningStateLight2015}
Zaki Leghtas, Steven Touzard, Ioan~M. Pop, Angela Kou, Brian Vlastakis, Andrei
  Petrenko, Katrina~M. Sliwa, Anirudh Narla, Shyam Shankar, Michael~J.
  Hatridge, Matthew Reagor, Luigi Frunzio, Robert~J. Schoelkopf, Mazyar
  Mirrahimi, and Michel~H. Devoret.
\newblock Confining the state of light to a quantum manifold by engineered
  two-photon loss.
\newblock \emph{Science}, 347\penalty0 (6224):\penalty0 853--857, February
  2015.
\newblock ISSN 0036-8075, 1095-9203.
\newblock \doi{10.1126/science.aaa2085}.

\bibitem[Mark et~al.(2020)Mark, Flannigan, Meinert, D'Incao, Daley, and
  N{\"a}gerl]{markInterplayCoherentDissipative2020}
M.~J. Mark, S.~Flannigan, F.~Meinert, J.~P. D'Incao, A.~J. Daley, and H.-C.
  N{\"a}gerl.
\newblock Interplay between coherent and dissipative dynamics of bosonic
  doublons in an optical lattice.
\newblock \emph{Physical Review Research}, 2\penalty0 (4):\penalty0 043050,
  October 2020.
\newblock ISSN 2643-1564.
\newblock \doi{10.1103/PhysRevResearch.2.043050}.

\bibitem[{Kimchi-Schwartz} et~al.(2016){Kimchi-Schwartz}, Martin, Flurin, Aron,
  Kulkarni, Tureci, and
  Siddiqi]{kimchi-schwartzStabilizingEntanglementSymmetrySelective2016}
M.~E. {Kimchi-Schwartz}, L.~Martin, E.~Flurin, C.~Aron, M.~Kulkarni, H.~E.
  Tureci, and I.~Siddiqi.
\newblock Stabilizing {{Entanglement}} via {{Symmetry-Selective Bath
  Engineering}} in {{Superconducting Qubits}}.
\newblock \emph{Physical Review Letters}, 116\penalty0 (24):\penalty0 240503,
  June 2016.
\newblock ISSN 0031-9007, 1079-7114.
\newblock \doi{10.1103/PhysRevLett.116.240503}.

\bibitem[Jacobs and
  Steck(2006)]{jacobsStraightforwardIntroductionContinuous2006}
Kurt Jacobs and Daniel~A. Steck.
\newblock A {{Straightforward Introduction}} to {{Continuous Quantum
  Measurement}}.
\newblock \emph{Contemporary Physics}, 47\penalty0 (5):\penalty0 279--303,
  September 2006.
\newblock ISSN 0010-7514, 1366-5812.
\newblock \doi{10.1080/00107510601101934}.

\bibitem[Korotkov(2001{\natexlab{b}})]{korotkovOutputSpectrumDetector2001}
Alexander~N. Korotkov.
\newblock Output spectrum of a detector measuring quantum oscillations.
\newblock \emph{Physical Review B}, 63\penalty0 (8):\penalty0 085312, February
  2001{\natexlab{b}}.
\newblock \doi{10.1103/PhysRevB.63.085312}.

\bibitem[Trotter(1959)]{trotterProductSemiGroupsOperators1959}
H.~F. Trotter.
\newblock On the {{Product}} of {{Semi-Groups}} of {{Operators}}.
\newblock \emph{Proceedings of the American Mathematical Society}, 10\penalty0
  (4):\penalty0 545--551, 1959.
\newblock ISSN 0002-9939.
\newblock \doi{10.2307/2033649}.

\bibitem[Schulman(1998)]{schulmanContinuousPulsedObservations1998}
L.~S. Schulman.
\newblock Continuous and pulsed observations in the quantum {{Zeno}} effect.
\newblock \emph{Physical Review A}, 57\penalty0 (3):\penalty0 1509--1515, March
  1998.
\newblock ISSN 1050-2947, 1094-1622.
\newblock \doi{10.1103/PhysRevA.57.1509}.

\bibitem[Pellegrini and
  Petruccione(2009)]{pellegriniDiffusionApproximationStochastic2009}
C.~Pellegrini and F.~Petruccione.
\newblock Diffusion approximation of stochastic master equations with jumps.
\newblock \emph{Journal of Mathematical Physics}, 50\penalty0 (12):\penalty0
  122101, December 2009.
\newblock ISSN 0022-2488, 1089-7658.
\newblock \doi{10.1063/1.3263941}.

\bibitem[Chantasri et~al.(2013)Chantasri, Dressel, and
  Jordan]{chantasriActionPrincipleContinuous2013}
A.~Chantasri, J.~Dressel, and A.~N. Jordan.
\newblock Action principle for continuous quantum measurement.
\newblock \emph{Physical Review A}, 88\penalty0 (4):\penalty0 042110, October
  2013.
\newblock ISSN 1050-2947, 1094-1622.
\newblock \doi{10.1103/PhysRevA.88.042110}.

\bibitem[Singh et~al.(2026)Singh, Bhandari, S{\'a}nchez, and
  Jordan]{singhCapturingElectronVirtual2025}
Alok~Nath Singh, Bibek Bhandari, Rafael S{\'a}nchez, and Andrew~N. Jordan.
\newblock Capturing an {{Electron}} in the {{Virtual State}}.
\newblock \emph{New Journal of Physics}, 28\penalty0 (7):\penalty0 074503, July
  2026.
\newblock \doi{10.1088/1367-2630/ae846a}.

\bibitem[Auff{\`e}ves et~al.(2010)Auff{\`e}ves, Gerace, G{\'e}rard, Santos,
  Andreani, and Poizat]{auffevesControllingDynamicsCoupled2010}
A.~Auff{\`e}ves, D.~Gerace, J.-M. G{\'e}rard, M.~Fran{\c c}a Santos, L.~C.
  Andreani, and J.-P. Poizat.
\newblock Controlling the dynamics of a coupled atom-cavity system by pure
  dephasing.
\newblock \emph{Physical Review B}, 81\penalty0 (24):\penalty0 245419, June
  2010.
\newblock ISSN 1098-0121, 1550-235X.
\newblock \doi{10.1103/PhysRevB.81.245419}.

\bibitem[Sete et~al.(2014)Sete, Gambetta, and
  Korotkov]{setePurcellEffectMicrowave2014}
Eyob~A. Sete, Jay~M. Gambetta, and Alexander~N. Korotkov.
\newblock Purcell effect with microwave drive: {{Suppression}} of qubit
  relaxation rate.
\newblock \emph{Physical review. B, Condensed matter and materials physics},
  89\penalty0 (10), 2014.
\newblock ISSN 1098-0121.
\newblock \doi{10.1103/PhysRevB.89.104516}.

\bibitem[Burgarth et~al.(2014)Burgarth, Facchi, Giovannetti, Nakazato,
  Pascazio, and Yuasa]{burgarthExponentialRiseDynamical2014}
Daniel~Klaus Burgarth, Paolo Facchi, Vittorio Giovannetti, Hiromichi Nakazato,
  Saverio Pascazio, and Kazuya Yuasa.
\newblock Exponential rise of dynamical complexity in quantum computing through
  projections.
\newblock \emph{Nature Communications}, 5\penalty0 (1):\penalty0 5173, October
  2014.
\newblock ISSN 2041-1723.
\newblock \doi{10.1038/ncomms6173}.

\bibitem[Alvarez and Suter(2011)]{alvarezDynamicalDecouplingNoise2011}
Gonzalo~A. Alvarez and Dieter Suter.
\newblock Measuring the spectrum of colored noise by dynamical decoupling.
\newblock \emph{Physical Review Letters}, 107\penalty0 (23):\penalty0 230501,
  November 2011.
\newblock ISSN 0031-9007, 1079-7114.
\newblock \doi{10.1103/PhysRevLett.107.230501}.

\bibitem[Yuan et~al.(2022)Yuan, Li, Zhang, Liu, Zhu, Qin, Vitanov, Lin, and
  Du]{yuanPreservingMultilevelQuantum2022}
Xinxing Yuan, Yue Li, Mengxiang Zhang, Chang Liu, Mingdong Zhu, Xi~Qin,
  Nikolay~V. Vitanov, Yiheng Lin, and Jiangfeng Du.
\newblock Preserving multilevel quantum coherence by dynamical decoupling.
\newblock \emph{Physical Review A}, 106\penalty0 (2):\penalty0 022412, August
  2022.
\newblock \doi{10.1103/PhysRevA.106.022412}.

\bibitem[Tripathi et~al.(2025)Tripathi, Goss, Vezvaee, Nguyen, Siddiqi, and
  Lidar]{tripathiQuditDynamicalDecoupling2025}
Vinay Tripathi, Noah Goss, Arian Vezvaee, Long~B. Nguyen, Irfan Siddiqi, and
  Daniel~A. Lidar.
\newblock Qudit {{Dynamical Decoupling}} on a {{Superconducting Quantum
  Processor}}.
\newblock \emph{Physical Review Letters}, 134\penalty0 (5):\penalty0 050601,
  February 2025.
\newblock ISSN 0031-9007, 1079-7114.
\newblock \doi{10.1103/PhysRevLett.134.050601}.

\bibitem[Vezvaee et~al.(2026)Vezvaee, Tripathi, {Morford-Oberst}, Butt,
  Kasatkin, and Lidar]{vezvaeeDemonstrationHighfidelityEntangled2026}
Arian Vezvaee, Vinay Tripathi, Mario {Morford-Oberst}, Friederike Butt, Victor
  Kasatkin, and Daniel~A. Lidar.
\newblock Demonstration of high-fidelity entangled logical qubits using
  transmons.
\newblock \emph{Nature Communications}, 17\penalty0 (1):\penalty0 3281, 2026.
\newblock \doi{10.1038/s41467-026-70011-3}.

\bibitem[Burgarth et~al.(2020)Burgarth, Facchi, Nakazato, Pascazio, and
  Yuasa]{burgarthQuantumZenoDynamics2020}
Daniel Burgarth, Paolo Facchi, Hiromichi Nakazato, Saverio Pascazio, and Kazuya
  Yuasa.
\newblock Quantum {{Zeno Dynamics}} from {{General Quantum Operations}}.
\newblock \emph{Quantum}, 4:\penalty0 289, July 2020.
\newblock \doi{10.22331/q-2020-07-06-289}.

\bibitem[Facchi et~al.(2004)Facchi, Lidar, and
  Pascazio]{facchiUnificationDynamicalDecoupling2004}
P.~Facchi, D.~A. Lidar, and S.~Pascazio.
\newblock Unification of dynamical decoupling and the quantum {{Zeno}} effect.
\newblock \emph{Physical Review A}, 69\penalty0 (3):\penalty0 032314, March
  2004.
\newblock \doi{10.1103/PhysRevA.69.032314}.

\bibitem[Khezri et~al.(2015)Khezri, Dressel, and
  Korotkov]{khezriQubitMeasurementError2015}
Mostafa Khezri, Justin Dressel, and Alexander~N. Korotkov.
\newblock Qubit measurement error from coupling with a detuned neighbor in
  circuit {{QED}}.
\newblock \emph{Physical Review A}, 92\penalty0 (5):\penalty0 052306, November
  2015.
\newblock ISSN 1050-2947, 1094-1622.
\newblock \doi{10.1103/PhysRevA.92.052306}.

\bibitem[Gautier et~al.(2023)Gautier, Mirrahimi, and
  Sarlette]{gautierDesigningHighFidelityZeno2023}
Ronan Gautier, Mazyar Mirrahimi, and Alain Sarlette.
\newblock Designing {{High-Fidelity Zeno Gates}} for {{Dissipative Cat
  Qubits}}.
\newblock \emph{PRX Quantum}, 4\penalty0 (4):\penalty0 040316, October 2023.
\newblock ISSN 2691-3399.
\newblock \doi{10.1103/PRXQuantum.4.040316}.

\bibitem[Barontini et~al.(2013)Barontini, Labouvie, Stubenrauch, Vogler,
  Guarrera, and Ott]{barontiniControllingDynamicsOpen2013}
G.~Barontini, R.~Labouvie, F.~Stubenrauch, A.~Vogler, V.~Guarrera, and H.~Ott.
\newblock Controlling the dynamics of an open many-body quantum system with
  localized dissipation.
\newblock \emph{Physical Review Letters}, 110\penalty0 (3):\penalty0 035302,
  January 2013.
\newblock ISSN 0031-9007, 1079-7114.
\newblock \doi{10.1103/PhysRevLett.110.035302}.

\end{thebibliography}

\end{document}